\documentclass[prd,showkeys,twocolumn,amsmath,amssymb]{revtex4}
\usepackage{graphicx}
\usepackage{epstopdf}
\usepackage{subfigure}
\usepackage{bm}
\usepackage[colorlinks,citecolor=blue,anchorcolor=red,menucolor=red,linkcolor=red,filecolor=red,runcolor=red,urlcolor=blue,frenchlinks=red]{hyperref}
\usepackage{ulem}
\usepackage{color}
\usepackage{multirow}

\makeatletter
\renewcommand{\@thesubfigure}{\hskip\subfiglabelskip}
\makeatother

\def\f(s){\left[(\alpha+\beta)m_c^2-\alpha\beta s\right]}

\allowdisplaybreaks[3]
\begin{document}
\title{$S$- and $P$-wave fully-strange tetraquark states from QCD sum rules}
%

\author{Niu Su}
\author{Hua-Xing Chen}
\email{hxchen@seu.edu.cn}

\affiliation{
School of Physics, Southeast University, Nanjing 210094, China
}
\begin{abstract}
We apply the QCD sum rule method to systematically study the $S$- and $P$-wave fully-strange tetraquark states within the diquark-antidiquark picture. We systematically construct their interpolating currents by explicitly adding the covariant derivative operator. Our results suggest that the $f_0(2100)$, $X(2063)$, and $f_2(2010)$ may be explained as the $S$-wave $s s \bar s \bar s$ tetraquark states with the quantum numbers $J^{PC} = 0^{++}$, $1^{+-}$, and $2^{++}$, respectively. Our results also suggest that both the $X(2370)$ and $X(2500)$ may be explained as the $P$-wave $s s \bar s \bar s$ tetraquark states of $J^{PC} = 0^{-+}$, and both the $\phi(2170)$ and $X(2400)$ may be explained as the $P$-wave $s s \bar s \bar s$ tetraquark states of $J^{PC} = 1^{--}$. The masses of the $s s \bar s \bar s$ tetraquark states with the exotic quantum number $J^{PC} = 1^{-+}$ are extracted from two non-correlated currents to be $2.45^{+0.20}_{-0.25}$~GeV and $2.49^{+0.21}_{-0.25}$~GeV.
\end{abstract}
\maketitle
\pagenumbering{arabic}

\section{Introduction}
\label{sec:intro}

In the past twenty years a lot of exotic hadrons were observed in experiments~\cite{pdg}, which bring us the renaissance of the hadron spectroscopy~\cite{Chen:2016qju,Liu:2019zoy,Chen:2022asf,Lebed:2016hpi,Esposito:2016noz,Hosaka:2016pey,Guo:2017jvc,Ali:2017jda,Olsen:2017bmm,Karliner:2017qhf,Bass:2018xmz,Brambilla:2019esw,Guo:2019twa,Ketzer:2019wmd,Yang:2020atz,Roberts:2021nhw,Fang:2021wes,Jin:2021vct,JPAC:2021rxu,Meng:2022ozq,Brambilla:2022ura}. Some of them are good candidates for the fully-strange tetraquark states, which contain many strangeness components. Experimentally, their widths are possibly not too broad, so they are capable of being observed. Theoretically, their internal structures are simpler than other multiquark states due to the Pauli principle restricting on identical strangeness quarks/antiquarks. This limits their potential number, and makes them easier to be observed.

There have been some rich-strangeness signals observed at around 2.0~GeV, for examples,
\begin{itemize}

\item In 2006 the BaBar collaboration observed the $\phi(2170)/Y(2175)$ in the $e^+ e^- \to \phi f_0(980)$ process~\cite{BaBar:2006gsq}.

\item In 2010 the BESIII collaboration observed the $X(2120)$ and $X(2370)$ in the $\pi \pi \eta^\prime$ invariant mass spectrum of the $J/\psi \to \gamma \pi \pi \eta^\prime$ decay~\cite{BESIII:2010gmv}. Later in 2019 they confirmed the $X(2370)$ in the $K \bar K \eta^\prime$ invariant mass spectrum of the $J/\psi \to \gamma K \bar K \eta^\prime$ decay, but they did not observe the $X(2120)$ in this process~\cite{BESIII:2019wkp}. This suggests that the $X(2370)$ contains more strangeness components.

\item In 2016 the BESIII collaboration performed a partial wave analysis of the $J/\psi \to \gamma\phi\phi$ decay~\cite{BESIII:2016qzq}, where they observed one scalar resonance $f_0(2100)$, one pseudoscalar resonance $X(2500)$, as well as three tensor resonances $f_2(2010)$, $f_2(2300)$, and $f_2(2340)$ in the $\phi\phi$ invariant mass spectrum.

\item In 2018 the BESIII collaboration observed the $X(2063)$ in the $\phi\eta^\prime$ invariant mass spectrum of the $J/\psi \to \phi \eta \eta^\prime$ decay~\cite{BESIII:2018zbm}.

\end{itemize}
With a large amount of $J/\psi$ sample, the BESIII collaboration are still carefully examining the physics happening at around 2.0~GeV, and more rich-strangeness signals are expected in the coming future. Such experiments can also be performed by the Belle-II, COMPASS, and GlueX collaborations, etc.

In the past years we have applied the QCD sum rule method to study the $s s \bar s \bar s$ tetraquark states, separately for the states with the quantum numbers $J^{PC} = 0^{-+}/1^{+-}/1^{--}/4^{+-}$~\cite{Chen:2008ej,Chen:2018kuu,Cui:2019roq,Dong:2020okt,Dong:2022otb}. Relevant QCD sum rule studies and quark model calculations can be found in Refs.~\cite{Wang:2006ri,Wang:2019nln,Agaev:2019coa,Azizi:2019ecm,Pimikov:2022brd,Liu:2020lpw,Lu:2019ira,Deng:2010zzd,Drenska:2008gr,Ebert:2008id}. Especially, a thorough quark model calculation was performed in Ref.~\cite{Liu:2020lpw} to systematically study the $1S$-, $1P$-, and $2S$-wave $s s \bar s \bar s$ tetraquark states. We find it useful to perform a similar QCD sum rule study, so in this paper we shall systematically study the $1S$- and $1P$-wave $s s \bar s \bar s$ tetraquark states using the QCD sum rule method. For simplicity, we shortly denote them as the $S$- and $P$-wave states.

\begin{figure}[hbt]
\begin{center}
\includegraphics[width=0.2\textwidth]{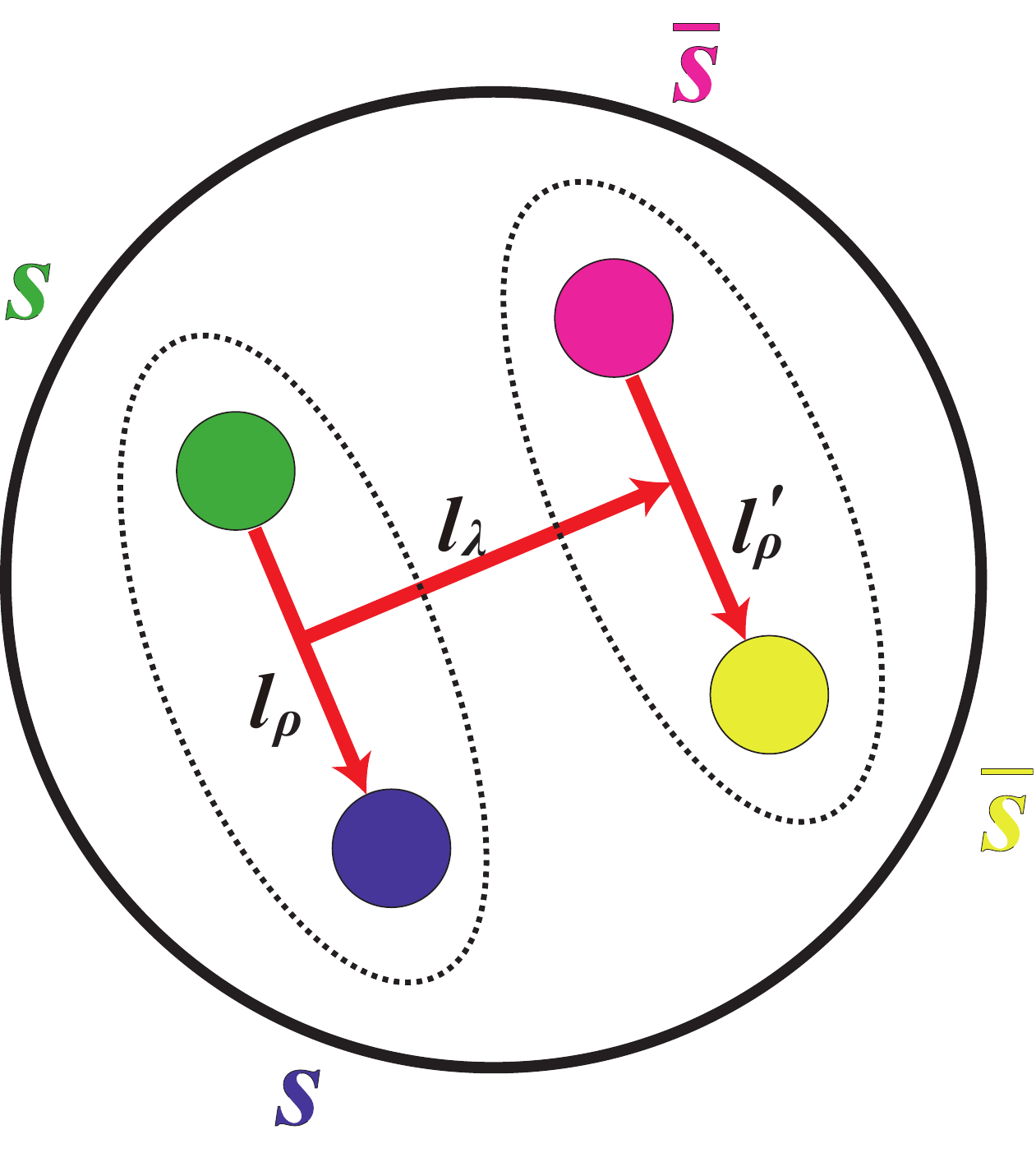}
\caption{Relative coordinates $\vec \lambda$ and $\vec \rho/\vec \rho^{\,\prime}$ for the diquark-antidiquark system. We use $l_\lambda$ to denote the orbital angular momentum between the diquark and antidiquark, and $l_\rho/l_\rho^\prime$ to denote the orbital angular momentum inside the diquark/antidiquark.}
\label{fig:orbit}
\end{center}
\end{figure}

In this paper we shall work within the diquark-antidiquark picture, where the orbital angular momentum can be between the diquark and antidiquark, or it can also be inside the diquark/antidiquark, as depicted in Fig.~\ref{fig:orbit}. We call the former $\lambda$-mode excitation and the latter $\rho$-mode excitation. As already classified in Ref.~\cite{Liu:2020lpw}, there are altogether four $S$-wave $ss \bar s \bar s$ states, eight $P$-wave states of the $\lambda$-mode, and twelve $P$-wave states of the $\rho$-mode. We shall systematically construct their interpolating currents by explicitly adding the covariant derivative operator, based on which we shall perform a systematical QCD sum rule study. Note that the tetraquark currents without derivatives were used in our previous QCD sum rule studies~\cite{Chen:2008ej,Chen:2018kuu,Cui:2019roq,Dong:2020okt}.

Before doing this, we note that the $s s \bar s \bar s$ tetraquark states are just possible explanations for the rich-strangeness signals observed at around 2.0~GeV, and there exist many other possibilities, for examples,
\begin{itemize}

\item The $\phi(2170)/Y(2175)$ was explained in Refs.~\cite{Napsuciale:2007wp,GomezAvila:2007ru,MartinezTorres:2008gy,AlvarezRuso:2009xn,Coito:2009na,Xie:2010ig} as a dynamically generated state in the $\phi K \bar K/\phi \pi \pi$ systems. Besides, it can also be explained as the $2^3D_1$ $s\bar{s}$ meson~\cite{Ding:2007pc,Pang:2019ttv,Pang:2019ovr,Wang:2021abg}, the hidden-strange baryonium state~\cite{Abud:2009rk,Zhao:2013ffn,Zhu:2019ibc}, and the strangeonium hybrid state~\cite{Ding:2006ya}, etc.

\item The $X(2370)$ can be explained as the fourth radial excitation of $\eta(548)/\eta^\prime(958)$~\cite{Yu:2011ta}, the pseudoscalar glueball~\cite{Eshraim:2012jv,Gui:2019dtm}, and their mixing~\cite{Liu:2010tr}. It can also be explained as the compact hexaquark state of $I^GJ^{PC} = 0^+0^{-+}$~\cite{Deng:2012wi} and the hidden-strange baryonium state~\cite{Wan:2021vny}, etc.

\item The $X(2500)$ was explained in Refs.~\cite{Pan:2016bac,Wang:2017iai,Xue:2018jvi,Wang:2020due,Li:2020xzs} as the $4^1S_0$ or $5^1S_0$ $s\bar s$ state.

\item The $X(2063)$ was explained in Ref.~\cite{Wang:2019qyy} as the second radial excitation of $h_1(1380)$ with $I(J^P) = 0(1^+)$.

\end{itemize}
We refer to Refs.~\cite{Morningstar:1999rf,Chen:2005mg,Richards:2010ck,Dudek:2011bn,Gregory:2012hu,Eshraim:2016mds,Eshraim:2019sgr} for more lattice QCD studies and Refs.~\cite{Chen:2011cj,Wang:2012wa,Liang:2013yta,Kozhevnikov:2019lmy,Lebiedowicz:2019jru,Kozhevnikov:2019rma,Chen:2020szc,Sun:2021kka} for some dynamical analyses.

Another relevant exotic state is the $\eta_1(1855)$ recently observed by BESIII in the $\eta \eta^\prime$ invariant mass spectrum of the $J/\psi \to \gamma \eta \eta^\prime$ decay~\cite{BESIII:2022riz,BESIII:2022qzu}. This resonance has the exotic quantum number $I^GJ^{PC} = 0^+1^{-+}$, which can not be accessed by conventional $\bar q q$ mesons. It may be explained as the hybrid meson~\cite{Chen:2010ic,Huang:2010dc,Chen:2022qpd,Qiu:2022ktc,Shastry:2022mhk} and the $K \bar K_1(1400)$ hadronic molecule~\cite{Dong:2022cuw,Yang:2022lwq,Wan:2022xkx,Zhang:2019ykd}, etc. Besides, the $\eta_1(1855)$ may also be explained as the $qs\bar q \bar s$ ($q=u/d$) tetraquark state of $I^GJ^{PC} = 0^+1^{-+}$~\cite{Chen:2008ne,Chen:2008qw}. Based on this interpretation, one naturally expect the existence of the $s s \bar s \bar s$ tetraquark state with $I^GJ^{PC} = 0^+1^{-+}$, which we shall pay special attention to in the present study.

This paper is organized as follows. In Sec.~\ref{sec:current}, we systematically construct the $S$- and $P$-wave fully-strange tetraquark states as well as their corresponding interpolating currents. We use these currents to perform QCD sum rule analyses in Sec.~\ref{sec:sumrule}, and the obtained results are summarized and discussed in Sec.~\ref{sec:summary}.

\section{Phenomenological analyses}
\label{sec:current}

In this section we follow Ref.~\cite{Liu:2020lpw} to construct the $S$- and $P$-wave fully-strange tetraquark states. We shall also construct their corresponding fully-strange tetraquark currents by explicitly adding the covariant derivative operator $D_\alpha = \partial_\alpha + i g_s A_\alpha$, so that these currents behave well under the Lorentz transformation. We shall work within the diquark-antidiquark picture in the present study.

\begin{figure*}[hbt]
\begin{center}
\includegraphics[width=0.71\textwidth]{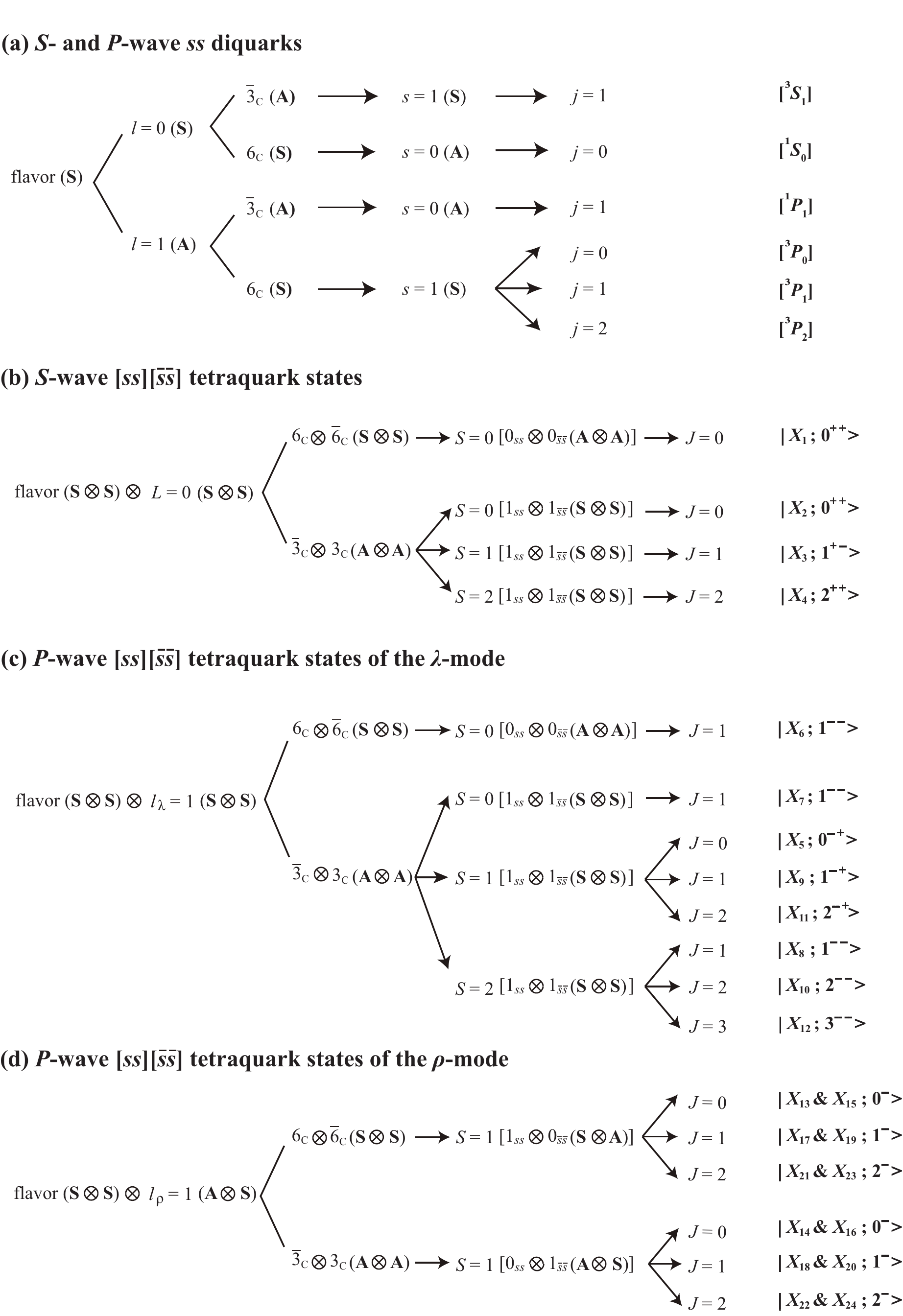}
\caption{Categorization of the $S$- and $P$-wave $ss$ diquarks as well as the $S$- and $P$-wave $s s \bar s \bar s$ tetraquark states. One needs to further reorganize the $S$- and $P$-wave diquarks/antidiquarks in order to obtain the two states $|X_{13} ; J^{PC} = 0^{--}\rangle$ and $|X_{15} ; J^{PC} = 0^{-+}\rangle$, as well as all the other $P$-wave $s s \bar s \bar s$ tetraquark states of the $\rho$-mode.}
\label{fig:category}
\end{center}
\end{figure*}

To start with, we investigate the $ss$ diquark composed of two identical strange quarks with the $symmetric$ flavor structure. According to the Pauli principle, the two strange quarks should be totally $antisymmetric$. As depicted in Fig.~\ref{fig:category}, we can construct two $S$-wave $ss$ diquarks with the $symmetric$ orbital structure:
\begin{enumerate}

\item We use $^{2s+1}l_j =$ $^3S_1$ to denote the $S$-wave $ss$ diquark of $j^P = 1^+$ and the color representation ${\bf \bar 3}_c$, where $s$, $l$, and $j$ are its spin, orbital, and total angular momenta, respectively. The corresponding antidiquark is denote as $^{2{\bar s}+1}{\bar l}_{\bar j} =$ $^{\bar 3}{\bar S}_{\bar 1}$, where $\bar s$, $\bar l$, and $\bar j$ are its spin, orbital, and total angular momenta, respectively. This $^3S_1$ diquark has the $symmetric$ spin and $antisymmetric$ color structures, and its corresponding diquark field is
\begin{equation}
s_a^T C \gamma_\mu s_b \, ,
\end{equation}
where $a$ and $b$ are color indices, $C = i\gamma_2 \gamma_0$ is the charge-conjugation operator, and the sum over repeated indices is taken. The superscript $T$ represents the transpose of the Dirac index only, while the color index is not transposed.

\item We use $^1S_0$ to denote the $S$-wave $ss$ diquark of $j^P = 0^+$ and the color representation ${\bf 6}_c$. It has the $antisymmetric$ spin and $symmetric$ color structures, and its corresponding diquark field is
\begin{equation}
s_a^T C \gamma_5 s_b \, .
\end{equation}

\end{enumerate}
As depicted in Fig.~\ref{fig:category}, we can construct four $P$-wave $ss$ diquarks with the $antisymmetric$ orbital structure:
\begin{enumerate}
\setcounter{enumi}{2}

\item We use $^1P_1$ to denote the $P$-wave $ss$ diquark having $s=0$, $j^P = 1^-$, and the color representation ${\bf \bar 3}_c$. It has the $antisymmetric$ spin and $antisymmetric$ color structures, and its corresponding diquark field is
\begin{equation}
[s_a^T C \gamma_5 {\overset{\leftrightarrow}{D}}_\mu s_b] \, ,
\end{equation}
with $\big[ X {\overset{\leftrightarrow}{D}}_\mu Y \big] = X [D_\mu Y] - [D_\mu X] Y$. Note that the operator $D_\alpha$ carries color indices, {\it e.g.},
\begin{eqnarray}
D_\alpha s_a &=& \partial_\alpha s_a + i g_s A_\alpha s_a
\\ \nonumber &=& \partial_\alpha s_a + i g_s A_\alpha^n {\lambda^n_{ab} \over 2} s_b \, .
\end{eqnarray}
For simplicity, we shall still use the notation $D_\alpha s_a$ so that $D_\alpha [s_a^T C \gamma_\mu s_b] = [D_\alpha s_a]^T C \gamma_\mu s_b + s_a^T C \gamma_\mu [D_\alpha s_b]$.

\item We use $^3P_0$ to denote the $P$-wave $ss$ diquark having $s=1$, $j^P = 0^-$, and the color representation ${\bf 6}_c$. It has the $symmetric$ spin and $symmetric$ color structures, and its corresponding diquark field is
\begin{equation}
[s_a^T C \gamma^\mu {\overset{\leftrightarrow}{D}}_\mu s_b] \, .
\end{equation}

\item We use $^3P_1$ to denote the $P$-wave $ss$ diquark having $s=1$, $j^P = 1^-$, and the color representation ${\bf 6}_c$. It has the $symmetric$ spin and $symmetric$ color structures, and its corresponding diquark field is
\begin{equation}
[s_a^T C \gamma_\mu {\overset{\leftrightarrow}{D}}_\nu s_b] - \{ \mu \leftrightarrow \nu \} \, .
\end{equation}

\item We use $^3P_2$ to denote the $P$-wave $ss$ diquark having $s=1$, $j^P = 2^-$, and the color representation ${\bf 6}_c$. It has the $symmetric$ spin and $symmetric$ color structures, and its corresponding diquark field is
\begin{equation}
\mathcal{S}[s_a^T C \gamma_\mu {\overset{\leftrightarrow}{D}}_\nu s_b] \, ,
\end{equation}
where $\mathcal{S}$ denotes symmetrization and subtracting the trace term in the set $\{\mu\nu\}$.

\end{enumerate}

In the following subsections we shall use the above $S$- and $P$-wave $ss$ diquarks/antidiquarks to systematically construct the $S$- and $P$-wave $s s \bar s \bar s$ tetraquark states as well as their corresponding interpolating currents. Their color structure can be either
\begin{equation}
{\bf \bar 3}_{ss} \otimes {\bf 3}_{\bar s \bar s} \rightarrow {\bf 1}_{[ss][\bar s \bar s]} \, ,
\end{equation}
or
\begin{equation}
{\bf 6}_{ss} \otimes {\bf \bar 6}_{\bar s \bar s} \rightarrow {\bf 1}_{[ss][\bar s \bar s]} \, .
\end{equation}

\subsection{$S$-wave states}

In this subsection we use the $S$-wave $ss$ diquarks/antidiquarks to construct the $S$-wave $s s \bar s \bar s$ tetraquark states. We denote them as $|X;J^{PC}\rangle = | ^{2s+1}l_j,\, ^{2\bar s+1}{\bar l}_{\bar j};\, J \rangle$.

As depicted in Fig.~\ref{fig:category}, we can construct four $S$-wave $s s \bar s \bar s$ tetraquark states:
\begin{eqnarray}
\nonumber | X_1; 0^{++} \rangle &=& | ^1S_0,\, ^{\bar 1}{\bar S}_{\bar 0} ;\, J=0 \rangle \, ,
\\ | X_2; 0^{++} \rangle &=& | ^3S_1,\, ^{\bar 3}{\bar S}_{\bar 1} ;\, J=0 \rangle \, ,
\\ \nonumber | X_3; 1^{+-} \rangle &=& | ^3S_1,\, ^{\bar 3}{\bar S}_{\bar 1} ;\, J=1 \rangle \, ,
\\ \nonumber | X_4; 2^{++} \rangle &=& | ^3S_1,\, ^{\bar 3}{\bar S}_{\bar 1} ;\, J=2 \rangle \, .
\end{eqnarray}
Their corresponding interpolating currents are
\begin{eqnarray}
J_1^{0^{++}} &=& s_a^T C \gamma_5 s_b~\bar s_a \gamma_5 C \bar s_b^T \, ,
\label{def:current1}
\\ J_2^{0^{++}} &=& s_a^T C \gamma_\mu s_b~\bar s_a \gamma^\mu C \bar s_b^T \, ,
\label{def:current2}
\\ J_{3,\alpha}^{1^{+-}} &=& s_a^T C \gamma^\mu s_b ~ \bar s_a \sigma_{\alpha\mu} \gamma_5 C \bar s_b^T
\label{def:current3}
\\ \nonumber && ~~~~~~~~~~ - s_a^T C \sigma_{\alpha\mu} \gamma_5 s_b ~ \bar s_a \gamma^\mu C \bar s_b^T \, ,
\\ J_{4,\alpha_1\alpha_2}^{2^{++}} &=& g_{\alpha_1\mu}g_{\alpha_2\nu} \mathcal{S}[s_a^T C \gamma^\mu s_b~\bar s_a \gamma^\nu C \bar s_b^T] \, .
\label{def:current4}
\end{eqnarray}
We have used the tensor diquark field $s_a^T C \sigma_{\mu\nu} \gamma_5 s_b$ to construct the third current $J_{3,\alpha}^{1^{+-}}$. In principle, this tensor diquark field can couple to both the $j^P = 1^+$ and $1^-$ channels, but its positive-parity component $s_a^T C \sigma_{ij} \gamma_5 s_b$ ($i, j=1, 2, 3$) gives the dominant contribution to $J_{3,i}^{1^{+-}}$ ($i=1, 2, 3$). Therefore, the tetraquark current $J_{3,\alpha}^{1^{+-}}$ corresponds to the state $| X_3; 1^{+-} \rangle$.

\subsection{$P$-wave states of the $\lambda$-mode}

In this subsection we use the $S$-wave $ss$ diquarks/antidiquarks to construct the $P$-wave $s s \bar s \bar s$ tetraquark states of the $\lambda$-mode. These states have the total orbital excitation $L=1$ with $l_\lambda=1$ and $l_\rho=0$, so we call them $\lambda$-mode excitations. We denote them as $|X;J^{PC}\rangle = | ^{2s+1}l_j,\,^{2\bar s+1}{\bar l}_{\bar j};\, S, J, \lambda \rangle$.

As depicted in Fig.~\ref{fig:category}, we can construct eight $P$-wave $s s \bar s \bar s$ states of the $\lambda$-mode:
\begin{eqnarray}
\nonumber | X_5; 0^{-+} \rangle       &=& | ^3S_1,\, ^{\bar 3}{\bar S}_{\bar 1};\, S=1, J=0, \lambda \rangle \, ,
\\ \nonumber | X_6; 1^{--} \rangle    &=& | ^1S_0,\, ^{\bar 1}{\bar S}_{\bar 0};\, S=0, J=1, \lambda \rangle \, ,
\\ \nonumber | X_7; 1^{--} \rangle    &=& | ^3S_1,\, ^{\bar 3}{\bar S}_{\bar 1};\, S=0, J=1, \lambda \rangle \, ,
\\ | X_8; 1^{--} \rangle    &=& | ^3S_1,\, ^{\bar 3}{\bar S}_{\bar 1};\, S=2, J=1, \lambda \rangle \, ,
\\ \nonumber | X_9; 1^{-+} \rangle    &=& | ^3S_1,\, ^{\bar 3}{\bar S}_{\bar 1};\, S=1, J=1, \lambda \rangle \, ,
\\ \nonumber | X_{10}; 2^{--} \rangle &=& | ^3S_1,\, ^{\bar 3}{\bar S}_{\bar 1};\, S=2, J=2, \lambda \rangle \, ,
\\ \nonumber | X_{11}; 2^{-+} \rangle &=& | ^3S_1,\, ^{\bar 3}{\bar S}_{\bar 1};\, S=1, J=2, \lambda \rangle \, ,
\\ \nonumber | X_{12}; 3^{--} \rangle &=& | ^3S_1,\, ^{\bar 3}{\bar S}_{\bar 1};\, S=2, J=3, \lambda \rangle \, .
\end{eqnarray}
Their corresponding interpolating currents are
\begin{eqnarray}
J_5^{0^{-+}} &=& [ s_a^T C \gamma^\nu s_b ] {\overset{\leftrightarrow}{D^{\mu}}} [ \bar s_a \sigma_{\mu\nu} \gamma_5 C \bar s_b^T ]
\label{def:current5}
\\ \nonumber && ~~~~~~~~~~ - [ s_a^T C \sigma_{\mu\nu} \gamma_5 s_b ] {\overset{\leftrightarrow}{D^{\mu}}} [ \bar s_a \gamma^\nu C \bar s_b^T ] \, ,
\\ J_{6,\alpha}^{1^{--}} &=& [s_a^T C \gamma_5 s_b] {\overset{\leftrightarrow}{D}}_{\alpha} [\bar s_a  \gamma_5 C \bar s_b^T] \, ,
\label{def:current6}
\\ J_{7,\alpha}^{1^{--}} &=& [s_a^T C \gamma_\mu s_b] {\overset{\leftrightarrow}{D}}_{\alpha} [\bar s_a \gamma^\mu C \bar s_b^T] \, ,
\label{def:current7}
\\ J_{8,\alpha}^{1^{--}} &=& g_{\alpha\mu}g_{\nu\rho} \mathcal{S}[[s_a^T C \gamma^\mu s_b] {\overset{\leftrightarrow}{D^{\rho}}} [\bar s_a \gamma^\nu C \bar s_b^T]] \, ,
\label{def:current8}
\\ J_{9,\alpha}^{1^{-+}} &=& [s_a^T C \gamma_\alpha s_b] {\overset{\leftrightarrow}{D^{\mu}}} [\bar s_a \gamma_\mu C \bar s_b^T]
\label{def:current9}
\\ \nonumber && ~~~~~~~~~~ - [s_a^T C \gamma_\mu s_b] {\overset{\leftrightarrow}{D^{\mu}}} [\bar s_a \gamma_\alpha C \bar s_b^T] \, ,
\\ \nonumber J_{10,\alpha_1\alpha_2}^{2^{--}} &=& g_{\alpha_1\mu}g_{\alpha_2\nu} \mathcal{S}[[ s_a^T C \gamma^{\mu} s_b ] {\overset{\leftrightarrow}{D}}_{\rho} [ \bar s_a \sigma^{\nu\rho} \gamma_5 C \bar s_b^T ]
\\ && ~~~~~ + [ s_a^T C \sigma^{\nu\rho} \gamma_5 s_b ] {\overset{\leftrightarrow}{D}}_{\rho} [ \bar s_a \gamma^{\mu} C \bar s_b^T ]] \, ,
\label{def:current10}
\\ \nonumber J_{11,\alpha_1\alpha_2}^{2^{-+}} &=& g_{\alpha_1\mu}g_{\alpha_2\nu} \mathcal{S}[[ s_a^T C \gamma_{\rho} s_b ] {\overset{\leftrightarrow}{D^{\mu}}} [ \bar s_a \sigma^{\nu\rho} \gamma_5 C \bar s_b^T ]
\\ && ~~~~~ - [ s_a^T C \sigma^{\nu\rho} \gamma_5 s_b ] {\overset{\leftrightarrow}{D^{\mu}}} [ \bar s_a \gamma_{\rho} C \bar s_b^T ]] \, ,
\label{def:current11}
\\ J_{12,\alpha_1\alpha_2\alpha_3}^{3^{--}} &=& \mathcal{S^\prime}[[s_a^T C \gamma_{\alpha_1} s_b] {\overset{\leftrightarrow}{D}}_{\alpha_2} [\bar s_a \gamma_{\alpha_3} C \bar s_b^T]] \, ,
\label{def:current12}
\end{eqnarray}
where $\mathcal{S^\prime}$ denotes symmetrization and subtracting trace terms in the set $\{\alpha_1\alpha_2\alpha_3\}$.

\subsection{$P$-wave states of the $\rho$-mode}

In this subsection we use the $S$- and $P$-wave $ss$ diquarks/antidiquarks to construct the $P$-wave $s s \bar s \bar s$ tetraquark states of the $\rho$-mode. These states have the total orbital excitation $L=1$ with $l_\lambda=0$ and $l_\rho=1$, so we call them $\rho$-mode excitations. We denote them as $|X;J^{PC}\rangle = | ^{2s+1}l_j,\, ^{2\bar s+1}{\bar l}_{\bar j};\, S, J, \rho \rangle$.

As depicted in Fig.~\ref{fig:category}, we can construct twelve $P$-wave $s s \bar s \bar s$ states of the $\rho$-mode:
\begin{eqnarray}
\nonumber | X_{13}; 0^{--} \rangle    &=& | ^1S_0,\, ^{\bar 3}{\bar P}_{\bar 0};\, 1, 0, \rho \rangle - | ^{3}{P}_{0},\, ^{\bar 1}{\bar S}_{\bar 0};\, 1, 0, \rho \rangle \, ,
\\ \nonumber | X_{14}; 0^{--} \rangle &=& | ^3S_1,\, ^{\bar 1}{\bar P}_{\bar 1};\, 1, 0, \rho \rangle - | ^{1}{P}_{1},\, ^{\bar 3}{\bar S}_{\bar 1};\, 1, 0, \rho \rangle \, ,
\\ \nonumber | X_{15}; 0^{-+} \rangle &=& | ^1S_0,\, ^{\bar 3}{\bar P}_{\bar 0};\, 1, 0, \rho \rangle + | ^{3}{P}_{0},\, ^{\bar 1}{\bar S}_{\bar 0};\, 1, 0, \rho \rangle \, ,
\\ \nonumber | X_{16}; 0^{-+} \rangle &=& | ^3S_1,\, ^{\bar 1}{\bar P}_{\bar 1};\, 1, 0, \rho \rangle + | ^{1}{P}_{1},\, ^{\bar 3}{\bar S}_{\bar 1};\, 1, 0, \rho \rangle \, ,
\\ \nonumber | X_{17}; 1^{--} \rangle &=& | ^1S_0,\, ^{\bar 3}{\bar P}_{\bar 1};\, 1, 1, \rho \rangle - | ^{3}{P}_{1},\, ^{\bar 1}{\bar S}_{\bar 0};\, 1, 1, \rho \rangle \, ,
\\ \nonumber | X_{18}; 1^{--} \rangle &=& | ^3S_1,\, ^{\bar 1}{\bar P}_{\bar 1};\, 1, 1, \rho \rangle - | ^{1}{P}_{1},\, ^{\bar 3}{\bar S}_{\bar 1};\, 1, 1, \rho \rangle \, ,
\\ \nonumber | X_{19}; 1^{-+} \rangle &=& | ^1S_0,\, ^{\bar 3}{\bar P}_{\bar 1};\, 1, 1, \rho \rangle + | ^{3}{P}_{1},\, ^{\bar 1}{\bar S}_{\bar 0};\, 1, 1, \rho \rangle \, ,
\\ \nonumber | X_{20}; 1^{-+} \rangle &=& | ^3S_1,\, ^{\bar 1}{\bar P}_{\bar 1};\, 1, 1, \rho \rangle + | ^{1}{P}_{1},\, ^{\bar 3}{\bar S}_{\bar 1};\, 1, 1, \rho \rangle \, ,
\\ \nonumber | X_{21}; 2^{--} \rangle &=& | ^1S_0,\, ^{\bar 3}{\bar P}_{\bar 2};\, 1, 2, \rho \rangle - | ^{3}{P}_{2},\, ^{\bar 1}{\bar S}_{\bar 0};\, 1, 2, \rho \rangle \, ,
\\ \nonumber | X_{22}; 2^{--} \rangle &=& | ^3S_1,\, ^{\bar 1}{\bar P}_{\bar 1};\, 1, 2, \rho \rangle - | ^{1}{P}_{1},\, ^{\bar 3}{\bar S}_{\bar 1};\, 1, 2, \rho \rangle \, ,
\\ \nonumber | X_{23}; 2^{-+} \rangle &=& | ^1S_0,\, ^{\bar 3}{\bar P}_{\bar 2};\, 1, 2, \rho \rangle + | ^{3}{P}_{2},\, ^{\bar 1}{\bar S}_{\bar 0};\, 1, 2, \rho \rangle \, ,
\\ \nonumber | X_{24}; 2^{-+} \rangle &=& | ^3S_1,\, ^{\bar 1}{\bar P}_{\bar 1};\, 1, 2, \rho \rangle + | ^{1}{P}_{1},\, ^{\bar 3}{\bar S}_{\bar 1};\, 1, 2, \rho \rangle \, .
\\
\end{eqnarray}
Their corresponding interpolating currents are
\begin{eqnarray}
J_{13}^{0^{--}} &=& [ s_a^T C \gamma_5 s_b ] [ \bar s_a \gamma^\mu C {\overset{\leftrightarrow}{D}}_\mu \bar s_b^T ]
\label{def:current13}
\\ \nonumber && ~~~~~~~~~~ - [ s_a^T C \gamma^\mu {\overset{\leftrightarrow}{D}}_\mu s_b ] [ \bar s_a \gamma_5 C \bar s_b^T ] \, ,
\\ J_{14}^{0^{--}} &=& [ s_a^T C \gamma^\mu s_b ] [ \bar s_a \gamma_5 C {\overset{\leftrightarrow}{D}}_\mu \bar s_b^T ]
\label{def:current14}
\\ \nonumber && ~~~~~~~~~~ - [ s_a^T C \gamma_5 {\overset{\leftrightarrow}{D}}_\mu s_b ] [ \bar s_a \gamma^\mu C \bar s_b^T ] \, ,
\\ J_{15}^{0^{-+}} &=& [ s_a^T C \gamma_5 s_b ] [ \bar s_a \gamma^\mu C {\overset{\leftrightarrow}{D}}_\mu \bar s_b^T ]
\label{def:current15}
\\ \nonumber && ~~~~~~~~~~ + [ s_a^T C \gamma^\mu {\overset{\leftrightarrow}{D}}_\mu s_b ] [ \bar s_a \gamma_5 C \bar s_b^T ] \, ,
\\ J_{16}^{0^{-+}} &=& [ s_a^T C \gamma^\mu s_b ] [ \bar s_a \gamma_5 C {\overset{\leftrightarrow}{D}}_\mu \bar s_b^T ]
\label{def:current16}
\\ \nonumber && ~~~~~~~~~~ + [ s_a^T C \gamma_5 {\overset{\leftrightarrow}{D}}_\mu s_b ] [ \bar s_a \gamma^\mu C \bar s_b^T ] \, ,
\\ J_{17,\alpha\beta}^{1^{--}} &=& [ s_a^T C \gamma_5 s_b ] [ \bar s_a \gamma_\beta C {\overset{\leftrightarrow}{D}}_\alpha \bar s_b^T ]
\label{def:current17}
\\ \nonumber && - [ s_a^T C \gamma_\beta {\overset{\leftrightarrow}{D}}_\alpha s_b ] [ \bar s_a \gamma_5 C \bar s_b^T ] - \{ \alpha \leftrightarrow \beta \} \, ,
\\ J_{18,\alpha\beta}^{1^{--}} &=& [ s_a^T C \gamma_\beta s_b ] [ \bar s_a \gamma_5 C {\overset{\leftrightarrow}{D}}_\alpha \bar s_b^T ]
\label{def:current18}
\\ \nonumber && - [ s_a^T C \gamma_5 {\overset{\leftrightarrow}{D}}_\alpha s_b ] [ \bar s_a \gamma_\beta C \bar s_b^T ] - \{ \alpha \leftrightarrow \beta \} \, ,
\\ J_{19,\alpha\beta}^{1^{-+}} &=& [ s_a^T C \gamma_5 s_b ] [ \bar s_a \gamma_\beta C {\overset{\leftrightarrow}{D}}_\alpha \bar s_b^T ]
\label{def:current19}
\\ \nonumber && + [ s_a^T C \gamma_\beta {\overset{\leftrightarrow}{D}}_\alpha s_b ] [ \bar s_a \gamma_5 C \bar s_b^T ] - \{ \alpha \leftrightarrow \beta \} \, ,
\\ J_{20,\alpha\beta}^{1^{-+}} &=& [ s_a^T C \gamma_\beta s_b ] [ \bar s_a \gamma_5 C {\overset{\leftrightarrow}{D}}_\alpha \bar s_b^T ]
\label{def:current20}
\\ \nonumber && + [ s_a^T C \gamma_5 {\overset{\leftrightarrow}{D}}_\alpha s_b ] [ \bar s_a \gamma_\beta C \bar s_b^T ] - \{ \alpha \leftrightarrow \beta \} \, ,
\\ J_{21,\alpha_1\alpha_2}^{2^{--}} &=& g_{\alpha_1\mu}g_{\alpha_2\nu} \mathcal{S}[[ s_a^T C \gamma_5 s_b ] [ \bar s_a \gamma^\nu C {\overset{\leftrightarrow}{D^\mu}} \bar s_b^T ]
\label{def:current21}
\\ \nonumber && ~~~~~~~~~~ - [ s_a^T C \gamma^\nu {\overset{\leftrightarrow}{D^\mu}} s_b ] [ \bar s_a \gamma_5 C \bar s_b^T ]] \, ,
\\ J_{22,\alpha_1\alpha_2}^{2^{--}} &=& g_{\alpha_1\mu}g_{\alpha_2\nu} \mathcal{S}[[ s_a^T C \gamma^\nu s_b ] [ \bar s_a \gamma_5 C {\overset{\leftrightarrow}{D^\mu}} \bar s_b^T ]
\label{def:current22}
\\ \nonumber && ~~~~~~~~~~ - [ s_a^T C \gamma_5 {\overset{\leftrightarrow}{D^\mu}} s_b ] [ \bar s_a \gamma^\nu C \bar s_b^T ]] \, ,
\\ J_{23,\alpha_1\alpha_2}^{2^{-+}} &=& g_{\alpha_1\mu}g_{\alpha_2\nu} \mathcal{S}[[ s_a^T C \gamma_5 s_b ] [ \bar s_a \gamma^\nu C {\overset{\leftrightarrow}{D^\mu}} \bar s_b^T ]
\label{def:current23}
\\ \nonumber && ~~~~~~~~~~ + [ s_a^T C \gamma^\nu {\overset{\leftrightarrow}{D^\mu}} s_b ] [ \bar s_a \gamma_5 C \bar s_b^T ]] \, ,
\\ J_{24,\alpha_1\alpha_2}^{2^{-+}} &=& g_{\alpha_1\mu}g_{\alpha_2\nu} \mathcal{S}[[ s_a^T C \gamma^\nu s_b ] [ \bar s_a \gamma_5 C {\overset{\leftrightarrow}{D^\mu}} \bar s_b^T ]
\label{def:current24}
\\ \nonumber && ~~~~~~~~~~ + [ s_a^T C \gamma_5 {\overset{\leftrightarrow}{D^\mu}} s_b ] [ \bar s_a \gamma^\nu C \bar s_b^T ]] \, .
\end{eqnarray}

The four currents $J_{17\cdots20,\alpha\beta}^{\cdots}$ all have two antisymmetric Lorentz indices $\alpha$ and $\beta$, so they actually contain both $J^P = 1^-$ and $1^+$ components. In the present study we shall use these currents to study the four $J^P = 1^-$ states $| X_{17\cdots20}; 1^{-\pm} \rangle$ through:
\begin{equation}
\langle 0| J_{17\cdots20,\alpha\beta}^{\cdots} | X_{17\cdots20}(\epsilon,q) \rangle = i f_{X_{17\cdots20}} {\bm \epsilon}_{\alpha\beta\rho\sigma} \epsilon^\rho q^\sigma \, ,
\label{eq:coupling2}
\end{equation}
where $\epsilon^\rho$ is the polarization vector, ${\bm \epsilon}_{\alpha\beta\rho\sigma}$ is the totally antisymmetric tensor, and $f_{X_{17\cdots20}}$ are the decay constants. Note that these four currents can also couple to the four $J^P = 1^+$ states $| X^\prime_{17\cdots20}; 1^{+\pm} \rangle$ through:
\begin{equation}
\langle 0| J_{17\cdots20,\alpha\beta}^{\cdots} | X^\prime_{17\cdots20}(\epsilon,q) \rangle = i f_{X^\prime_{17\cdots20}} (q_\alpha\epsilon_\beta - q_\beta\epsilon_\alpha) \, .
\label{eq:coupling3}
\end{equation}
Technically, we can easily isolate $| X_{17\cdots20}; 1^{-\pm} \rangle$ at the hadron level by investigating the correlation function proportional to
\begin{eqnarray}
\nonumber && \langle 0 | J_{17\cdots20,\alpha\beta}^{\cdots} | X_{17\cdots20} \rangle \langle X_{17\cdots20} | J_{17\cdots20,\alpha^\prime\beta^\prime}^{\cdots,\dagger} | 0 \rangle
\\ \nonumber &=& f_{X_{17\cdots20}}^2 {\bm \epsilon}_{\alpha\beta\rho\sigma} \epsilon^\rho q^\sigma {\bm \epsilon}_{\alpha^\prime\beta^\prime\rho^\prime\sigma^\prime} \epsilon^{*\rho^\prime} q^{\sigma^\prime}
\\ &=& - f_{X_{17\cdots20}}^2 ~ q^2 ~ \left( g_{\alpha \alpha^\prime} g_{\beta \beta^\prime} - g_{\alpha \beta^\prime} g_{\beta \alpha^\prime} \right) + \cdots \, ,
\end{eqnarray}
given that the correlation function of $| X^\prime_{17\cdots20}; 1^{+\pm} \rangle$ dose not contain the above coefficient.

%
\section{QCD sum rule Analysis}
\label{sec:sumrule}
%

The method of QCD sum rules is a powerful and successful non-perturbative method~\cite{Shifman:1978bx,Reinders:1984sr,Shifman:1978by,Shifman:1978bw,Novikov:1983gd,Grozin:1994hd,Grozin:2007zz}. In this section we apply this method to study the twenty-four currents given in Eqs.~(\ref{def:current1}-\ref{def:current4}), Eqs.~(\ref{def:current5}-\ref{def:current12}), and Eqs.~(\ref{def:current13}-\ref{def:current24}).

The four currents $J_{17\cdots20,\alpha\beta}^{\cdots}$ couple to the states $|X_{17\cdots20};J^{PC}\rangle$ through Eq.~(\ref{eq:coupling2}). The other twenty currents $J^{\cdots}_{1\cdots16/21\cdots24,\alpha_1\cdots\alpha_J}$ of spin-$J$ couple to the states $|X_{1\cdots16/21\cdots24};J^{PC}\rangle$ through
\begin{eqnarray}
\nonumber && \langle 0| J^{\cdots}_{1\cdots16/21\cdots24,\alpha_1\cdots\alpha_J} | X_{1\cdots16/21\cdots24};J^{PC} \rangle
\\ &=& f_{X_{1\cdots16/21\cdots24}} \epsilon_{\alpha_1\cdots\alpha_J} \, .
\label{eq:coupling1}
\end{eqnarray}
Here $f_X$ is the decay constant, and $\epsilon_{\alpha_1\cdots\alpha_J}$ is the traceless and symmetric polarization tensor, satisfying:
\begin{equation}
\epsilon_{\alpha_1\cdots\alpha_J} \epsilon^*_{\beta_1\cdots\beta_J} = \mathcal{S}^{\prime\prime} [\tilde g_{\alpha_1 \beta_1} \cdots \tilde g_{\alpha_J \beta_J}] \, ,
\end{equation}
where $\tilde g_{\mu \nu} = g_{\mu \nu} - q_\mu q_\nu / q^2$, and $\mathcal{S}^{\prime\prime}$ denotes symmetrization and subtracting trace terms in the sets $\{\alpha_1\cdots\alpha_J\}$ and $\{\beta_1\cdots\beta_J\}$.

We use the current $J_1^{0^{++}}$ as an example, and study its two-point correlation function
%
\begin{equation}
\Pi(q^2) = i \int d^4x e^{iqx} \langle 0 | {\bf T}[ J_1^{0^{++}}(x) J_1^{0^{++},\dagger}(0)] | 0 \rangle \, ,
\label{def:pi}
\end{equation}
%
at both the hadron and quark-gluon levels.

At the hadron level we express Eq.~(\ref{def:pi}) through the dispersion relation as
%
\begin{equation}
\Pi(q^2) = \int^\infty_{16 m_s^2}\frac{\rho(s)}{s-q^2-i\varepsilon}ds \, ,
\label{eq:hadron}
\end{equation}
%
where $\rho(s) \equiv {\rm Im}\Pi(s)/\pi$ is the spectral density. We parameterize it as one pole dominance for the ground state $| X_1; 0^{++} \rangle$ and a continuum contribution:
%
\begin{eqnarray}
\nonumber \rho_{\rm phen}(s) &\equiv& \sum_n\delta(s-M^2_n) \langle 0| J_1^{0^{++}} | n\rangle \langle n| J_1^{0^{++},\dagger} |0 \rangle
\\ &=& f^2_X \delta(s-M^2_X) + \rm{continuum} \, .
\label{eq:rho}
\end{eqnarray}
%

At the quark-gluon level we apply the method of operator product expansion (OPE) to calculate Eq.~(\ref{def:pi}) and extract the OPE spectral density $\rho_{\rm OPE}(s)$. After performing the Borel transformation at both the hadron and quark-gluon levels, we approximate the continuum using $\rho_{\rm OPE}(s)$ above a threshold value $s_0$, and arrive at the sum rule equation
%
\begin{equation}
\Pi(s_0, M_B^2) \equiv f^2_X e^{-M_X^2/M_B^2} = \int^{s_0}_{16 m_s^2} e^{-s/M_B^2}\rho_{\rm OPE}(s)ds \, ,
\label{eq:fin}
\end{equation}
%
which can be used to calculate $M_X$ through
%
\begin{equation}
M^2_X(s_0, M_B) = \frac{\int^{s_0}_{16 m_s^2} e^{-s/M_B^2}s\rho_{\rm OPE}(s)ds}{\int^{s_0}_{16 m_s^2} e^{-s/M_B^2}\rho_{\rm OPE}(s)ds} \, .
\label{eq:LSR}
\end{equation}
%

\begin{figure*}[]
\begin{center}
\subfigure[(a)]{
\scalebox{0.15}{\includegraphics{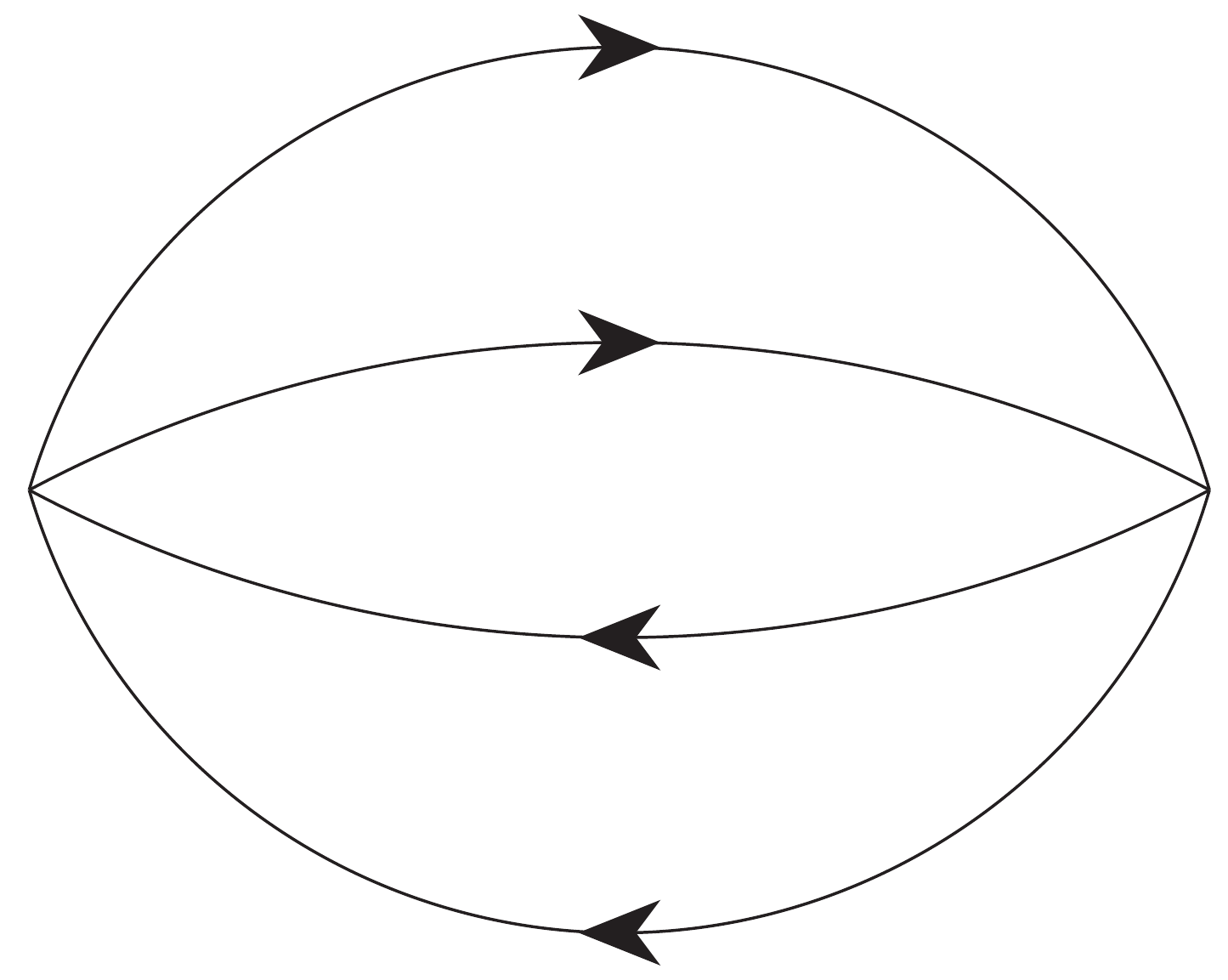}}}
\\[2mm]
\subfigure[(b--1)]{
\scalebox{0.15}{\includegraphics{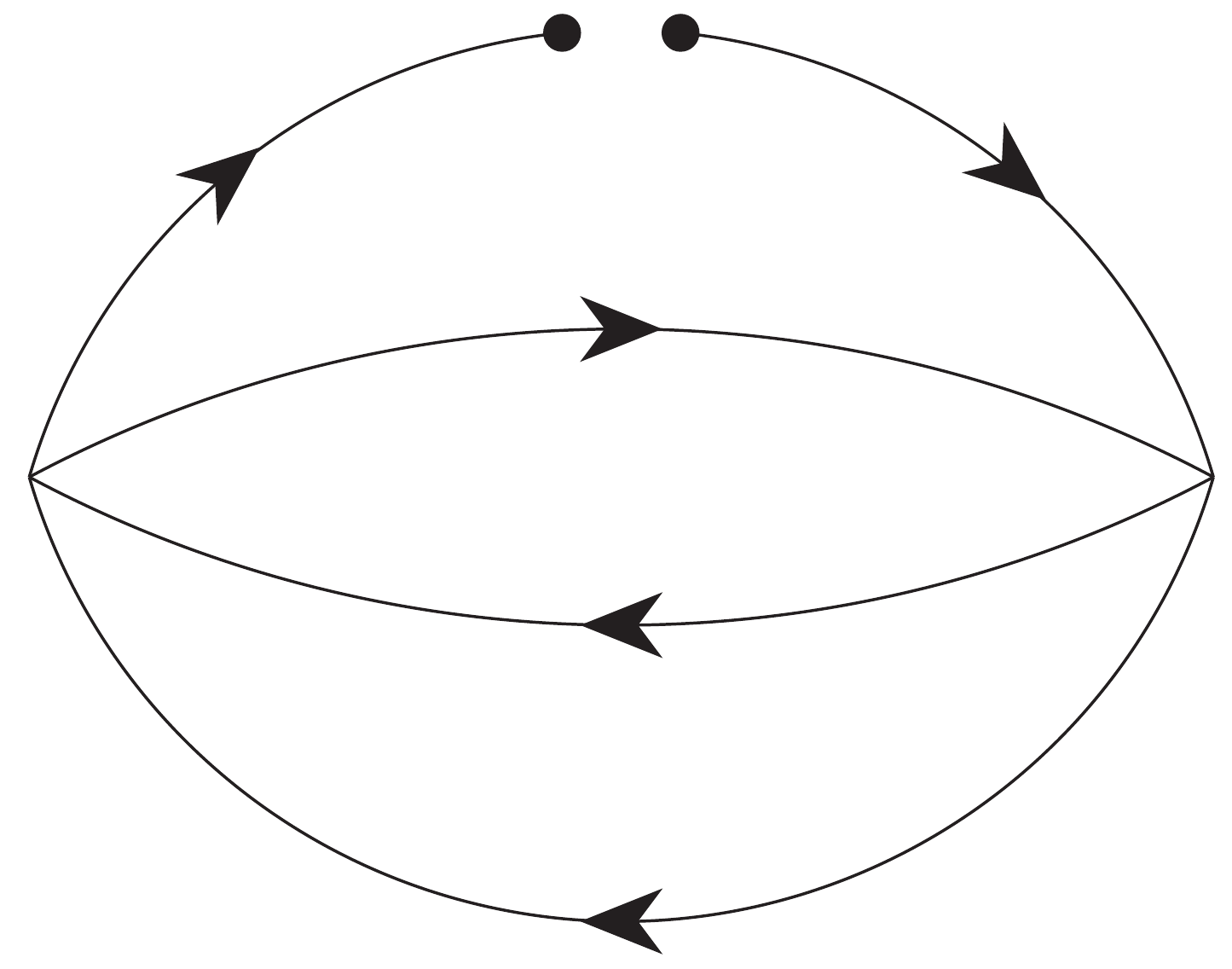}}}~~~~~
\subfigure[(b--2)]{
\scalebox{0.15}{\includegraphics{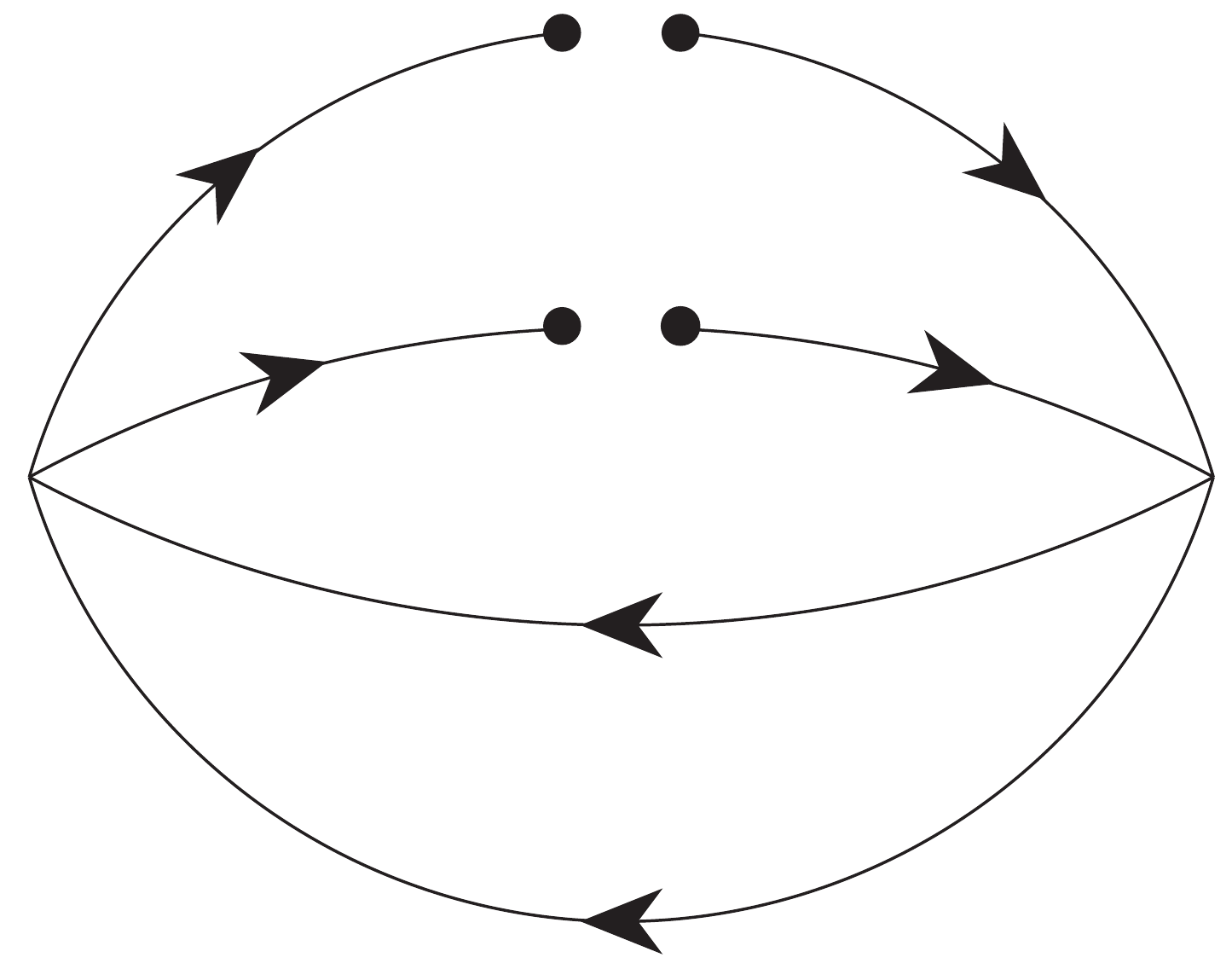}}}~~~~~
\subfigure[(b--3)]{
\scalebox{0.15}{\includegraphics{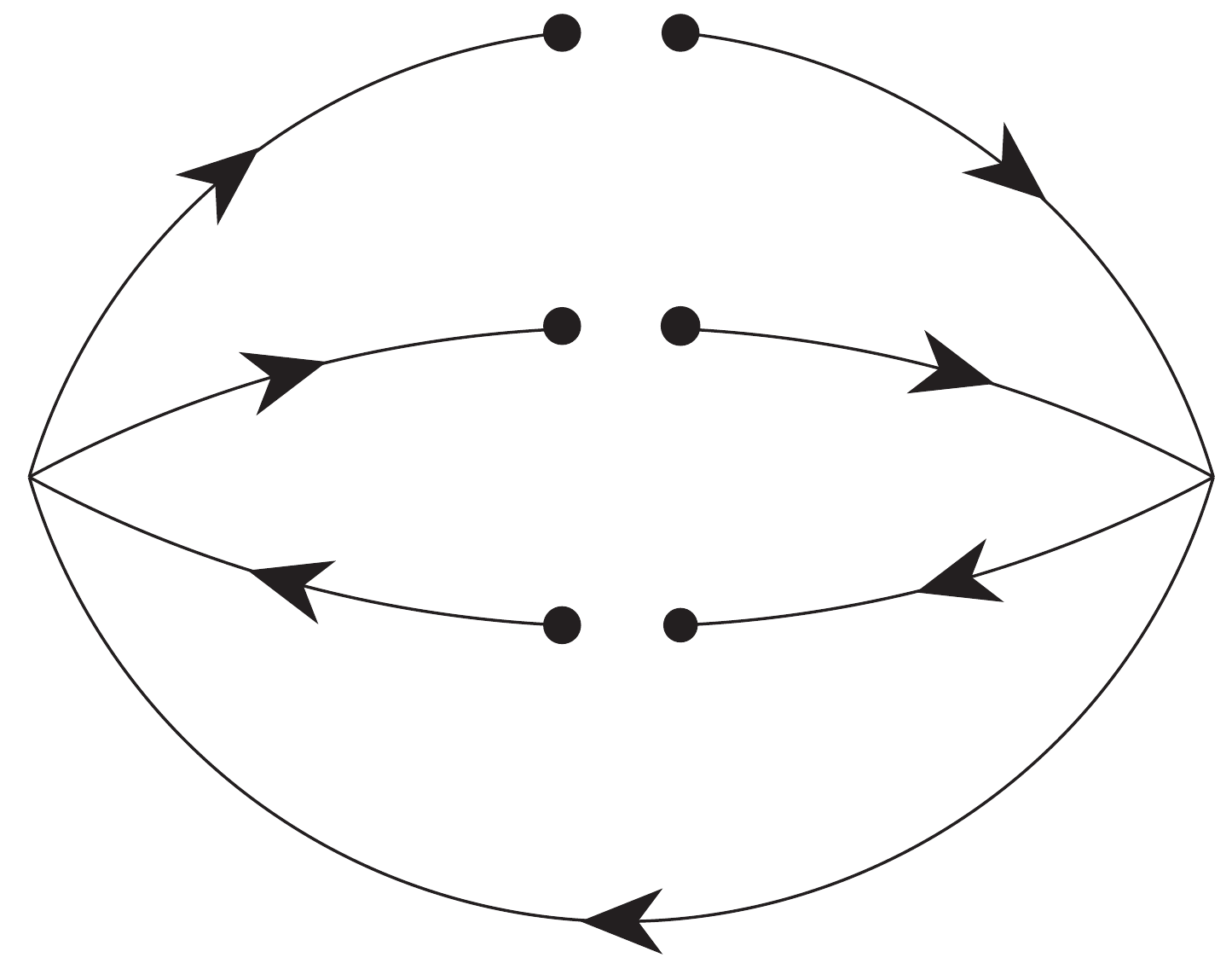}}}
\\[2mm]
\subfigure[(c--1)]{
\scalebox{0.15}{\includegraphics{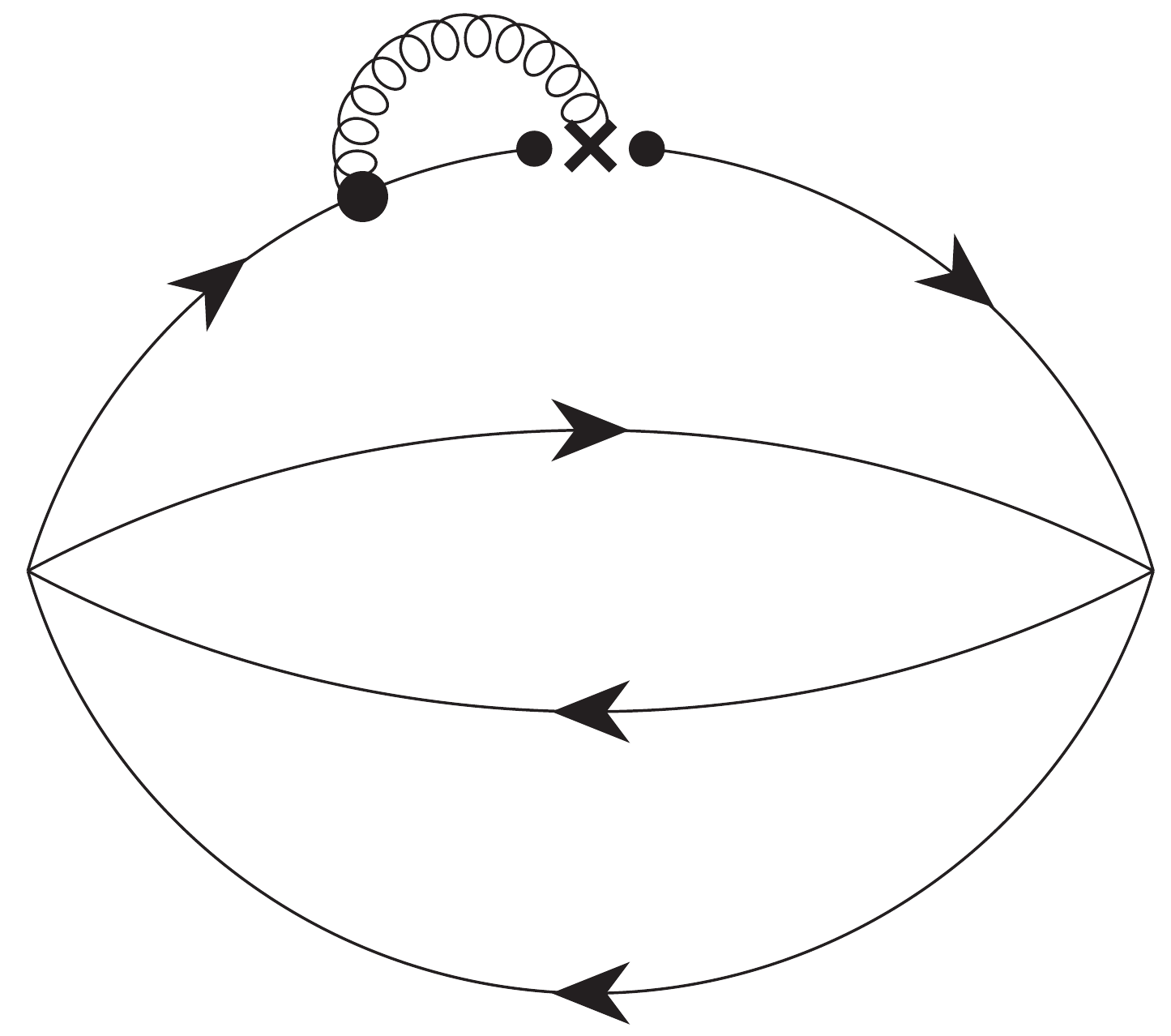}}}~~~~~
\subfigure[(c--2)]{
\scalebox{0.15}{\includegraphics{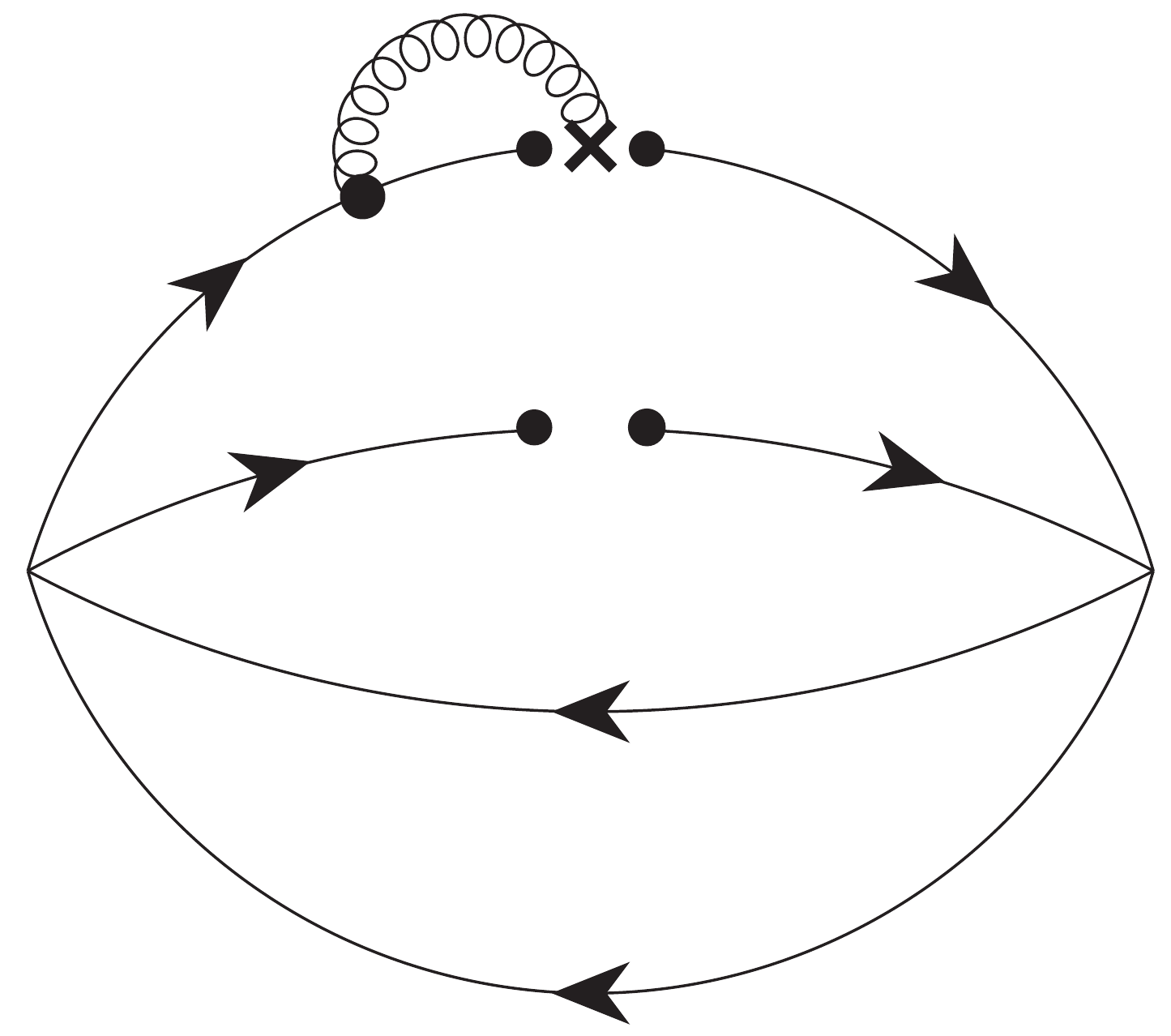}}}~~~~~
\subfigure[(c--3)]{
\scalebox{0.15}{\includegraphics{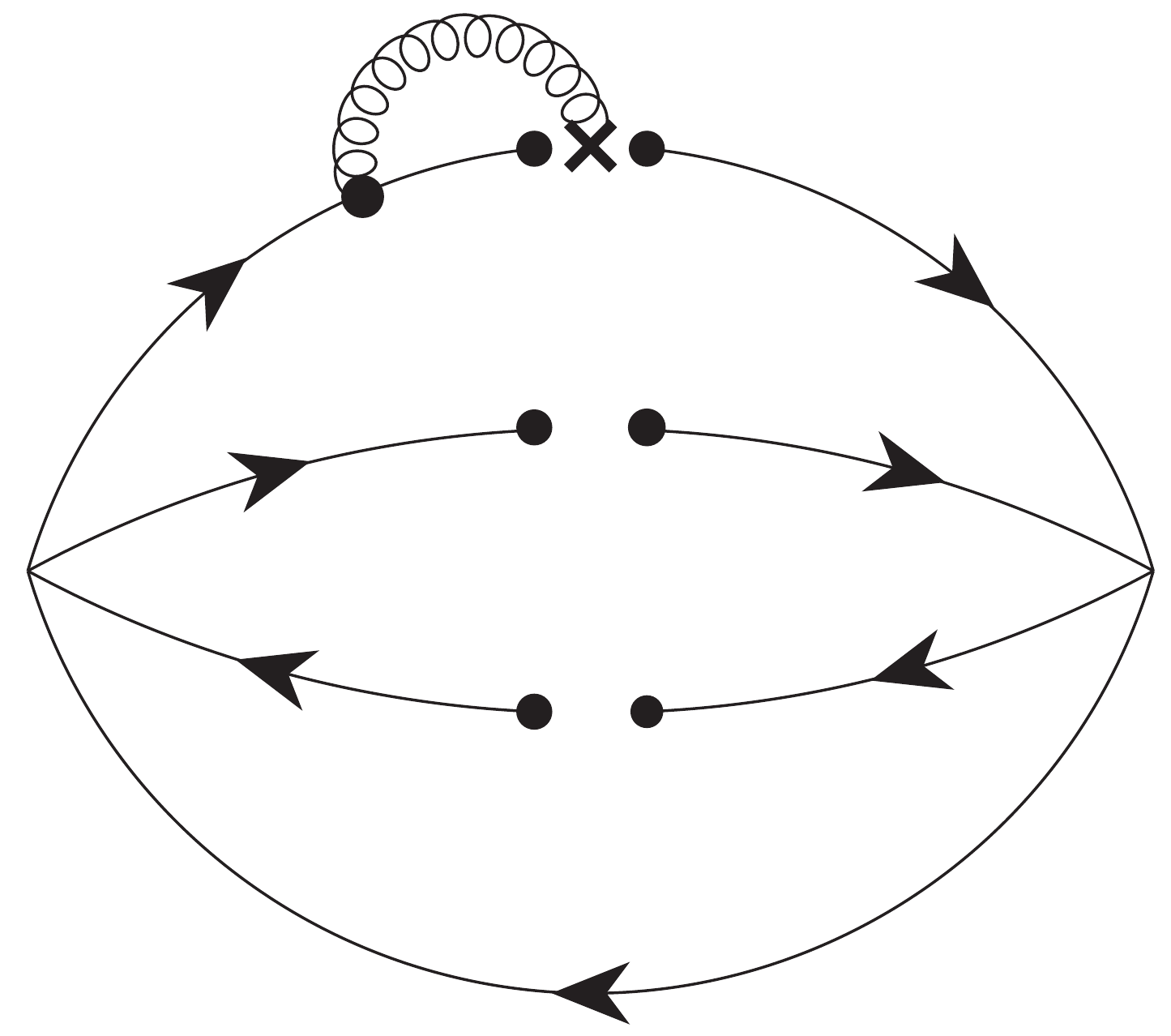}}}~~~~~
\subfigure[(c--4)]{
\scalebox{0.15}{\includegraphics{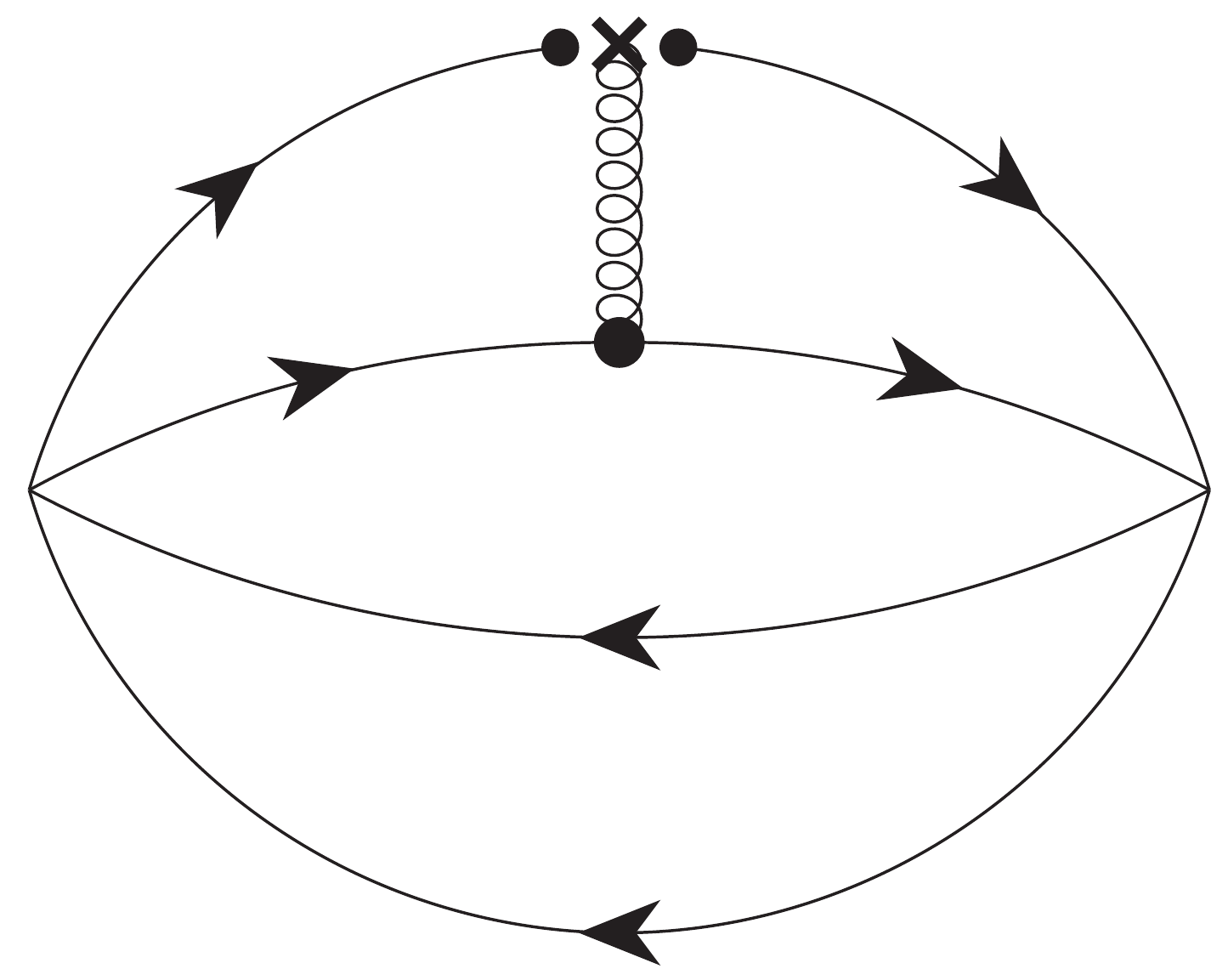}}}~~~~~
\subfigure[(c--5)]{
\scalebox{0.15}{\includegraphics{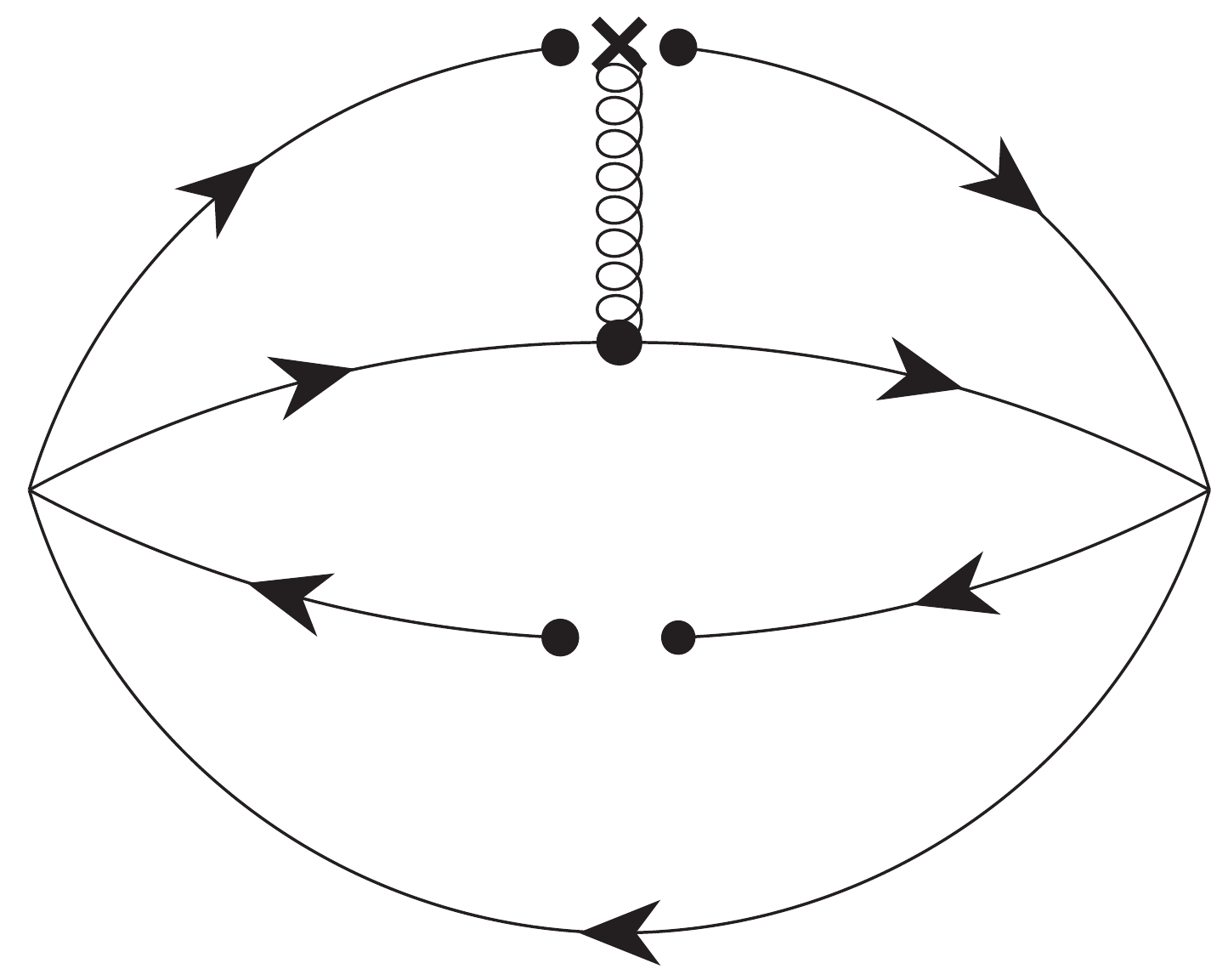}}}~~~~~
\subfigure[(c--6)]{
\scalebox{0.15}{\includegraphics{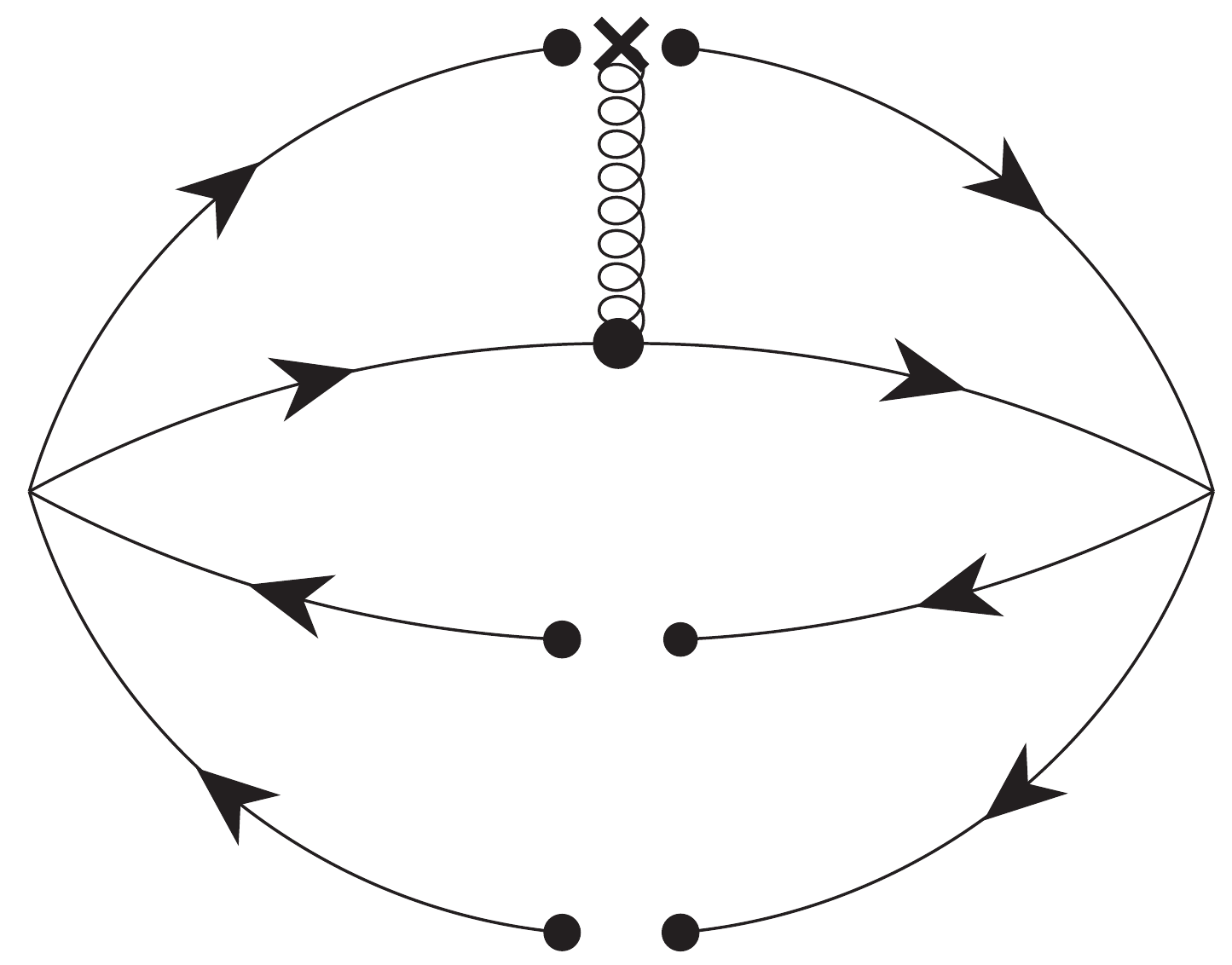}}}
\\[2mm]
\subfigure[(d--1)]{
\scalebox{0.15}{\includegraphics{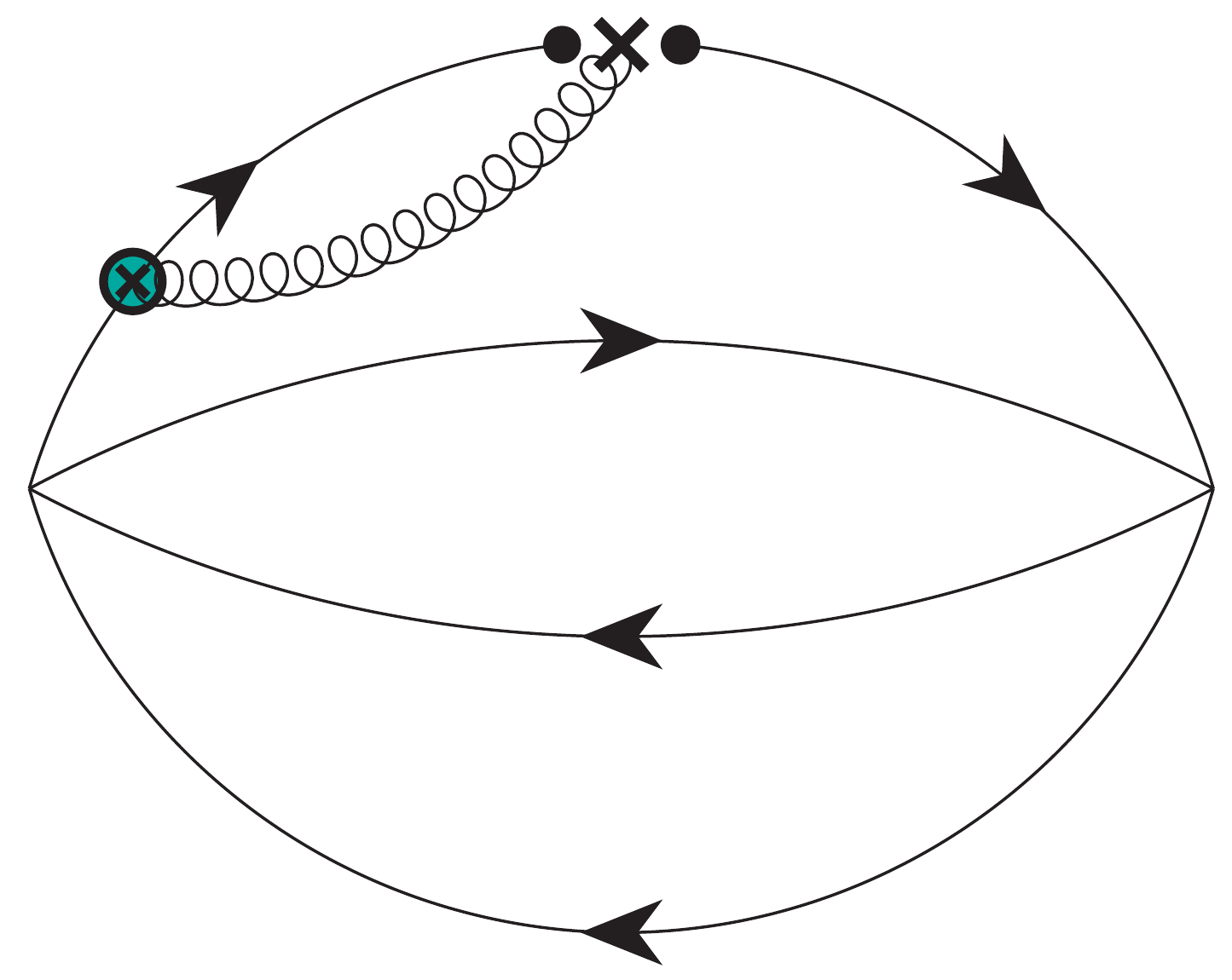}}}~~~~~
\subfigure[(d--2)]{
\scalebox{0.15}{\includegraphics{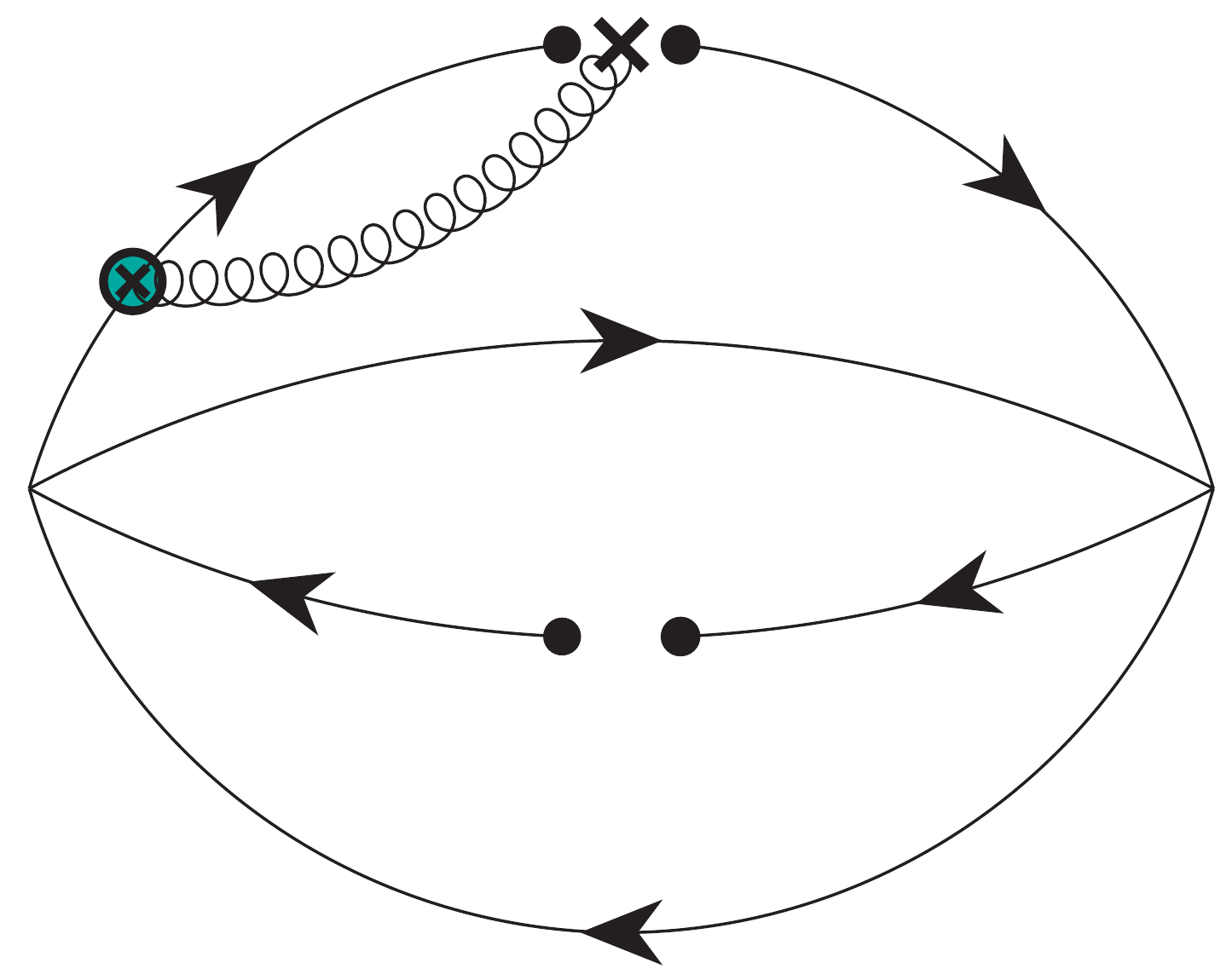}}}~~~~~
\subfigure[(d--3)]{
\scalebox{0.15}{\includegraphics{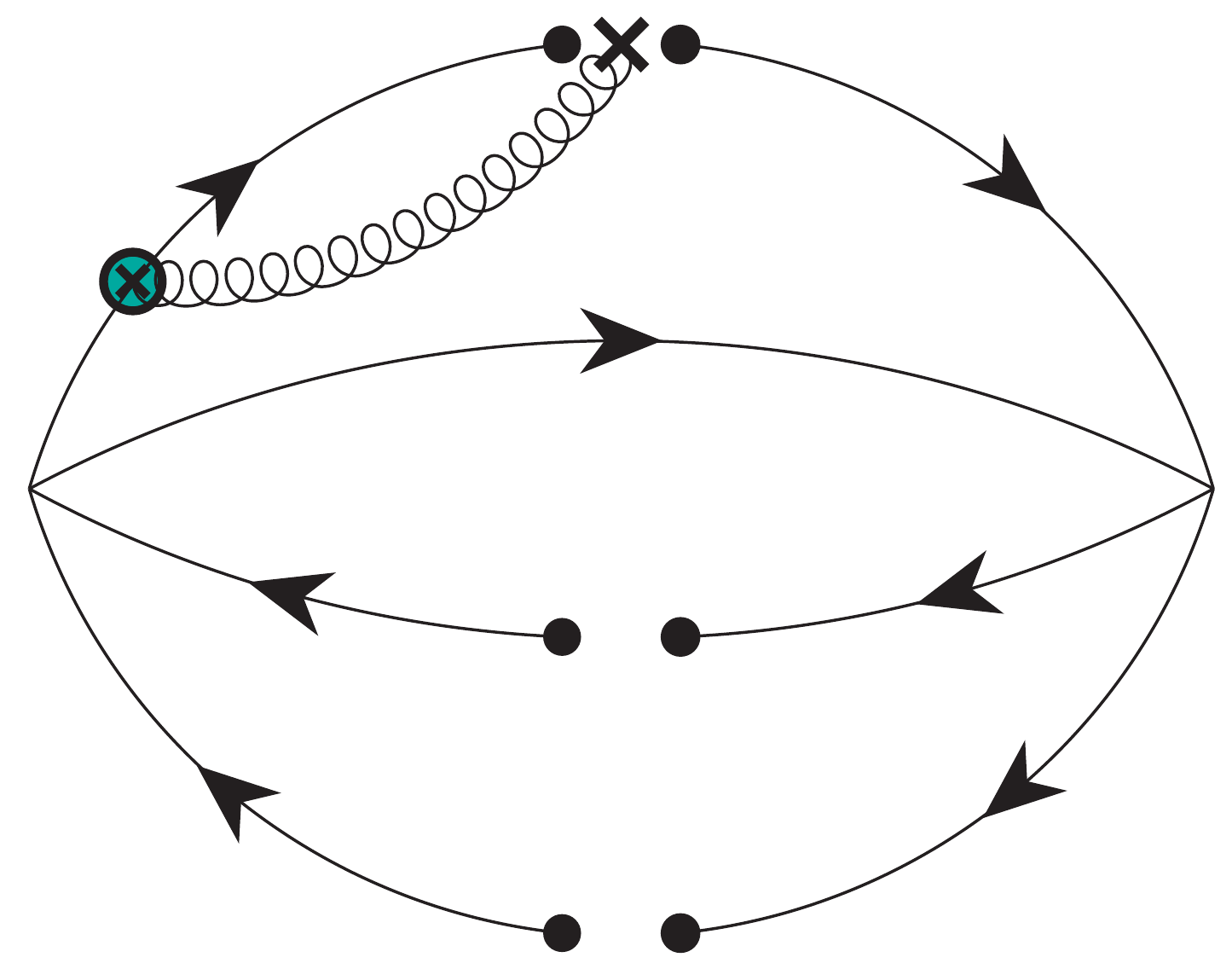}}}~~~~~
\subfigure[(d--4)]{
\scalebox{0.15}{\includegraphics{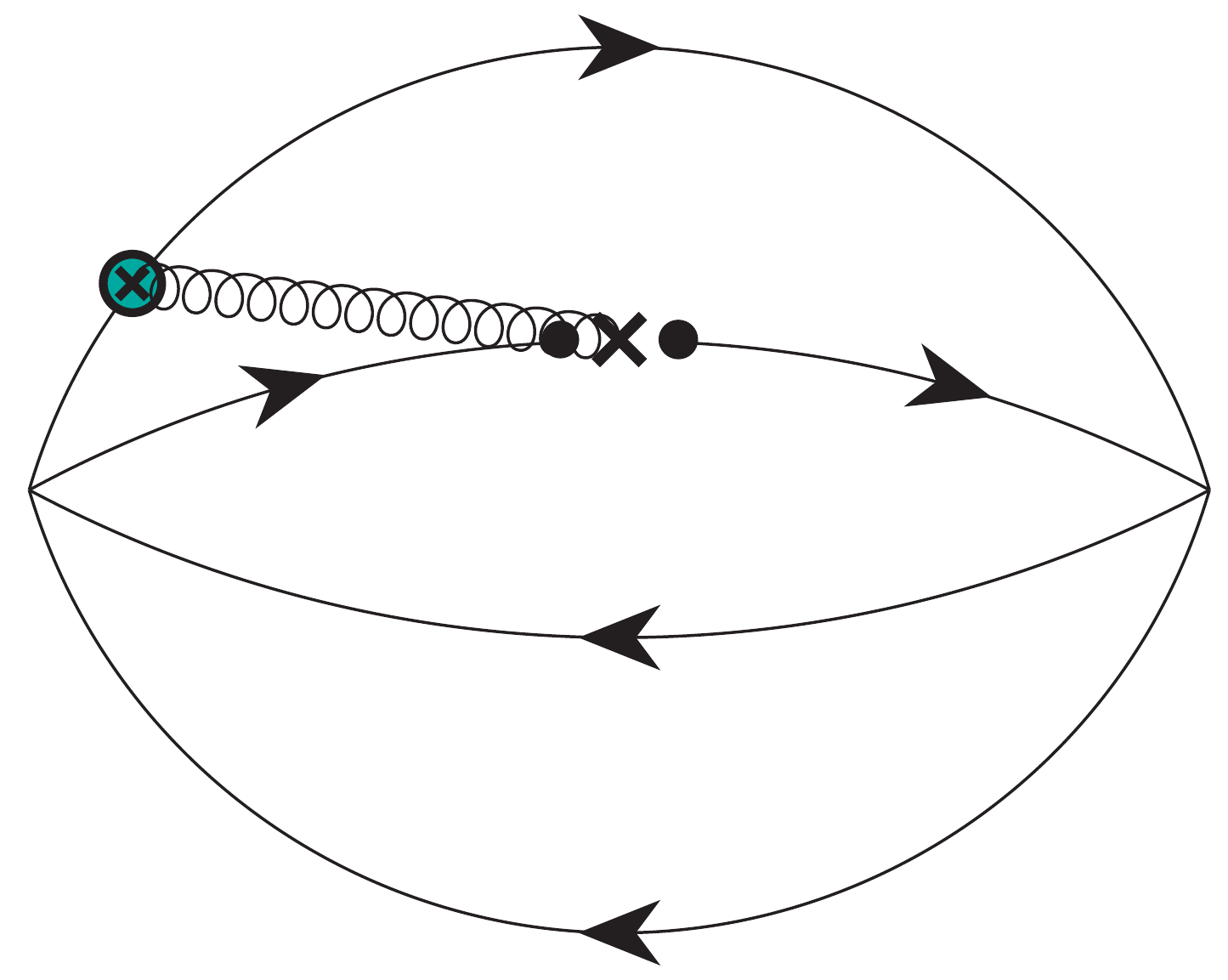}}}~~~~~
\subfigure[(d--5)]{
\scalebox{0.15}{\includegraphics{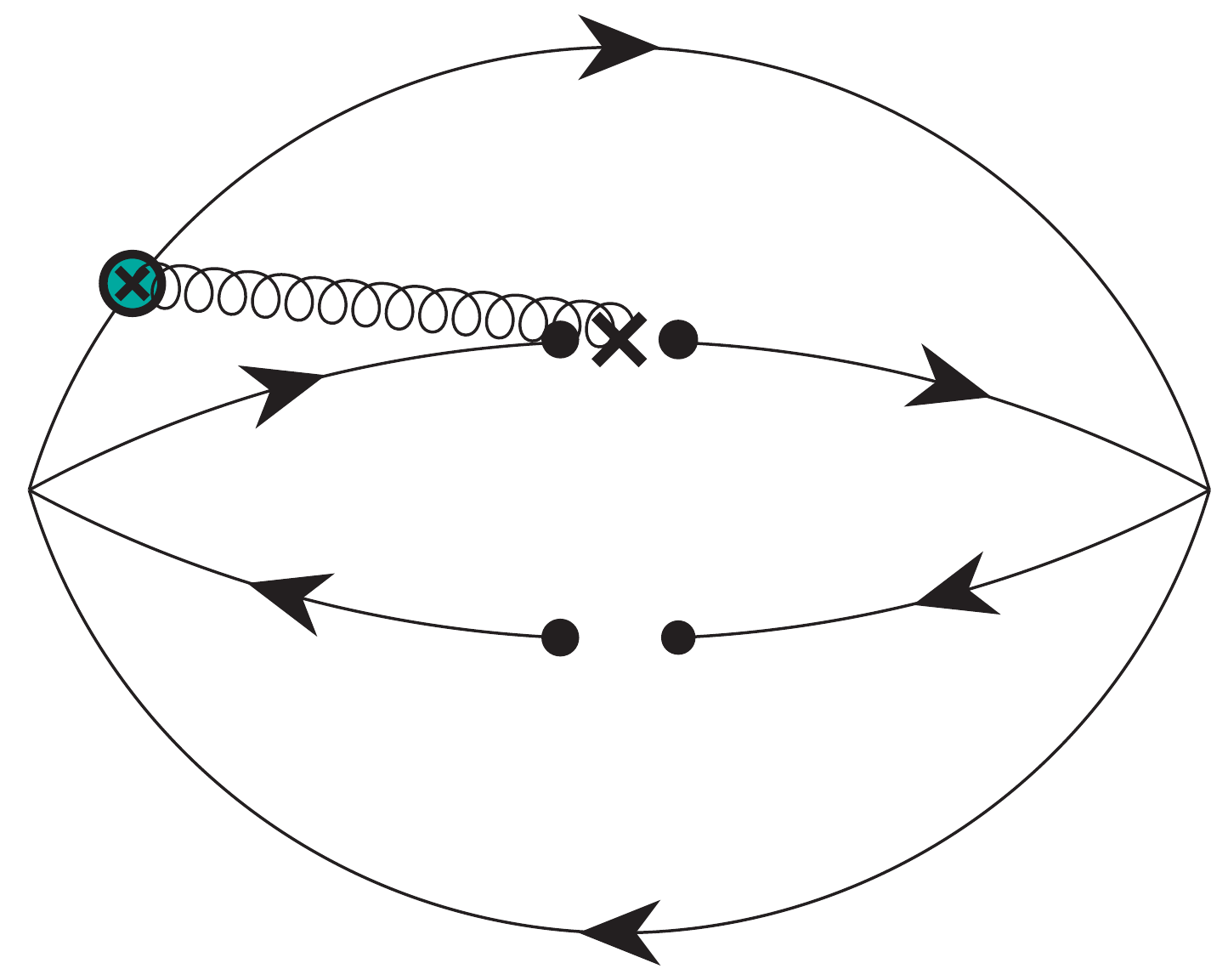}}}~~~~~
\subfigure[(d--6)]{
\scalebox{0.15}{\includegraphics{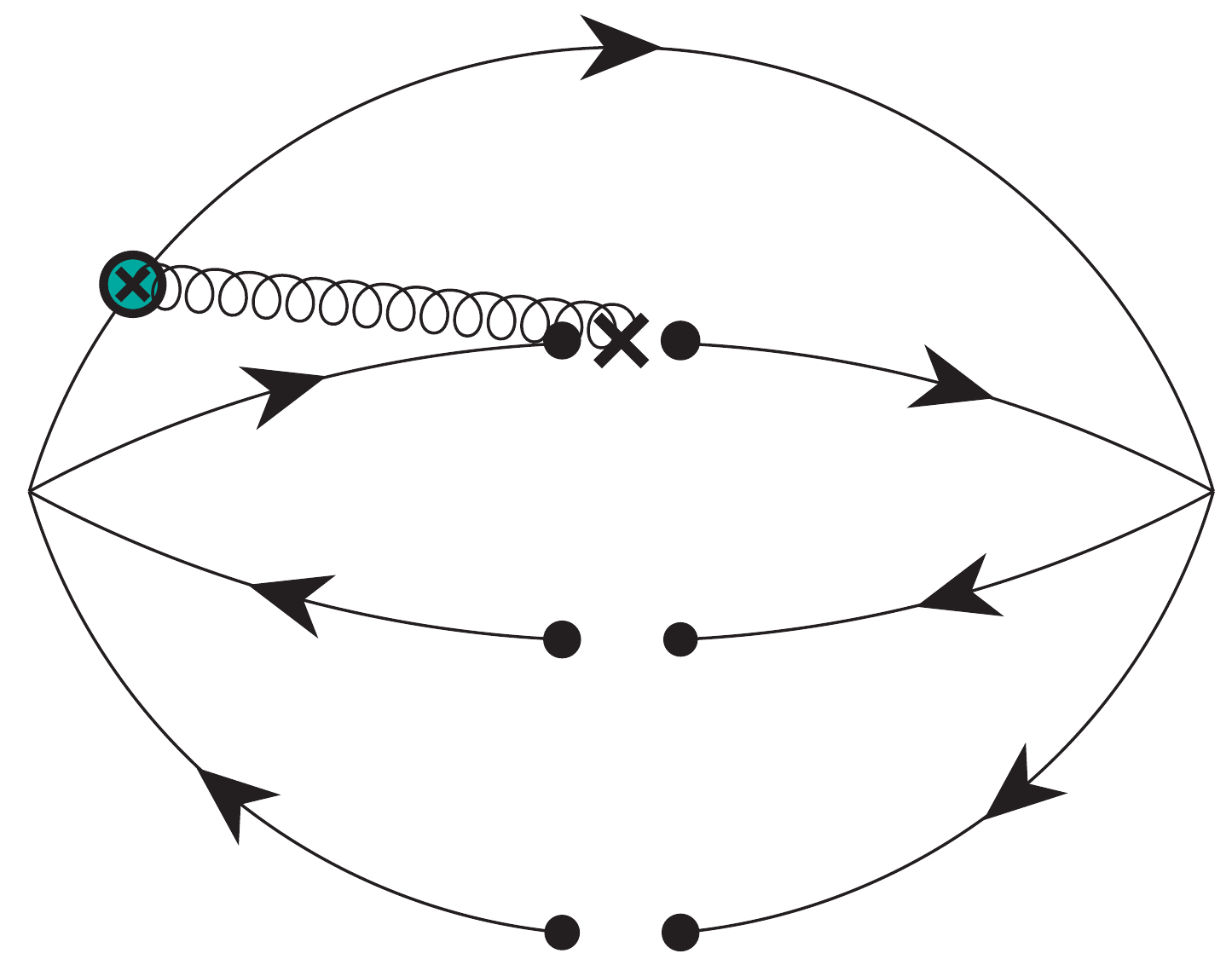}}}
\\[2mm]
\subfigure[(e--1)]{
\scalebox{0.15}{\includegraphics{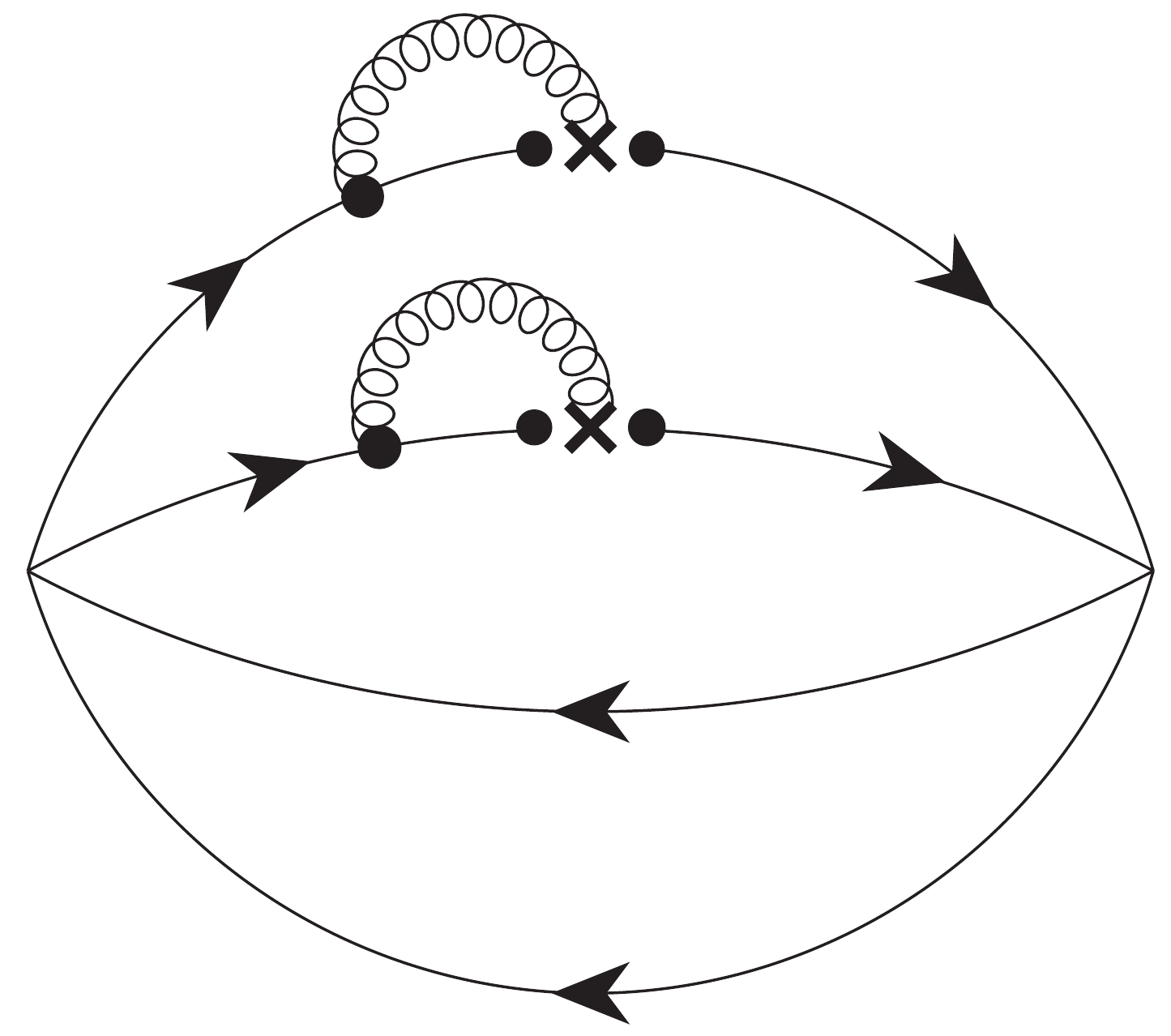}}}~~~~~
\subfigure[(e--2)]{
\scalebox{0.15}{\includegraphics{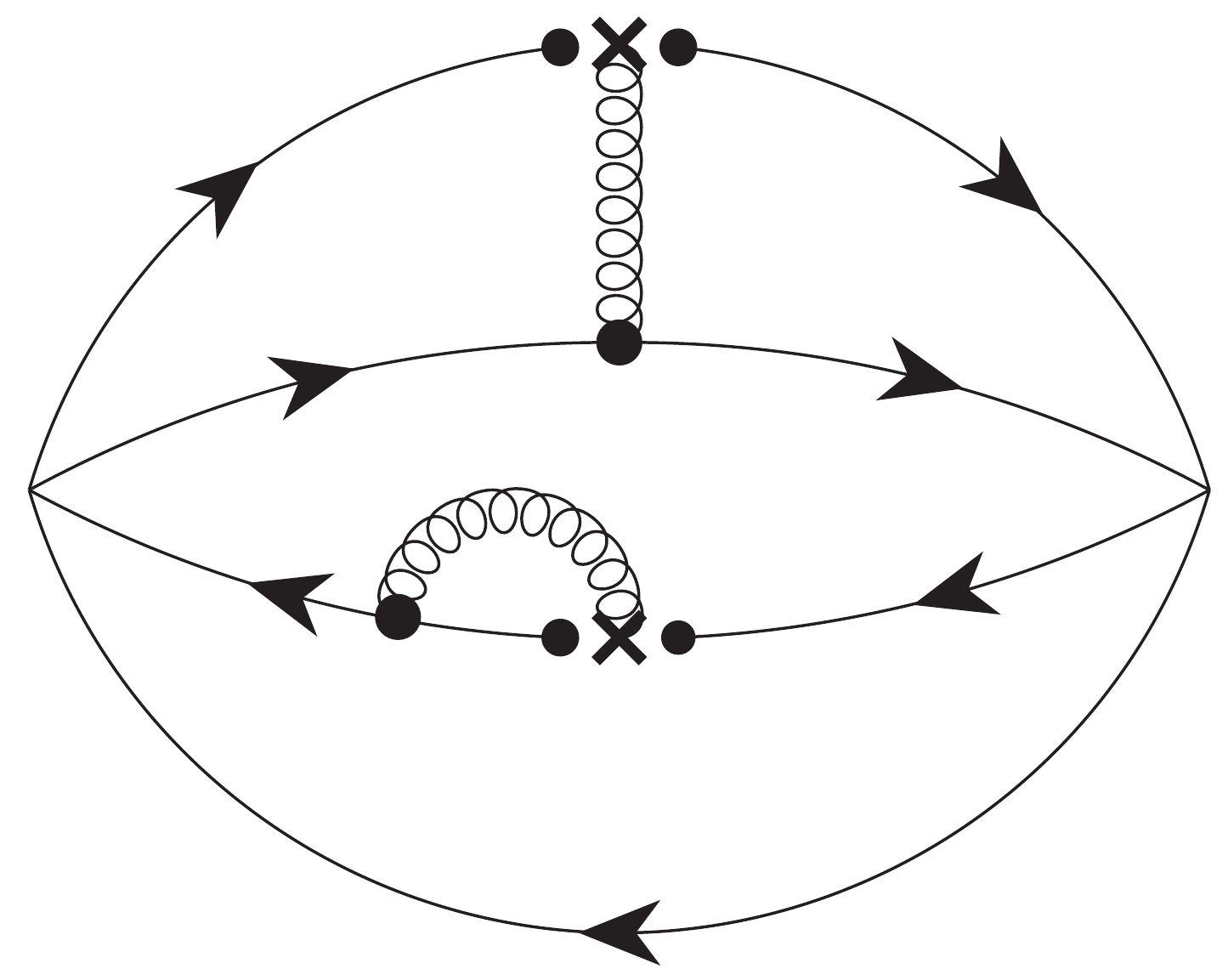}}}~~~~~
\subfigure[(e--3)]{
\scalebox{0.15}{\includegraphics{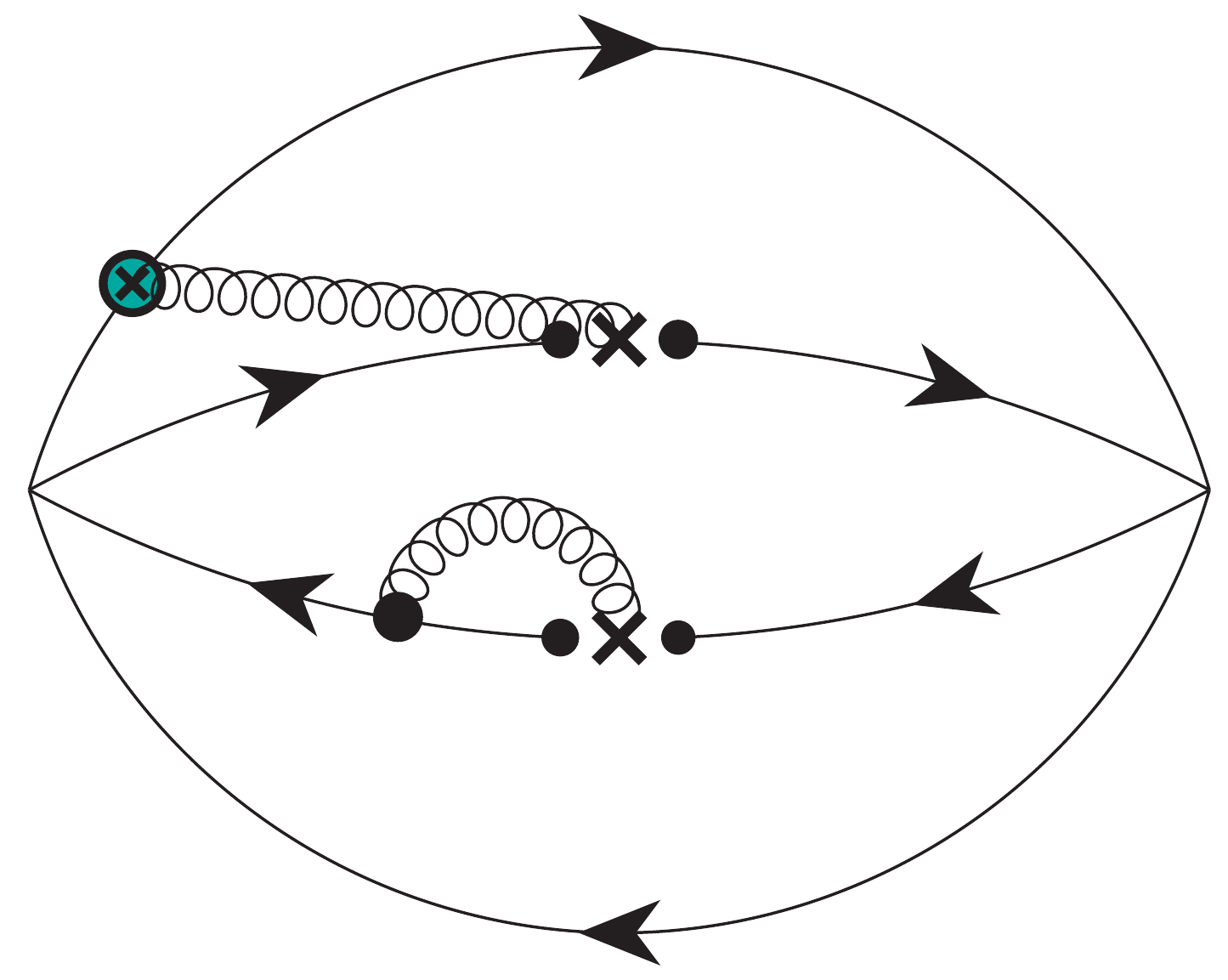}}}~~~~~
\subfigure[(e--4)]{
\scalebox{0.15}{\includegraphics{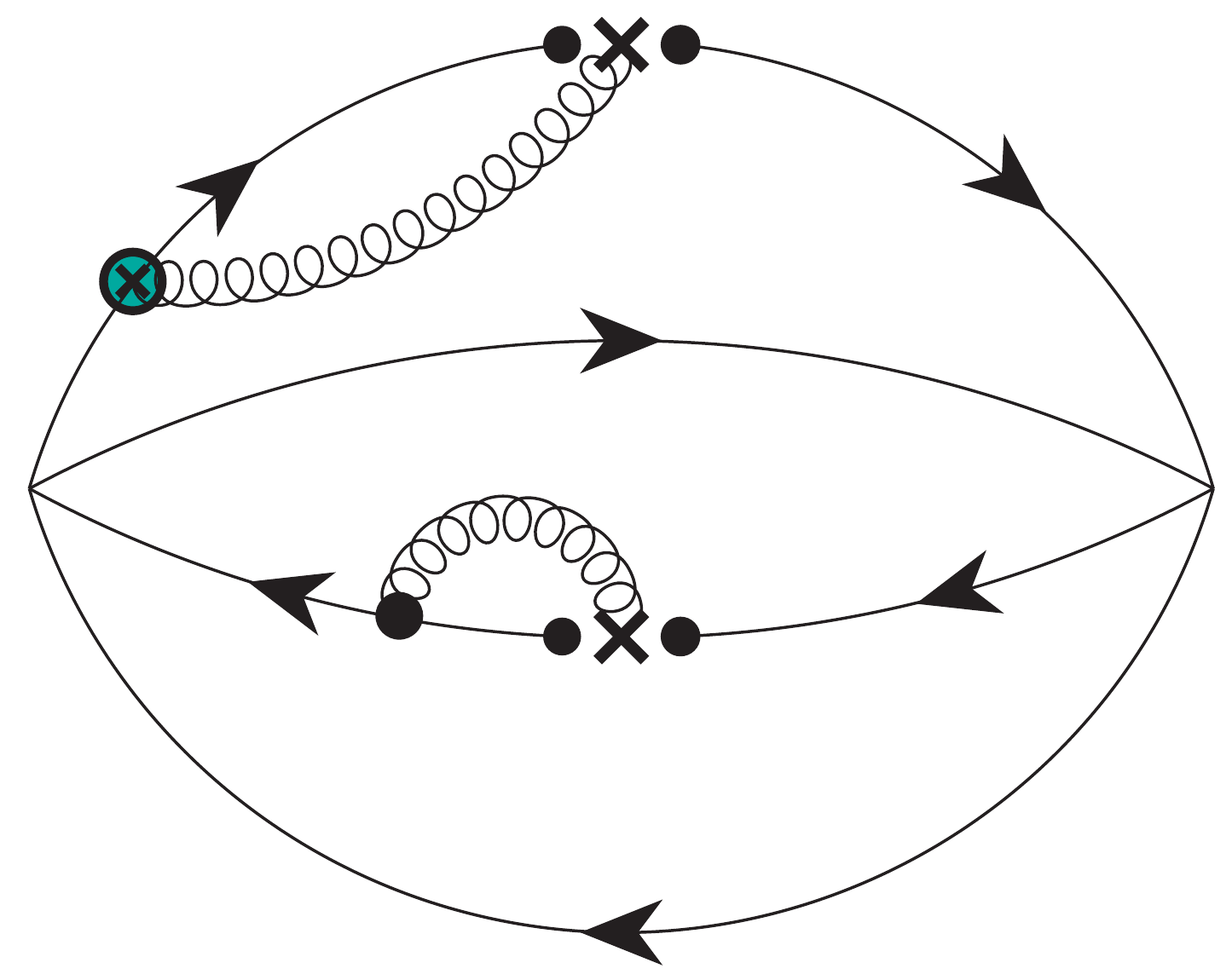}}}
\\[2mm]
\subfigure[(f--1)]{
\scalebox{0.15}{\includegraphics{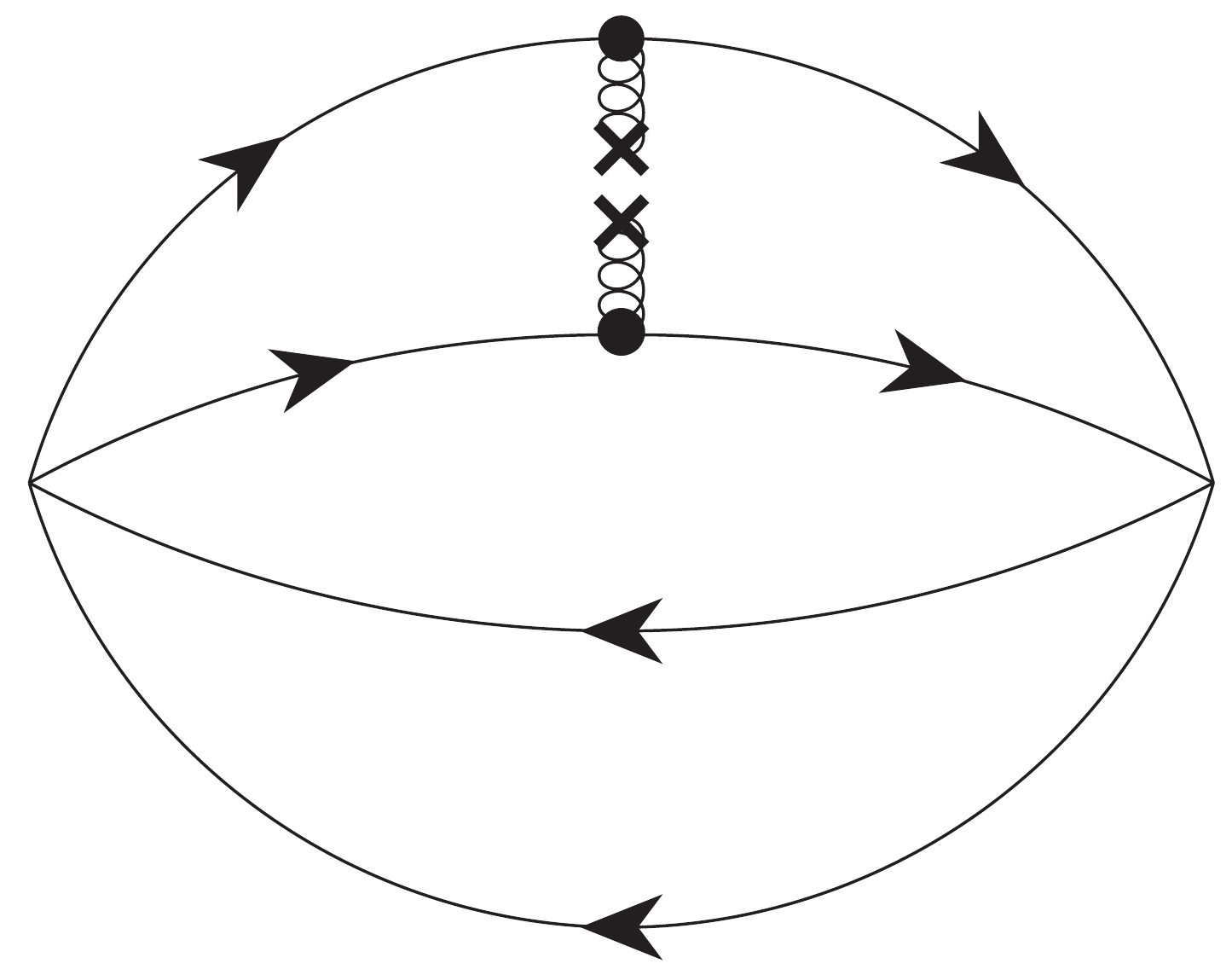}}}~~~~~
\subfigure[(f--2)]{
\scalebox{0.15}{\includegraphics{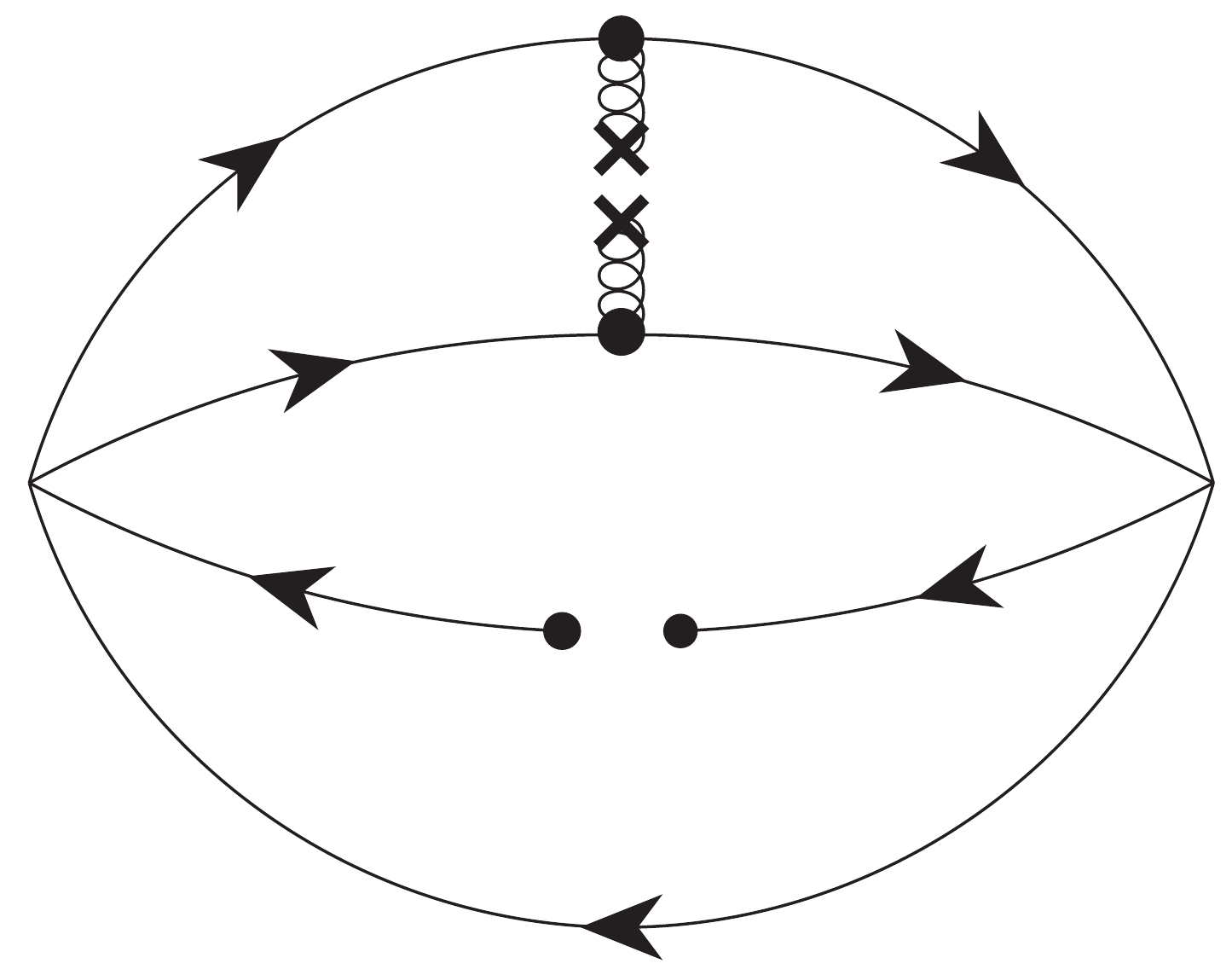}}}~~~~~
\subfigure[(f--3)]{
\scalebox{0.15}{\includegraphics{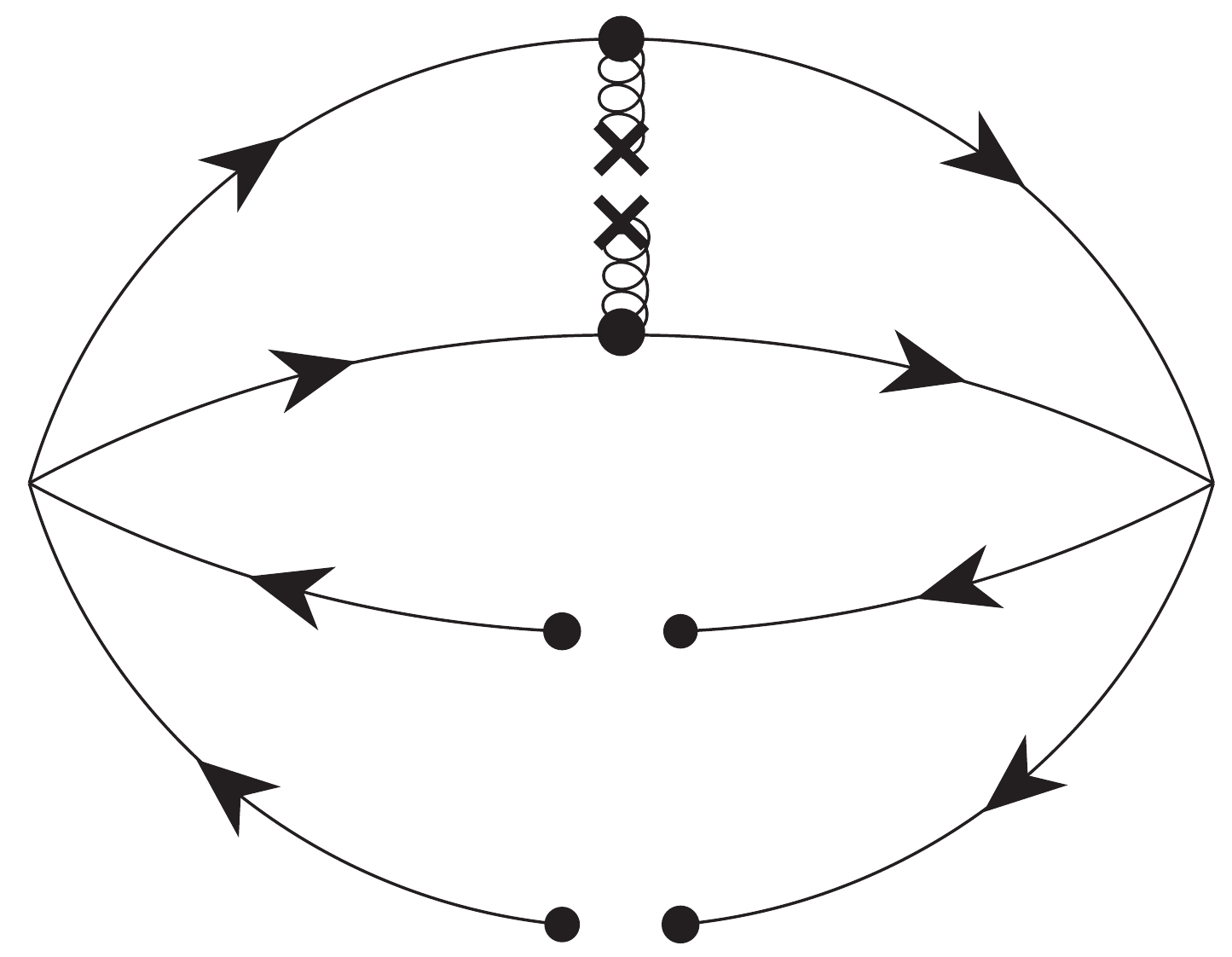}}}~~~~~
\subfigure[(f--4)]{
\scalebox{0.15}{\includegraphics{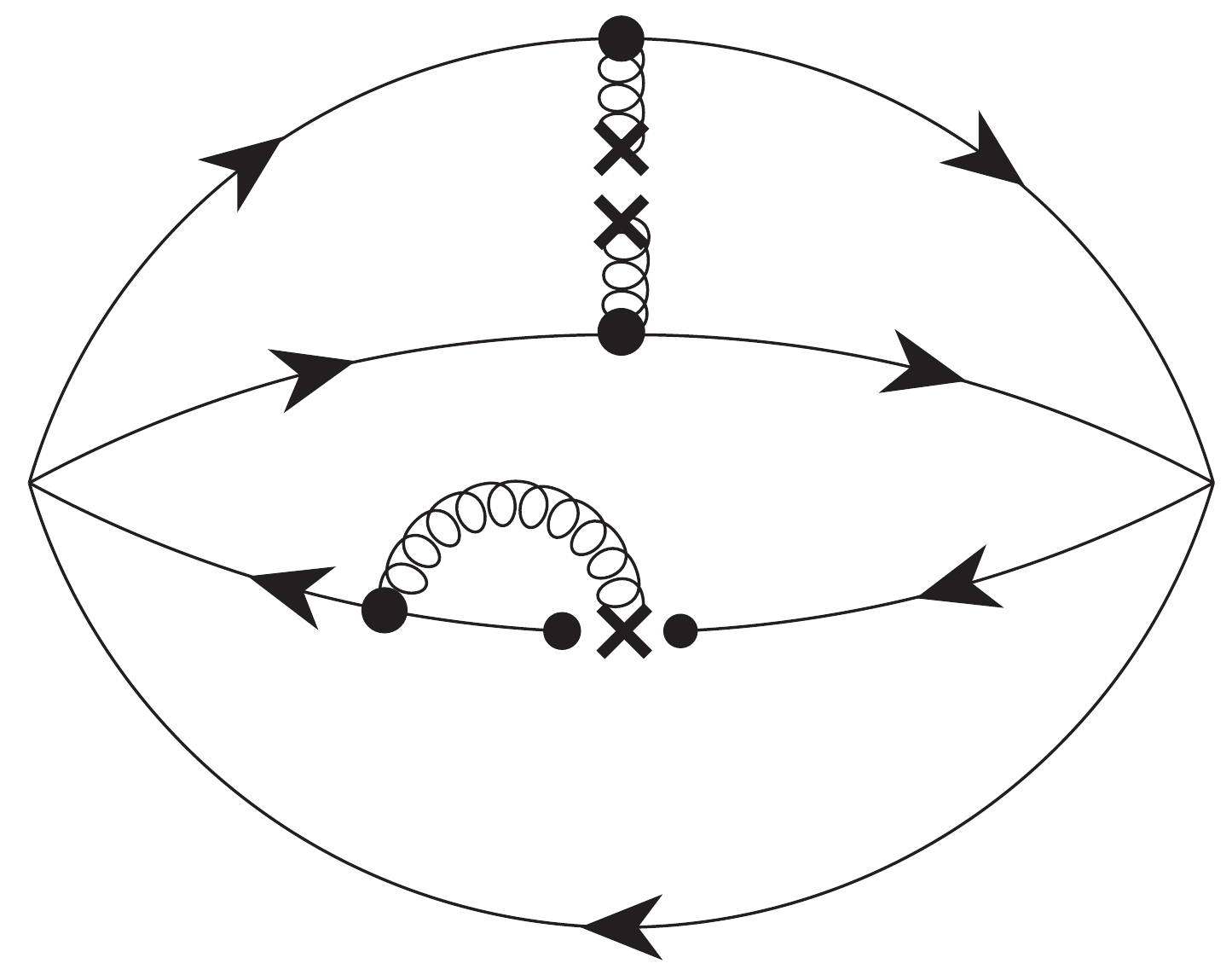}}}~~~~~
\subfigure[(f--5)]{
\scalebox{0.15}{\includegraphics{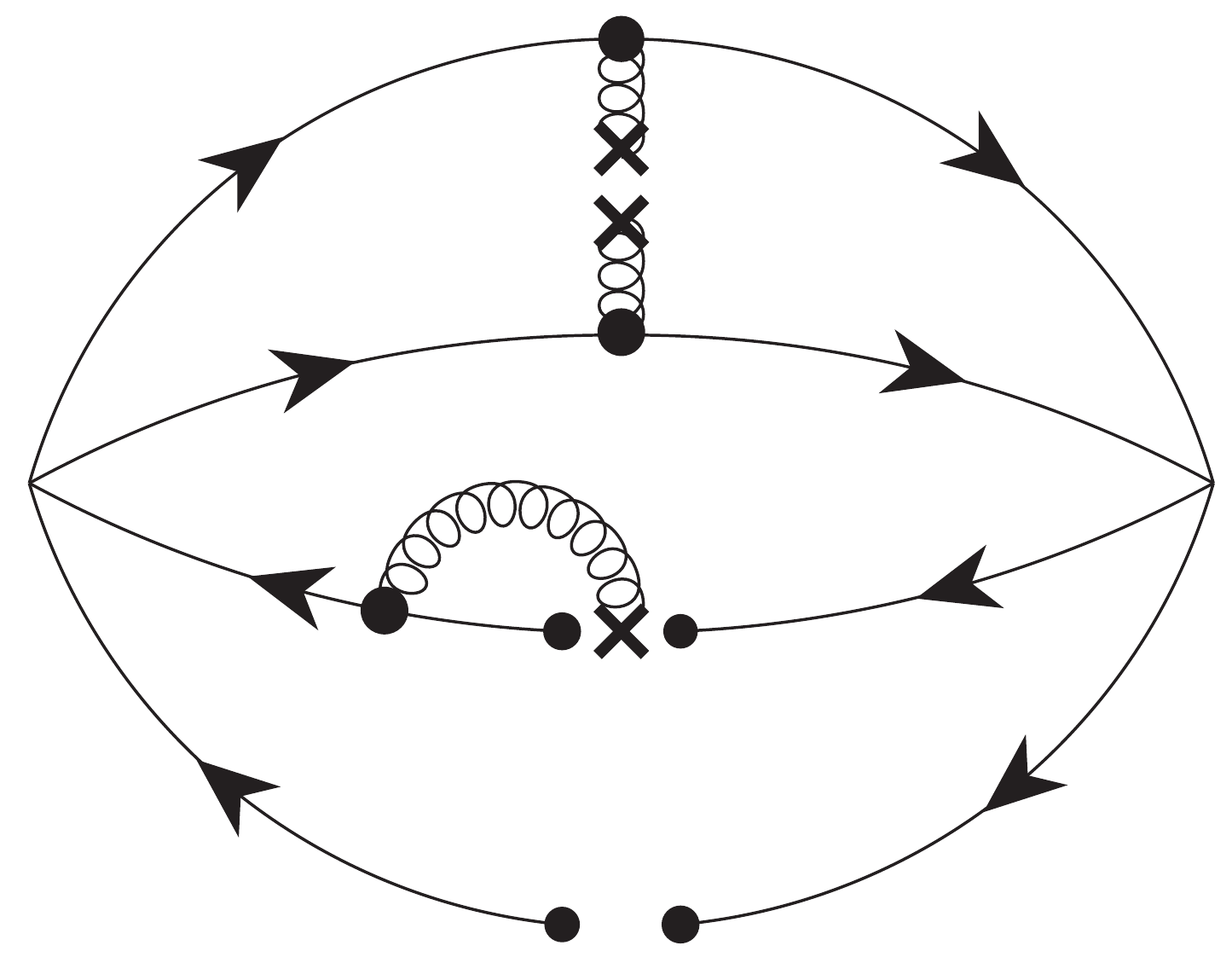}}}
\\[2mm]
\subfigure[(g--1)]{
\scalebox{0.15}{\includegraphics{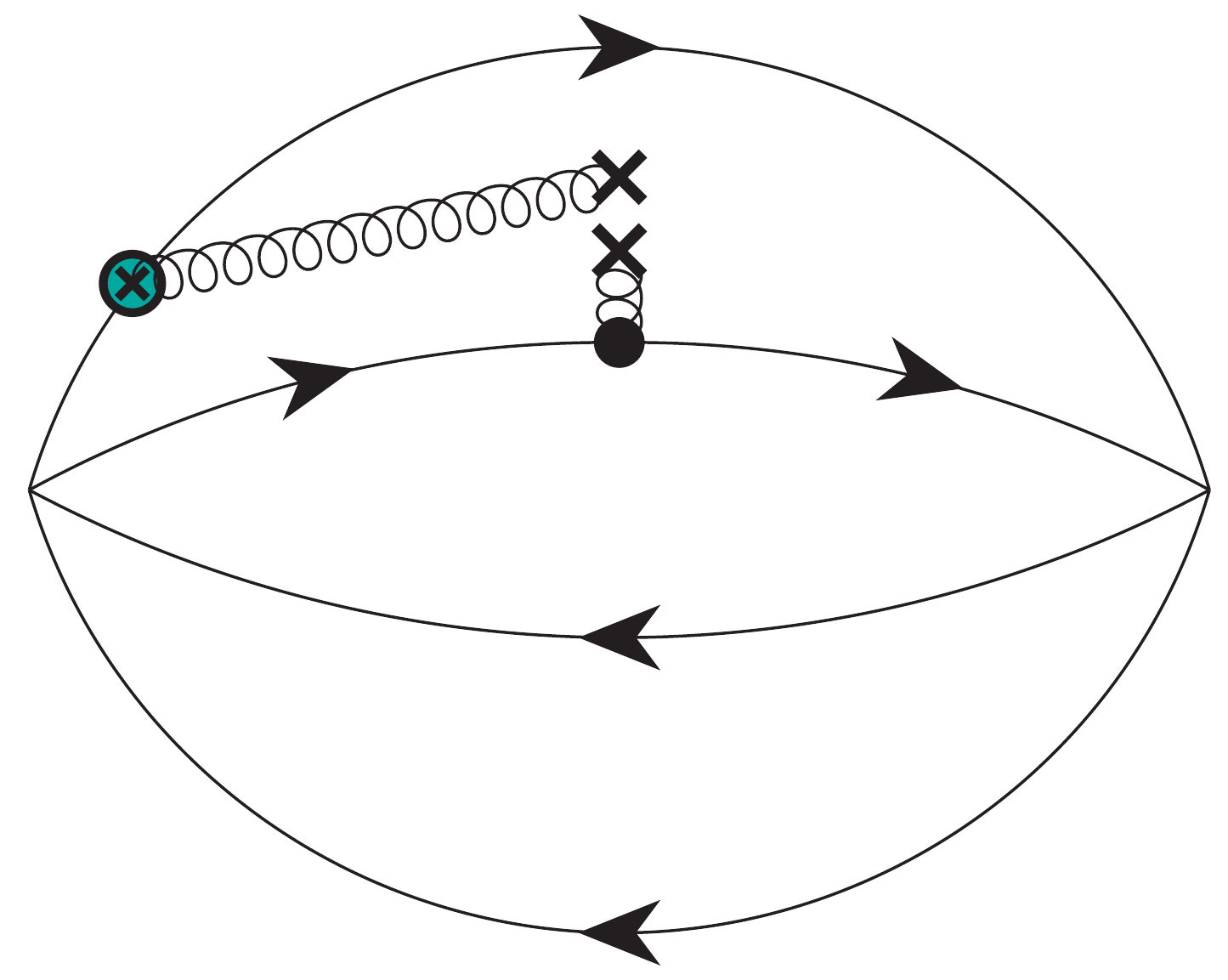}}}~~~~~
\subfigure[(g--2)]{
\scalebox{0.15}{\includegraphics{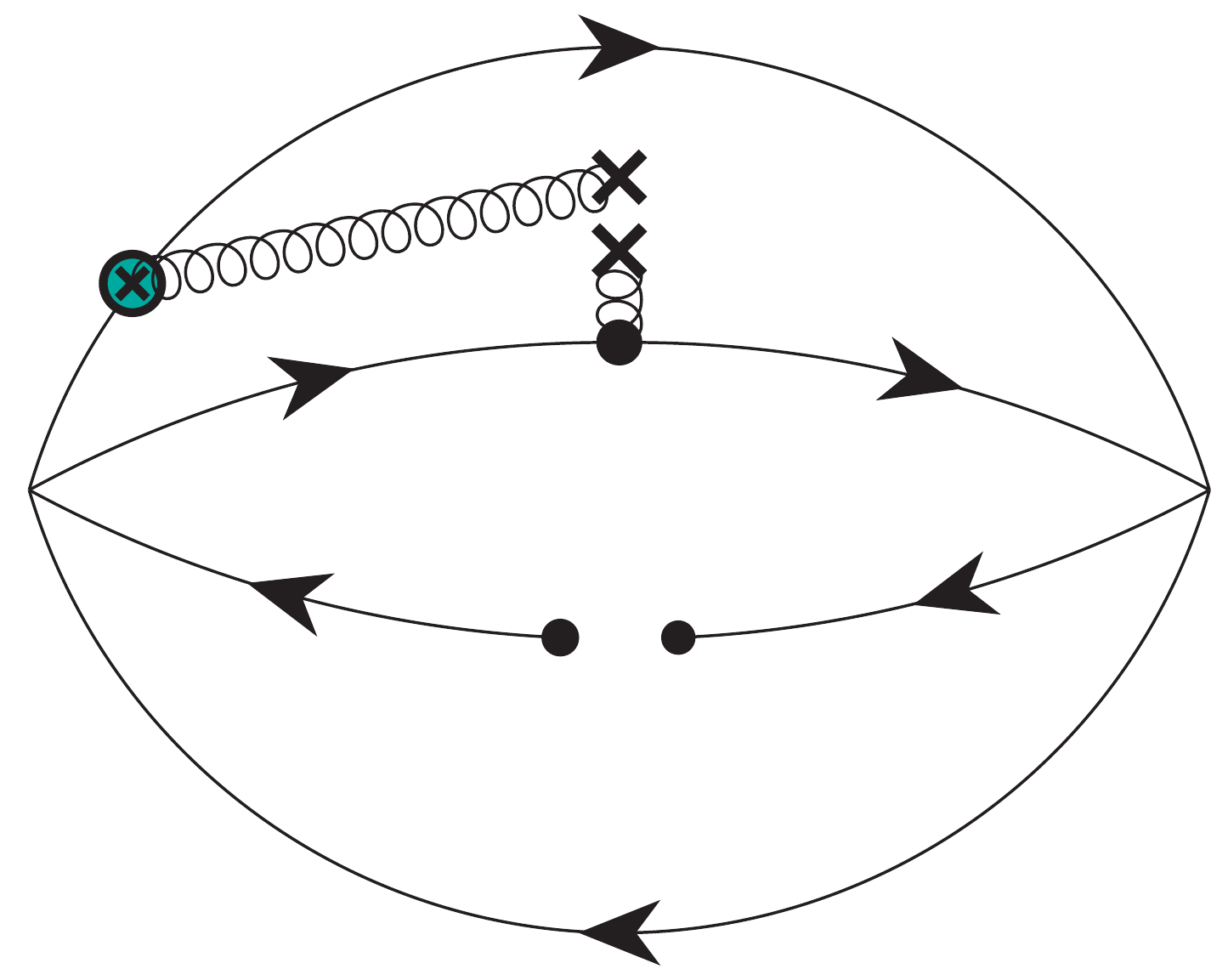}}}~~~~~
\subfigure[(g--3)]{
\scalebox{0.15}{\includegraphics{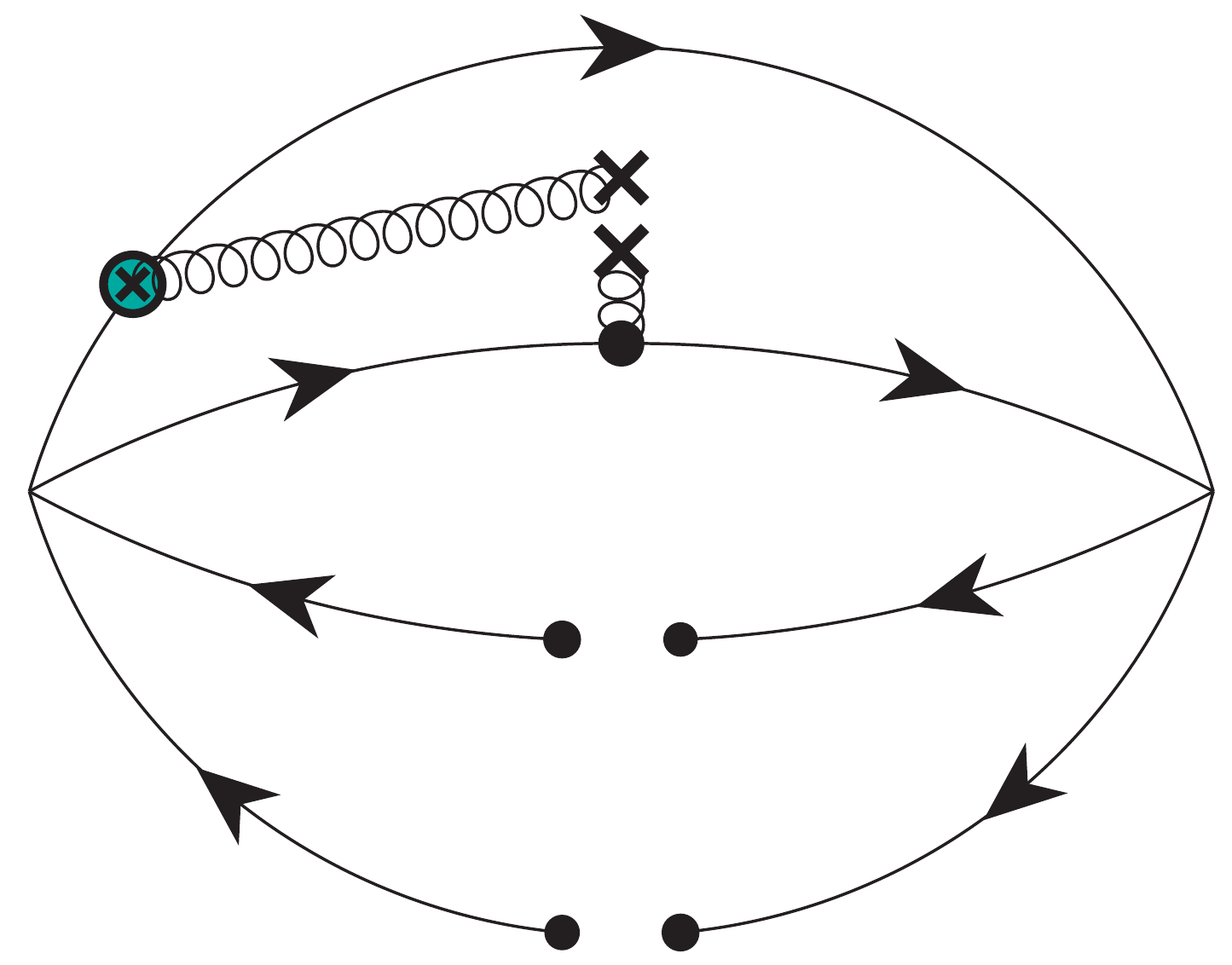}}}~~~~~
\subfigure[(g--4)]{
\scalebox{0.15}{\includegraphics{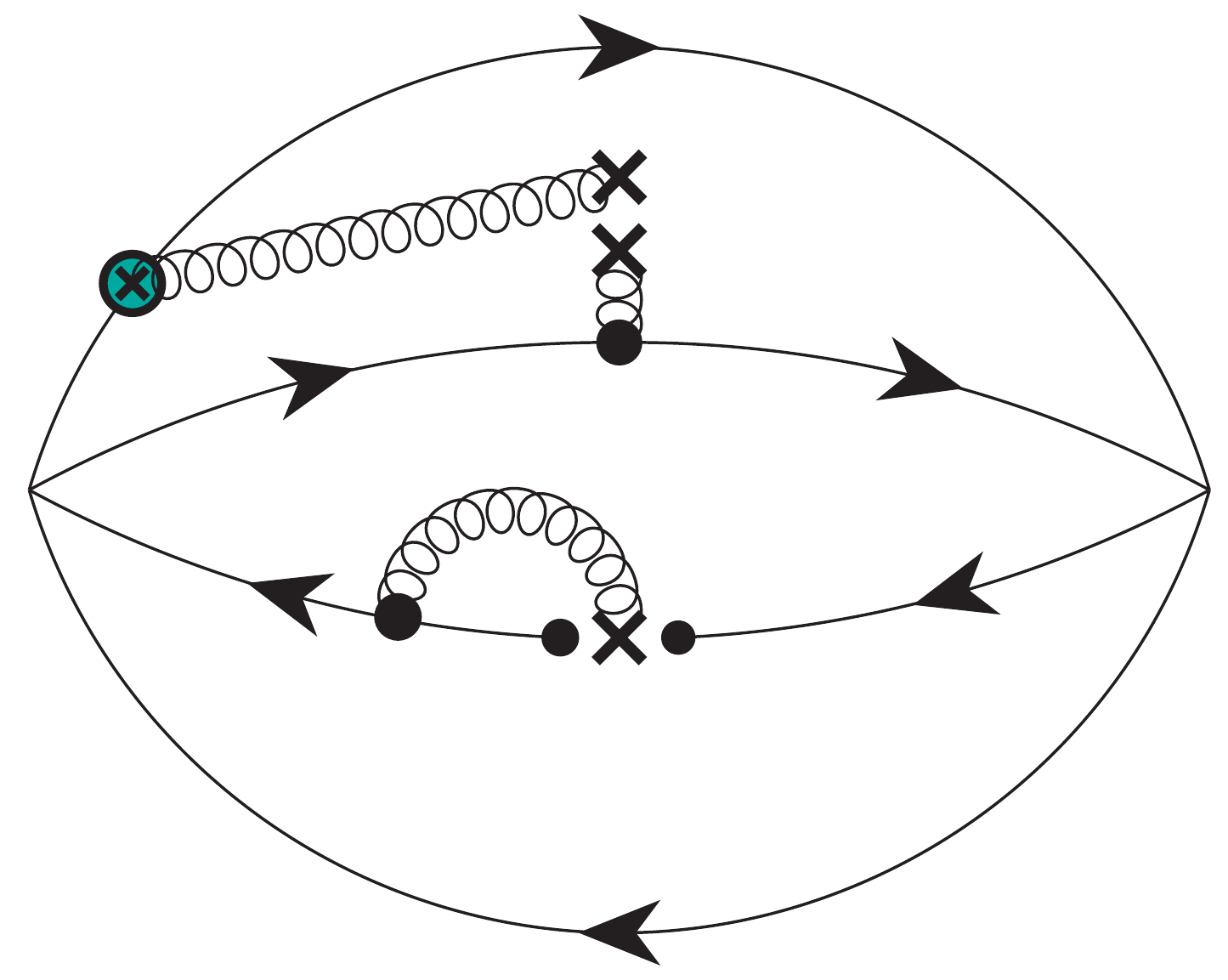}}}~~~~~
\subfigure[(g--5)]{
\scalebox{0.15}{\includegraphics{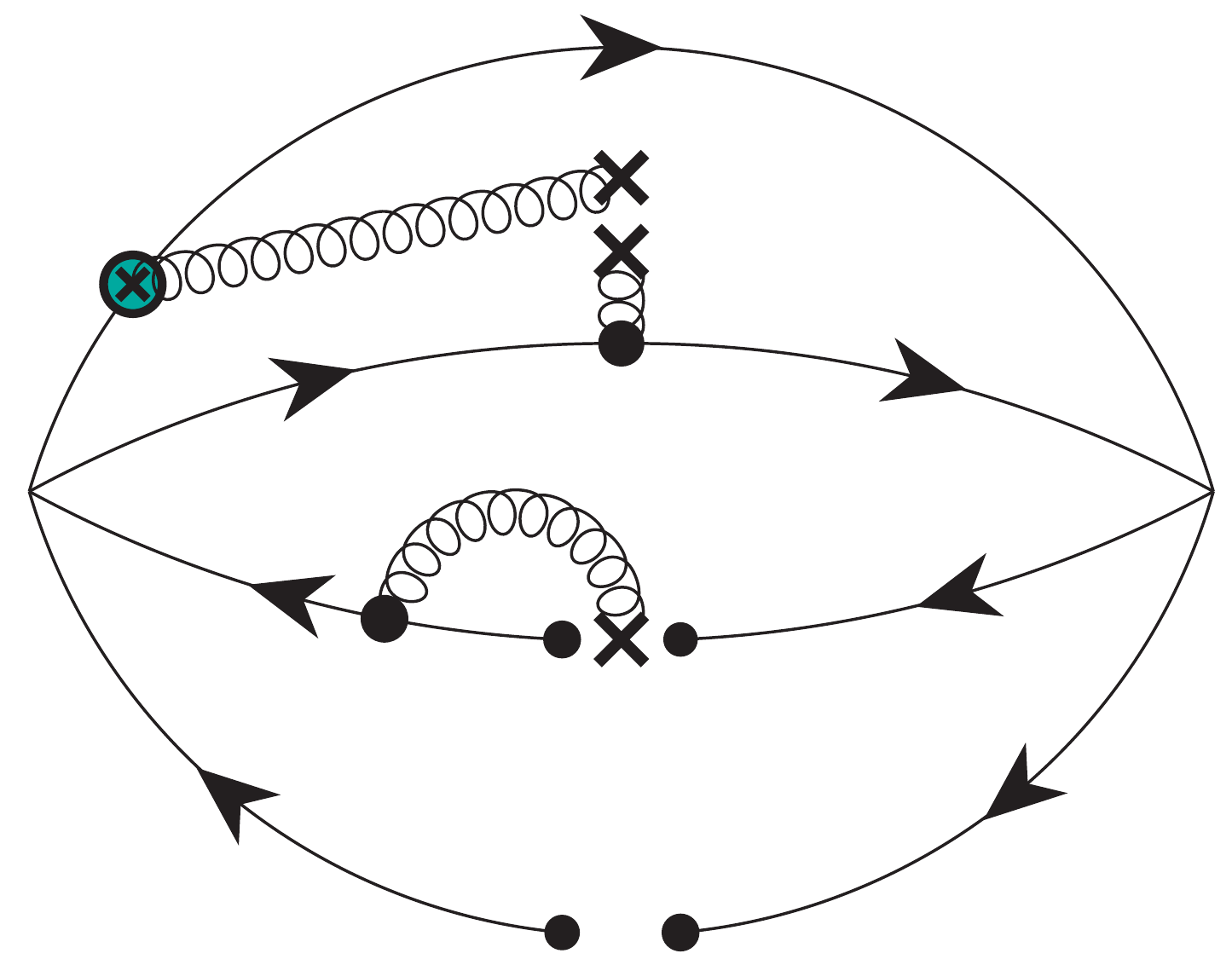}}}
\end{center}
\caption{Feynman diagrams for the fully-strange tetraquark states, including the perturbative term, the strange quark mass $m_s$, the quark condensate $\langle \bar s s \rangle$, the gluon condensate $\langle g_s^2 GG\rangle$, the quark-gluon mixed condensate $\langle \bar g_s s \sigma G s \rangle$, and their combinations. The diagrams (a) and (b--i) are proportional to $g_s^{N=0}$; the diagrams (c--i) and (d--i) are proportional to $g_s^{N=1}$; the diagrams (e--i), (f--i), and (g--i) are proportional to $g_s^{N\geq2}$.}
\label{fig:feynman}
\end{figure*}

In this study we take into account the Feynman diagrams depicted in Fig.~\ref{fig:feynman}. The covariant derivative operator $D_\alpha = \partial_\alpha + i g_s A_\alpha$ can be naturally separated into two parts, and we depict the latter term using a green vertex, {\it e.g.}, see the diagrams depicted in Fig.~\ref{fig:feynman}(d--i) and Fig.~\ref{fig:feynman}(g--i). In the calculations we have taken into account the perturbative term, the strange quark mass $m_s$, the quark condensate $\langle \bar s s \rangle$, the gluon condensate $\langle g_s^2 GG \rangle$, the quark-gluon mixed condensate $\langle g_s \bar s \sigma G s \rangle$, and their combinations. We have assumed the vacuum saturation for higher dimensional operators, {\it e.g.}, $\langle \bar s s \bar s s \rangle \approx \langle \bar s s \rangle^2$ and $\langle \bar s s g_s \bar s \sigma G s \rangle \approx \langle \bar s s \rangle \langle g_s \bar s \sigma G s \rangle$. Other condensates such as $\langle g_s^3G^3\rangle$ and $\langle g_s \bar s D_\mu G^{\mu\nu} \gamma_\nu s \rangle$ are not considered in the present study. We have taken into account all the diagrams proportional to $g_s^{N=0}$ and $g_s^{N=1}$. We find that the $D=6$ term $\langle \bar s s \rangle^2$ and the $D=8$ term $\langle \bar s s \rangle\langle g_s \bar s \sigma G s \rangle$ are important. We have only partly calculated the diagrams proportional to $g_s^{N\geq2}$, whose contributions are found to be small.

The sum rule equation from the current $J_1^{0^{++}}$ is
\begin{eqnarray}
\nonumber \Pi_{11} &=& \int^{s_0}_{16 m_s^2}  e^{-s/M^2} ds \times \Bigg [
{s^4 \over 30720 \pi^6}-{m_s^2 s^3 \over 768 \pi^6}
\\ \nonumber &&
+ \Big( -{\langle g_s^2 GG \rangle \over 6144 \pi^6}
-{m_s \langle \bar s s \rangle \over 48 \pi^4}\Big )s^2
\\ \nonumber &&
+ \Big ( {\langle \bar s s \rangle^2 \over6\pi^2}
- {m_s \langle g_s \bar s \sigma G s \rangle \over 16 \pi^4}
+ { \langle g_s^2 GG \rangle m_s^2 \over 1024 \pi^6 }\Big )s
\\ \nonumber &&
+{\langle \bar s s \rangle \langle g_s \bar s \sigma G s \rangle \over 6 \pi^2}
+{ \langle g_s^2 GG \rangle m_s \langle \bar s s \rangle \over 384 \pi^4}
-{m_s^2\langle \bar s s \rangle^2\over12\pi^2 } \Bigg ]
\\ \nonumber &&
+ \Big( {\langle g_s \bar s \sigma G s \rangle^2 \over 48\pi^2} -{\langle g_s^2 GG \rangle \langle \bar s s \rangle^2 \over 288\pi^2}
\\ \nonumber &&
~~~ + { \langle g_s^2 GG \rangle m_s \langle g_s \bar s \sigma G s \rangle \over 768\pi^4}
-{4 m_s \langle \bar s s \rangle^3 \over 9} \Big)
\\ \nonumber &&
+ {1 \over M_B^2} \Big( {\langle g_s^2 GG \rangle \langle \bar s s \rangle \langle g_s \bar s \sigma G s \rangle \over 576\pi^2} + { \langle g_s^2 GG \rangle m_s^2 \langle \bar s s \rangle^2 \over  1152 \pi^2}
\\ &&
~~~ -{m_s^2 \langle g_s \bar s \sigma G s \rangle^2\over48\pi^2}+{4m_s\langle \bar s s \rangle^2 \langle g_s \bar s \sigma G s \rangle \over9} \Big) \, .
\label{eq:piJ1}
\end{eqnarray}
Sum rule equations extracted from other currents are given in Appendix~\ref{app:ope}, which will be used to perform numerical analyses using the following values for various QCD parameters~\cite{Yang:1993bp,Narison:2002pw,Gimenez:2005nt,Jamin:2002ev,Ioffe:2002be,Ovchinnikov:1988gk,Ellis:1996xc,pdg}:
%
\begin{eqnarray}
\nonumber m_s(2\mbox{ GeV}) &=& 93 ^{+11}_{-5} \mbox{ MeV} \, ,
\\ \nonumber  \langle g_s^2GG\rangle &=& (0.48\pm 0.14) \mbox{ GeV}^4 \, ,
\\ \langle\bar qq \rangle &=& -(0.240 \pm 0.010)^3 \mbox{ GeV}^3 \, ,
\label{condensates}
\\ \nonumber \langle\bar ss \rangle &=& (0.8\pm 0.1)\times \langle\bar qq \rangle \, ,
\\
\nonumber \langle g_s\bar s\sigma G s\rangle &=& - M_0^2\times\langle\bar ss\rangle \, ,
\\
\nonumber M_0^2 &=& (0.8 \pm 0.2) \mbox{ GeV}^2 \, .
\end{eqnarray}
%

We use the spectral density given in Eq.~(\ref{eq:piJ1}) as an example to perform numerical analyses. It is extracted from the current $J_1^{0^{++}}$, and corresponds to the state $| X_1; 0^{++} \rangle = | ^1S_0,\, ^{\bar 1}{\bar S}_{\bar 0} ;\, J=0 \rangle$. As shown in Eq.~(\ref{eq:LSR}), its mass $M_X$ depends on two free parameters, the threshold value $s_0$ and the Borel mass $M_B$. We analyze three aspects to determine their working regions: a) the convergence of OPE, b) the pole contribution, and c) the mass dependence on these two parameters.

\begin{figure}[hbt]
\begin{center}
\includegraphics[width=0.45\textwidth]{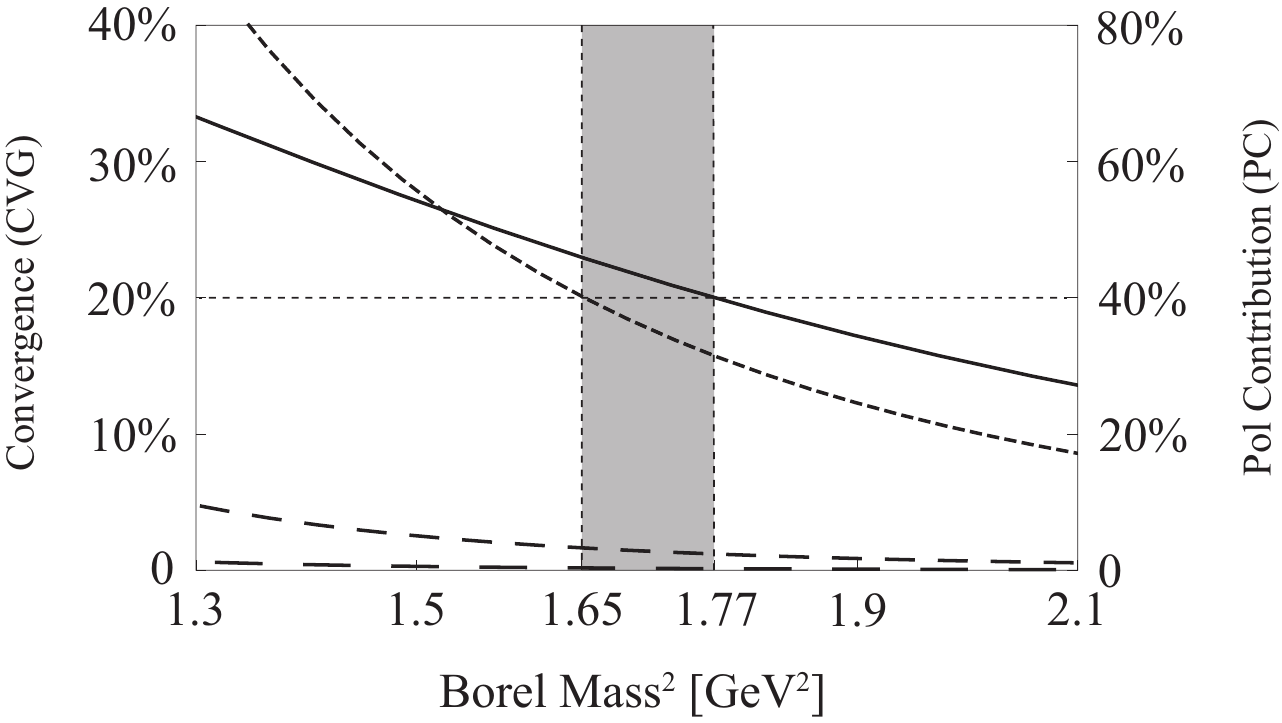}
\caption{CVG$_{12}$ (long dashed curve defined in Eq.~(\ref{eq:convergence12})), CVG$_{10}$ (middle dashed curve defined in Eq.~(\ref{eq:convergence10})), CVG$_{8}$ (short dashed curve defined in Eq.~(\ref{eq:convergence8})), and the pole contribution (solid curve defined in Eq.~(\ref{eq:pole})) as functions of the Borel mass $M_B$. These curves are obtained using the current $J_1^{0^{++}}$ with $s_0 = 6.5$~GeV$^2$.}
\label{fig:cvgpole}
\end{center}
\end{figure}

Firstly, we investigate the convergence of OPE, and require the $D=12/10/8$ terms to be less than 5\%/10\%/20\%, respectively:
\begin{eqnarray}
\mbox{CVG}_{12} &\equiv& \left|\frac{ \Pi_{11}^{D=12}(\infty, M_B^2) }{ \Pi_{11}(\infty, M_B^2) }\right| \leq 5\% \, ,
\label{eq:convergence12}
\\
\mbox{CVG}_{10} &\equiv& \left|\frac{ \Pi_{11}^{D=10}(\infty, M_B^2) }{ \Pi_{11}(\infty, M_B^2) }\right| \leq 10\% \, ,
\label{eq:convergence10}
\\
\mbox{CVG}_{8} &\equiv& \left|\frac{ \Pi_{11}^{D=8}(\infty, M_B^2) }{ \Pi_{11}(\infty, M_B^2) }\right| \leq 20\% \, .
\label{eq:convergence8}
\end{eqnarray}
This is the cornerstone of a reliable QCD sum rule analysis. As shown in Fig.~\ref{fig:cvgpole} using three dashed curves, we determine the lower bound of the Borel mass to be $M_B^2 > 1.65$~GeV$^2$. Since we have only partly calculated the diagrams proportional to $g_s^{N\geq2}$, it is useful to see how large these terms are:
\begin{eqnarray}
\left|\frac{ \Pi_{11}^{g_s^{N=1}}(\infty, M_B^2) }{ \Pi_{11}(\infty, M_B^2) }\right| &\leq& 31.0\% \, ,
\\
\left|\frac{ \Pi_{11}^{g_s^{N\geq2}}(\infty, M_B^2) }{ \Pi_{11}(\infty, M_B^2) }\right| &\leq& 4.7\% \, .
\end{eqnarray}

Secondly, we investigate the one-pole-dominance assumption, and require the pole contribution to be larger than 40\%:
\begin{equation}
\mbox{Pole contribution} \equiv \left|\frac{ \Pi_{11}(s_0, M_B^2) }{ \Pi_{11}(\infty, M_B^2) }\right| \geq 40\% \, .
\label{eq:pole}
\end{equation}
As shown in Fig.~\ref{fig:cvgpole} using the solid curve, we determine the upper bound of the Borel mass to be $M_B^2 < 1.77$~GeV$^2$, when setting $s_0$ to be $s_0 = 6.5$~GeV$^2$. Altogether the Borel window is extracted to be $1.65$~GeV$^2 < M_B^2 < 1.77$~GeV$^2$ for $s_0 = 6.5$~GeV$^2$. We repeat the same procedures by changing $s_0$, and find that there are non-vanishing Borel windows as long as $s_0 > s_{0}^{min} = 6.0$~GeV$^2$.

\begin{figure*}[]
\begin{center}
\includegraphics[width=0.45\textwidth]{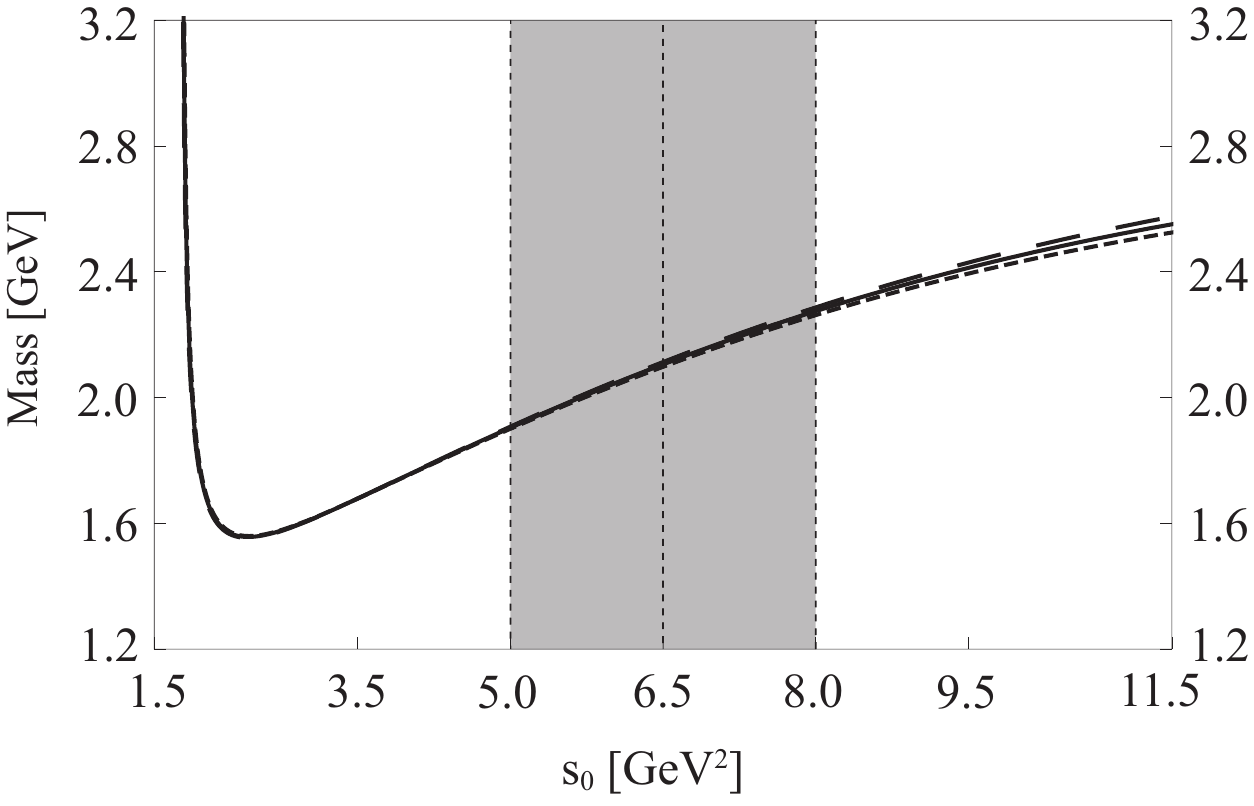}
~~~~~
\includegraphics[width=0.45\textwidth]{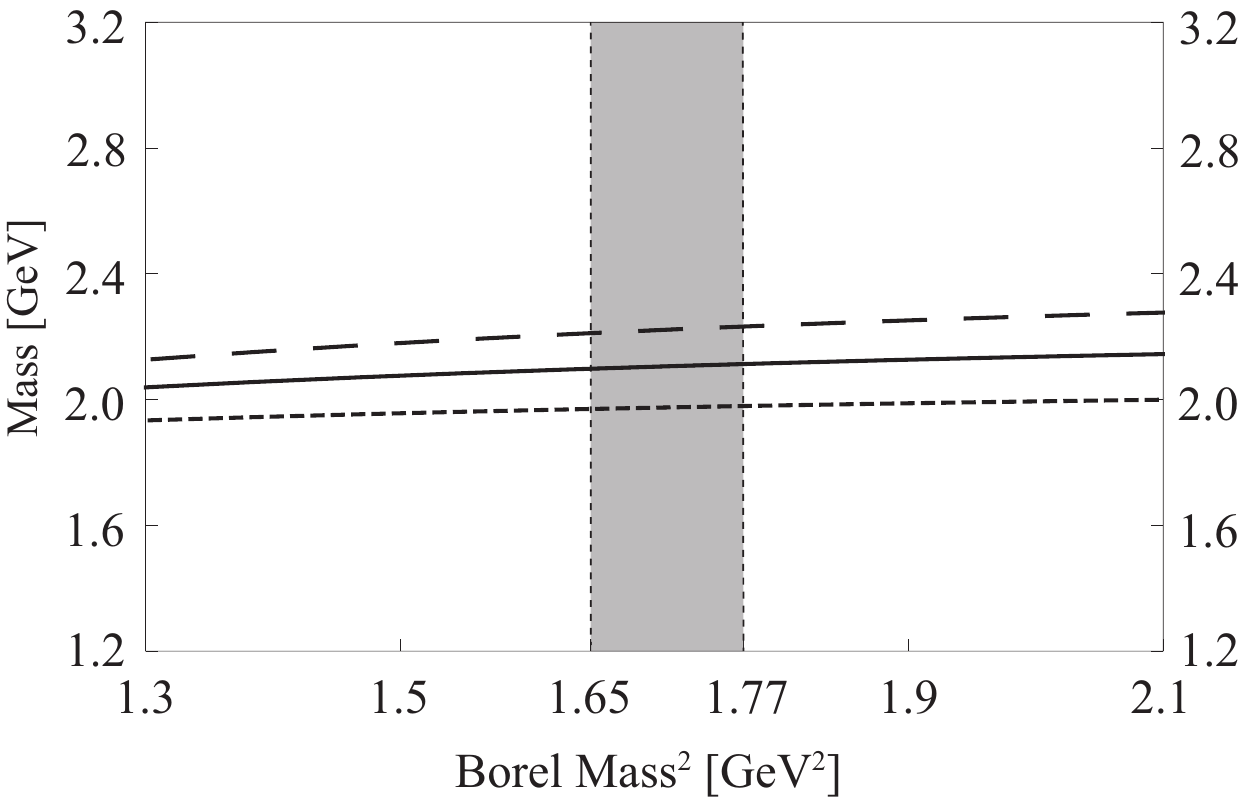}
\caption{
Mass of $| X_1; 0^{++} \rangle$ extracted from the current $J_1^{0^{++}}$, as a function of the threshold value $s_0$ (left) and the Borel mass $M_B$ (right).
In the left panel the short-dashed/solid/long-dashed curves are plotted by setting $M_B^2 = 1.65/1.71/1.77$ GeV$^2$, respectively.
In the right panel the short-dashed/solid/long-dashed curves are plotted by setting $s_0 = 5.5/6.5/7.5$ GeV$^2$, respectively.}
\label{fig:J1mass}
\end{center}
\end{figure*}

Thirdly, we investigate the mass dependence on $s_0$ and $M_B$. We show $M_X$ in Fig.~\ref{fig:J1mass} with respective to these two parameters. Its dependence on $s_0$ is moderate around $s_0 \sim 6.5$~GeV$^2$, and its dependence on $M_B$ is weak inside the Borel window $1.65$~GeV$^2 < M_B^2 < 1.77$~GeV$^2$. Accordingly, we choose the working regions of $s_0$ and $M_B$ to be $5.0$~GeV$^2< s_0 < 8.0$~GeV$^2$ and $1.65$~GeV$^2 < M_B^2 < 1.77$~GeV$^2$, where the mass of $| X_1; 0^{++} \rangle$ is evaluated to be
\begin{equation}
M_{| X_1; 0^{++} \rangle} = 2.11^{+0.19}_{-0.21}{\rm~GeV} \, .
\end{equation}
Its central value corresponds to $s_0=6.5$~GeV$^2$ and $M_B^2 = 1.71$~GeV$^2$, and its uncertainty comes from the threshold value $s_0$, the Borel mass $M_B$, and various QCD parameters listed in Eqs.~(\ref{condensates}).

We repeat the same procedures to study the other twenty-three currents defined in Eqs.~(\ref{def:current2}-\ref{def:current4}), Eqs.~(\ref{def:current5}-\ref{def:current12}), and Eqs.~(\ref{def:current13}-\ref{def:current24}). The obtained results are summarized in Table~\ref{tab:results}, where we choose $s_0 = 6.5$~GeV$^2$ for all the $S$-wave $s s \bar s \bar s$ tetraquark states, and $s_0 = 8.5 \sim 9.0$~GeV$^2$ for some of the $P$-wave $s s \bar s \bar s$ tetraquark states. We shall use these results to draw conclusions in the next section. The minimum threshold value $s_0^{min}$ is larger than $9.0$~GeV$^2$ for the tetraquark currents $J_{6,8,9,13,15}^{\cdots}$, suggesting that their sum rule results may not be very well. Therefore, we shall not use them to draw conclusions, but note that this does not indicate the non-existence of their corresponding tetraquark states $|X_{6,8,9,13,15};J^{PC}\rangle$.

%
\section{Summary and Discussions}
\label{sec:summary}

\begin{table*}[hpt]
\begin{center}
\renewcommand{\arraystretch}{1.5}
\caption{QCD sum rule results of the $S$- and $P$-wave fully-strange tetraquark states, with possible experimental candidates given for comparisons.}
\begin{tabular}{c|c|c|c|c|c|c|c}
\hline\hline
~~\multirow{2}{*}{Currents}~~&~~~~~~~~~~~~\multirow{2}{*}{Configuration}~~~~~~~~~~~~&~$s_0^{min}$~& \multicolumn{2}{c|}{Working Regions}&~~\multirow{2}{*}{Pole~[\%]}~~&~\multirow{2}{*}{~Mass~[GeV]~}~&\multirow{2}{*}{Candidate}
\\ \cline{4-5}
&&~~$[{\rm GeV}^2]$~~&~~$M_B^2~[{\rm GeV}^2]$~~&~~$s_0~[{\rm GeV}^2]$~~&&&
\\ \hline\hline
$J_{1}^{0^{++}}$  & $|X_1;0^{++}\rangle\sim| ^1S_0,\, ^{\bar 1}{\bar S}_{\bar 0} ;\, J=0 \rangle$              &  6.0   &  $1.65$--$1.77$   &  $6.5\pm1.5$  &  $40$--$46$  &  $2.11^{+0.19}_{-0.21}$  &  $f_0(2100)$
\\
$J_{2}^{0^{++}}$  & $|X_2;0^{++}\rangle\sim| ^3S_1,\, ^{\bar 3}{\bar S}_{\bar 1} ;\, J=0 \rangle$              &  2.9   &  $1.27$--$1.76$   &  $6.5\pm1.5$  &  $40$--$70$  &  $1.99^{+0.19}_{-0.24}$  &  --
\\
$J_{3}^{1^{+-}}$  & $|X_3;1^{+-}\rangle\sim| ^3S_1,\, ^{\bar 3}{\bar S}_{\bar 1} ;\, J=1 \rangle $             &  5.4   &  $1.69$--$1.96$   &  $6.5\pm1.5$  &  $40$--$53$  &  $2.06^{+0.18}_{-0.20}$  &  $X(2063)$
\\
$J_{4}^{2^{++}}$  & $|X_4;2^{++}\rangle\sim| ^3S_1,\, ^{\bar 3}{\bar S}_{\bar 1} ;\, J=2 \rangle$              &  5.4   &  $1.53$--$1.77$   &  $6.5\pm1.5$  &  $40$--$53$  &  $2.09^{+0.19}_{-0.22}$  &  $f_2(2010)$
\\ \hline
$J_{5}^{0^{-+}}$  & $|X_5;0^{-+}\rangle\sim| ^3S_1,\, ^{\bar 3}{\bar S}_{\bar 1};\, 1, 0, \lambda \rangle$     &  3.5   &  $1.48$--$1.99$   &  $8.5\pm2.0$  &  $40$--$70$  &  $2.31^{+0.21}_{-0.26}$  &  $X(2370)$
\\
$J_{6}^{1^{--}}$  & $|X_6;1^{--}\rangle\sim| ^1S_0,\, ^{\bar 1}{\bar S}_{\bar 0};\, 0, 1, \lambda \rangle$     &  9.5   &  --               &  --           &  --          &  --                      &  --
\\
$J_{7}^{1^{--}}$  & $|X_7;1^{--}\rangle\sim| ^3S_1,\, ^{\bar 3}{\bar S}_{\bar 1};\, 0, 1, \lambda \rangle$     &  7.3   &  $1.59$--$1.77$   &  $8.5\pm2.0$  &  $40$--$52$  &  $2.34^{+0.23}_{-0.30}$  &  $\phi(2170)$
\\
$J_{8}^{1^{--}}$  & $|X_8;1^{--}\rangle\sim| ^3S_1,\, ^{\bar 3}{\bar S}_{\bar 1};\, 2, 1, \lambda \rangle$     &  10.1  &  --               &  --           &  --          &  --                      &  --
\\
$J_{9}^{1^{-+}}$  & $|X_9;1^{-+}\rangle\sim| ^3S_1,\, ^{\bar 3}{\bar S}_{\bar 1};\, 1, 1, \lambda \rangle$     &  10.5  &  --               &  --           &  --          &  --                      &  --
\\
$J_{10}^{2^{--}}$ & $|X_{10};2^{--}\rangle\sim| ^3S_1,\, ^{\bar 3}{\bar S}_{\bar 1};\, 2, 2, \lambda \rangle$  &  6.0   &  $1.38$--$1.76$   &  $8.5\pm2.0$  &  $40$--$66$  &  $2.32^{+0.23}_{-0.30}$  &  --
\\
$J_{11}^{2^{-+}}$ & $|X_{11};2^{-+}\rangle\sim| ^3S_1,\, ^{\bar 3}{\bar S}_{\bar 1};\, 1, 2, \lambda \rangle$  &  6.3   &  $1.51$--$1.93$   &  $8.5\pm2.0$  &  $40$--$64$  &  $2.40^{+0.20}_{-0.25}$  &  --
\\
$J_{12}^{3^{--}}$ & $|X_{12};3^{--}\rangle\sim| ^3S_1,\, ^{\bar 3}{\bar S}_{\bar 1};\, 2, 3, \lambda \rangle$  &  9.0   &  $1.94$--$1.94$   &  $9.0\pm2.0$  &  $40$--$40$  &  $2.41^{+0.25}_{-0.30}$ &  --
\\ \hline
$J_{13}^{0^{--}}$ & $|X_{13};0^{--}\rangle\sim| ^1S_0,$$^{\bar 3}{\bar P}_{\bar 0};1, 0, \rho \rangle$         &  11.5  &  --               &  --           &  --          &  --                      &  --
\\
$J_{14}^{0^{--}}$ & $|X_{14};0^{--}\rangle\sim| ^3S_1,\, ^{\bar 1}{\bar P}_{\bar 1};\, 1, 0, \rho \rangle$     &  7.4   &  $1.51$--$1.74$   &  $8.5\pm2.0$  &  $40$--$53$  &  $2.50^{+0.21}_{-0.24}$  &  --
\\
$J_{15}^{0^{-+}}$ & $|X_{15};0^{-+}\rangle\sim| ^1S_0,\, ^{\bar 3}{\bar P}_{\bar 0};\, 1, 0, \rho \rangle$     &  11.5  &  --               &  --           &  --          &  --                      &  --
\\
$J_{16}^{0^{-+}}$ & $|X_{16};0^{-+}\rangle\sim| ^3S_1,\, ^{\bar 1}{\bar P}_{\bar 1};\, 1, 0, \rho \rangle$     &  8.1   &  $1.63$--$1.73$   &  $8.6\pm2.0$  &  $40$--$46$  &  $2.55^{+0.21}_{-0.23}$  &  $X(2500)$
\\
$J_{17}^{1^{--}}$ & $|X_{17};1^{--}\rangle\sim| ^1S_0,\, ^{\bar 3}{\bar P}_{\bar 1};\, 1, 1, \rho \rangle$     &  6.6   &  $1.46$--$1.82$   &  $8.5\pm2.0$  &  $40$--$61$  &  $2.43^{+0.20}_{-0.24}$  &  \multirow{2}{*}{$X(2400)$}
\\
$J_{18}^{1^{--}}$ & $|X_{18};1^{--}\rangle\sim| ^3S_1,\, ^{\bar 1}{\bar P}_{\bar 1};\, 1, 1, \rho \rangle$     &  7.4   &  $1.67$--$1.88$   &  $8.5\pm2.0$  &  $40$--$51$  &  $2.44^{+0.20}_{-0.25}$
\\
$J_{19}^{1^{-+}}$ & $|X_{19};1^{-+}\rangle\sim| ^1S_0,\, ^{\bar 3}{\bar P}_{\bar 1};\, 1, 1, \rho \rangle$     &  8.0   &  $1.68$--$1.77$   &  $8.5\pm2.0$  &  $40$--$45$  &  $2.49^{+0.21}_{-0.25}$  &  --
\\
$J_{20}^{1^{-+}}$ & $|X_{20};1^{-+}\rangle\sim| ^3S_1,\, ^{\bar 1}{\bar P}_{\bar 1};\, 1, 1, \rho \rangle$     &  7.9   &  $1.75$--$1.87$   &  $8.5\pm2.0$  &  $40$--$46$  &  $2.45^{+0.20}_{-0.25}$  &  --
\\
$J_{21}^{2^{--}}$ & $|X_{21};2^{--}\rangle\sim| ^1S_0,\, ^{\bar 3}{\bar P}_{\bar 2};\, 1, 2, \rho \rangle$     &  4.2   &  $1.39$--$1.89$   &  $8.5\pm2.0$  &  $40$--$70$  &  $2.36^{+0.20}_{-0.26}$  &  --
\\
$J_{22}^{2^{--}}$ & $|X_{22};2^{--}\rangle\sim| ^3S_1,\, ^{\bar 1}{\bar P}_{\bar 1};\, 1, 2, \rho \rangle$     &  8.1   &  $1.75$--$1.85$   &  $8.6\pm2.0$  &  $40$--$45$  &  $2.49^{+0.20}_{-0.24}$  &  --
\\
$J_{23}^{2^{-+}}$ & $|X_{23};2^{-+}\rangle\sim| ^1S_0,\, ^{\bar 3}{\bar P}_{\bar 2};\, 1, 2, \rho \rangle$     &  6.4   &  $1.49$--$1.86$   &  $8.5\pm2.0$  &  $40$--$62$  &  $2.38^{+0.20}_{-0.27}$  &  --
\\
$J_{24}^{2^{-+}}$ & $|X_{24};2^{-+}\rangle\sim| ^3S_1,\, ^{\bar 1}{\bar P}_{\bar 1};\, 1, 2, \rho \rangle$     &  8.5   &  $1.81$--$1.92$   &  $9.0\pm2.0$  &  $40$--$45$  &  $2.55^{+0.20}_{-0.24}$  &  --
\\ \hline\hline
\end{tabular}
\label{tab:results}
\end{center}
\end{table*}

\begin{table*}[hpt]
\begin{center}
\renewcommand{\arraystretch}{1.5}
\caption{Masses of the $S$- and $P$-wave fully-strange tetraquark states, in units of MeV. Our QCD sum rule results are listed in the 3rd column, and the quark model calculations taken from Refs.~\cite{Liu:2020lpw,Lu:2019ira,Deng:2010zzd,Drenska:2008gr,Ebert:2008id} are listed in the 4th-8th columns.}
\begin{tabular}{c|c|c|c|c|c|c|c}
\hline\hline
~~Currents~~&~~~~~~~~~~~~Configuration~~~~~~~~~~~~&~QCD sum rules~&~~Ref.~\cite{Liu:2020lpw}~~&~~Ref.~\cite{Lu:2019ira}~~&~~Ref.~\cite{Deng:2010zzd}~~&~~Ref.~\cite{Drenska:2008gr}~~&~~Ref.~\cite{Ebert:2008id}~~
\\ \hline\hline
$J_{1}^{0^{++}}$  &  $|X_1;0^{++}\rangle\sim| ^1S_0,\, ^{\bar 1}{\bar S}_{\bar 0} ;\, J=0 \rangle$              &  $2.11^{+0.19}_{-0.21}$  &  2365      &    --        &      1925      &  --       &   --
\\
$J_{2}^{0^{++}}$  &  $|X_2;0^{++}\rangle\sim| ^3S_1,\, ^{\bar 3}{\bar S}_{\bar 1} ;\, J=0 \rangle$              &  $1.99^{+0.19}_{-0.24}$  &  2293      &1716          &--              &--         &2203
\\
$J_{3}^{1^{+-}}$  &  $|X_3;1^{+-}\rangle\sim| ^3S_1,\, ^{\bar 3}{\bar S}_{\bar 1} ;\, J=1 \rangle$              &  $2.06^{+0.18}_{-0.20}$  &  2323      &1960          &--              &--         &2267
\\
$J_{4}^{2^{++}}$  &  $|X_4;2^{++}\rangle\sim| ^3S_1,\, ^{\bar 3}{\bar S}_{\bar 1} ;\, J=2 \rangle$              &  $2.09^{+0.19}_{-0.22}$  &  2378      &2255          &--              &--         &2357
\\ \hline
$J_{5}^{0^{-+}}$  &  $|X_5;0^{-+}\rangle\sim| ^3S_1,\, ^{\bar 3}{\bar S}_{\bar 1};\, 1, 0, \lambda \rangle$     &  $2.31^{+0.21}_{-0.26}$  &  2576      &2450          &--              &--         &--
\\
$J_{6}^{1^{--}}$  &  $|X_6;1^{--}\rangle\sim| ^1S_0,\, ^{\bar 1}{\bar S}_{\bar 0};\, 0, 1, \lambda \rangle$     &  --                      &  2889      &--            &2290            &--          &--
\\
$J_{7}^{1^{--}}$  &  $|X_7;1^{--}\rangle\sim| ^3S_1,\, ^{\bar 3}{\bar S}_{\bar 1};\, 0, 1, \lambda \rangle$     &  $2.34^{+0.23}_{-0.30}$  &  2636      &2574          &2188            &2090/2333   &--
\\
$J_{8}^{1^{--}}$  &  $|X_8;1^{--}\rangle\sim| ^3S_1,\, ^{\bar 3}{\bar S}_{\bar 1};\, 2, 1, \lambda \rangle$     &  --                      &  2584      &2468          &--              &2000/2243   &--
\\
$J_{9}^{1^{-+}}$  &  $|X_9;1^{-+}\rangle\sim| ^3S_1,\, ^{\bar 3}{\bar S}_{\bar 1};\, 1, 1, \lambda \rangle$     &  --                      &  2633      &2581          &--              &--          &--
\\
$J_{10}^{2^{--}}$ &  $|X_{10};2^{--}\rangle\sim| ^3S_1,\, ^{\bar 3}{\bar S}_{\bar 1};\, 2, 2, \lambda \rangle$  &  $2.32^{+0.23}_{-0.30}$  &  2665       &2622         &--              &--          &--
\\
$J_{11}^{2^{-+}}$ &  $|X_{11};2^{-+}\rangle\sim| ^3S_1,\, ^{\bar 3}{\bar S}_{\bar 1};\, 1, 2, \lambda \rangle$  &  $2.40^{+0.20}_{-0.25}$  &  2673       &2619         &--              &--          &--
\\
$J_{12}^{3^{--}}$ &  $|X_{12};3^{--}\rangle\sim| ^3S_1,\, ^{\bar 3}{\bar S}_{\bar 1};\, 2, 3, \lambda \rangle$  &  $2.41^{+0.25}_{-0.30}$  &  2719       &2660         &--              &--          &--
\\ \hline
$J_{13}^{0^{--}}$ &  $|X_{13};0^{--}\rangle\sim| ^1S_0,\, ^{\bar 3}{\bar P}_{\bar 0};1, 0, \rho \rangle$        &  --                      &  2635       &--           &--              &--          &--
\\
$J_{14}^{0^{--}}$ &  $|X_{14};0^{--}\rangle\sim| ^3S_1,\, ^{\bar 1}{\bar P}_{\bar 1};\, 1, 0, \rho \rangle$     &  $2.50^{+0.21}_{-0.24}$  &  2694       &2004         &--              &--          &--
\\
$J_{15}^{0^{-+}}$ &  $|X_{15};0^{-+}\rangle\sim| ^1S_0,\, ^{\bar 3}{\bar P}_{\bar 0};\, 1, 0, \rho \rangle$     &  --                      &  2616       &--           &--              &--           &--
\\
$J_{16}^{0^{-+}}$ &  $|X_{16};0^{-+}\rangle\sim| ^3S_1,\, ^{\bar 1}{\bar P}_{\bar 1};\, 1, 0, \rho \rangle$     &  $2.55^{+0.21}_{-0.23}$  &  2685       &2004         &--              &--           &--
\\
$J_{17}^{1^{--}}$ &  $|X_{17};1^{--}\rangle\sim| ^1S_0,\, ^{\bar 3}{\bar P}_{\bar 1};\, 1, 1, \rho \rangle$     &  $2.43^{+0.20}_{-0.24}$  &  2585       & --          &--              &--            &--
\\
$J_{18}^{1^{--}}$ &  $|X_{18};1^{--}\rangle\sim| ^3S_1,\, ^{\bar 1}{\bar P}_{\bar 1};\, 1, 1, \rho \rangle$     &  $2.44^{+0.20}_{-0.25}$  &  2694       &2227         &--              &--            &--
\\
$J_{19}^{1^{-+}}$ &  $|X_{19};1^{-+}\rangle\sim| ^1S_0,\, ^{\bar 3}{\bar P}_{\bar 1};\, 1, 1, \rho \rangle$     &  $2.49^{+0.21}_{-0.25}$  &  2628       &--           &--              &--            &--
\\
$J_{20}^{1^{-+}}$ &  $|X_{20};1^{-+}\rangle\sim| ^3S_1,\, ^{\bar 1}{\bar P}_{\bar 1};\, 1, 1, \rho \rangle$     &  $2.45^{+0.20}_{-0.25}$  &  2712       &2227         &--              &--            &--
\\
$J_{21}^{2^{--}}$ &  $|X_{21};2^{--}\rangle\sim| ^1S_0,\, ^{\bar 3}{\bar P}_{\bar 2};\, 1, 2, \rho \rangle$     &  $2.36^{+0.20}_{-0.26}$  &  2620       &--           &--              &--            &--
\\
$J_{22}^{2^{--}}$ &  $|X_{22};2^{--}\rangle\sim| ^3S_1,\, ^{\bar 1}{\bar P}_{\bar 1};\, 1, 2, \rho \rangle$     &  $2.49^{+0.20}_{-0.24}$  &  2725       &2497         &--              &--            &--
\\
$J_{23}^{2^{-+}}$ &  $|X_{23};2^{-+}\rangle\sim| ^1S_0,\, ^{\bar 3}{\bar P}_{\bar 2};\, 1, 2, \rho \rangle$     &  $2.38^{+0.20}_{-0.27}$  &  2638       &--           &--              &--            &--
\\
$J_{24}^{2^{-+}}$ &  $|X_{24};2^{-+}\rangle\sim| ^3S_1,\, ^{\bar 1}{\bar P}_{\bar 1};\, 1, 2, \rho \rangle$     &  $2.55^{+0.20}_{-0.24}$  &  2733       &2497         &--              &--             &--
\\ \hline\hline
\end{tabular}
\label{tab:comparison}
\end{center}
\end{table*}
In this paper we apply the QCD sum rule method to systematically study the $S$- and $P$-wave fully-strange tetraquark states within the diquark-antidiquark picture. For the $P$-wave states, the orbital angular momentum can be between the diquark and antidiquark, or it can also be inside the diquark/antidiquark. We call the former $\lambda$-mode excitation and the latter $\rho$-mode excitation. There are altogether four $S$-wave $ss \bar s \bar s$ states, eight $P$-wave states of the $\lambda$-mode, and twelve $P$-wave states of the $\rho$-mode. We systematically construct their corresponding interpolating currents by explicitly adding the covariant derivative operator. We use these currents to perform QCD sum rule analyses, and the obtained results are summarized in Table~\ref{tab:results}.

We compare our QCD sum rule results in Table~\ref{tab:comparison} with those obtained in Refs.~\cite{Liu:2020lpw,Lu:2019ira,Deng:2010zzd,Drenska:2008gr,Ebert:2008id} using various quark models. Our QCD sum rule results are generally smaller than, but still more or less consistent with, the quark model calculation of Ref.~\cite{Liu:2020lpw}. Note that there can be significant mixing among the states with the same quantum number. This mixing effect has been systematically investigated in Ref.~\cite{Liu:2020lpw} through the nonrelativistic quark model, and it has also been partly investigated in Refs.~\cite{Chen:2008ne,Chen:2018kuu,Cui:2019roq,Dong:2020okt} for the $J^{PC}=0^{-+}/1^{-\pm}$ tetraquark states through the QCD sum rule method. However, a complete QCD sum rule study of the mixing effect is still not easy, so we do not systematically take it into account in the present study neither.

\begin{figure}[hbt]
\begin{center}
\includegraphics[width=0.5\textwidth]{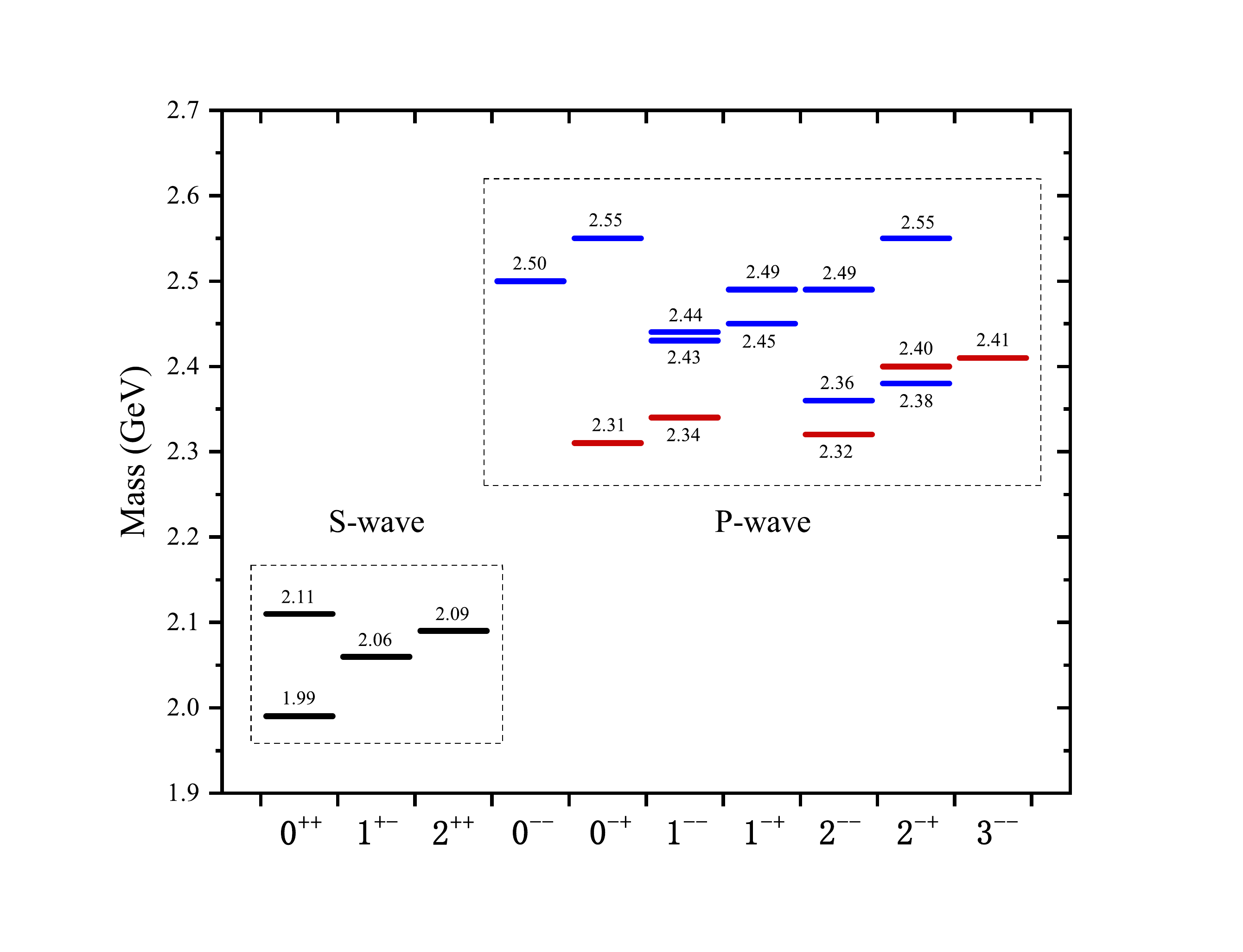}
\caption{Mass spectrum of the fully-strange tetraquark states, including the $S$-wave states (black) as well as the $P$-wave states of the $\lambda$-mode (red) and the $\rho$-mode (blue).}
\label{fig:mass}
\end{center}
\end{figure}

Generally speaking, our results suggest that the $S$-wave $s s \bar s \bar s$ tetraquark states lie in the mass range of $1.99\sim2.11$~GeV, and the $P$-wave states lie in the mass range of $2.31\sim2.55$~GeV, as depicted in Fig.~\ref{fig:mass}. The $S$-wave $s s \bar s \bar s$ tetraquark states decay into the $\eta^\prime\eta^\prime/\eta^\prime\phi/\phi\phi$ channels through $S$-wave, while the $P$-wave states decay into these channels through $P$-wave, so the widths of the latter might be smaller than the former. As already discussed in Sec.~\ref{sec:intro}, there are some rich-strangeness signals at around 2.0~GeV, which are related to the fully-strange tetraquark states with the quantum numbers $J^{PC} = 0^{++}$, $2^{++}$, $1^{+-}$, $0^{-+}$, $1^{--}$, and $1^{-+}$. We separately discuss them as follows.

\subsection{$S$-wave states of $J^{PC} = 0^{++}$ and $2^{++}$}

The $S$-wave $s s \bar s \bar s$ tetraquark states of $J^{PC} = 0^{++}$ and $2^{++}$ decay into the $\phi\phi$ channel through $S$-wave. In 2016 the BESIII collaboration performed a partial wave analysis of the $J/\psi \to \gamma\phi\phi$ decay, and observed one scalar resonance $f_0(2100)$ as well as three tensor resonances $f_2(2010)$, $f_2(2300)$, and $f_2(2340)$ in the $\phi\phi$ invariant mass spectrum~\cite{BESIII:2016qzq}. Their masses and widths were measured to be:
\begin{eqnarray}
f_0(2100) &:&     M \approx 2101~{\rm MeV} \, ,
\\ \nonumber && \Gamma \approx 224~{\rm MeV} \, ;
\\
f_2(2010) &:&     M \approx 2011~{\rm MeV} \, ,
\\ \nonumber && \Gamma \approx 202~{\rm MeV} \, ;
\\
f_2(2300) &:&     M \approx 2297~{\rm MeV} \, ,
\\ \nonumber && \Gamma \approx 149~{\rm MeV} \, ;
\\
f_2(2340) &:&     M \approx 2339~{\rm MeV} \, ,
\\ \nonumber && \Gamma \approx 319~{\rm MeV} \, .
\end{eqnarray}

As depicted in Fig.~\ref{fig:mass}, there are two $S$-wave $s s \bar s \bar s$ states of $J^{PC} = 0^{++}$, whose masses are calculated to be $1.99^{+0.19}_{-0.24}$~GeV and $2.11^{+0.19}_{-0.21}$~GeV. The latter one can be used to explain the scalar resonance $f_0(2100)$ as the $S$-wave $s s \bar s \bar s$ tetraquark state of $J^{PC} = 0^{++}$. There is one $S$-wave $s s \bar s \bar s$ state of $J^{PC} = 2^{++}$, whose mass is calculated to be $2.09^{+0.19}_{-0.22}$. It can be used to explain the tensor resonance $f_2(2010)$ as the $S$-wave $s s \bar s \bar s$ tetraquark state of $J^{PC} = 2^{++}$.

\subsection{$S$-wave state of $J^{PC} = 1^{+-}$}

The $S$-wave $s s \bar s \bar s$ tetraquark state of $J^{PC} = 1^{+-}$ decays into the $\phi\eta^\prime$ channel through $S$-wave. In 2018 the BESIII collaboration observed the $X(2063)$ resonance in the $\phi\eta^\prime$ invariant mass spectrum of the $J/\psi \to \phi \eta \eta^\prime$ decay~\cite{BESIII:2018zbm}. Its mass and width were measured to be:
\begin{eqnarray}
X(2063) &:&     M = 2062.8 \pm 13.1 \pm 7.2~{\rm MeV} \, ,
\\ \nonumber && \Gamma = 177 \pm 36 \pm 35 ~{\rm MeV} \, .
\end{eqnarray}

As depicted in Fig.~\ref{fig:mass}, there is one $S$-wave $s s \bar s \bar s$ state of $J^{PC} = 1^{+-}$. Previously in Ref.~\cite{Cui:2019roq}, we used the current $J_{3,\alpha}^{1^{+-}}$ defined in Eq.~(\ref{def:current3}), and applied the QCD sum rule method to calculate its mass to be $2.00^{+0.10}_{-0.09}$~GeV. In the present study we use the same current and calculate its mass to be $2.06^{+0.18}_{-0.20}$~GeV. These two results are well consistent with each other, both of which support the interpretation of the $X(2063)$ as the $S$-wave $s s \bar s \bar s$ tetraquark state of $J^{PC} = 1^{+-}$.

\subsection{$P$-wave states of $J^{PC} = 0^{-+}$}

The $P$-wave $s s \bar s \bar s$ tetraquark states of $J^{PC} = 0^{-+}$ decay into the $\phi\phi$ channel through $P$-wave. In 2016 the BESIII collaboration observed the $X(2500)$ resonance in the $\phi \phi$ invariant mass spectrum of the $J/\psi \to \gamma \phi \phi$ decay~\cite{BESIII:2016qzq}. Besides, in 2010 the BESIII collaboration observed two resonances $X(2120)$ and $X(2370)$ in the $\pi \pi \eta^\prime$ invariant mass spectrum of the $J/\psi \to \gamma \pi \pi \eta^\prime$ decay~\cite{BESIII:2010gmv}. Later in 2019 they confirmed the $X(2370)$ in the $K \bar K \eta^\prime$ invariant mass spectrum of the $J/\psi \to \gamma K \bar K \eta^\prime$ decay, but they did not observe the $X(2120)$ in this process~\cite{BESIII:2019wkp}. This suggests that the $X(2370)$ contains more strangeness components. The experimental parameters of the $X(2370)$ and $X(2500)$ were measured to be:
\begin{eqnarray}
X(2500)  &:& M      = 2470\,^{+15}_{-19}\,^{+101}_{-23}~{\rm MeV} \, ,
\\ \nonumber && \Gamma = 230\,^{+64}_{-35}\,^{+56}_{-33}~{\rm MeV} \, ;
\\ X(2370)     &:& M      = 2341.6 \pm 6.5 \pm 5.7~{\rm MeV} \, ,
\\ \nonumber && \Gamma = 117 \pm 10 \pm 8~{\rm MeV} \, .
\end{eqnarray}

Previously in Ref.~\cite{Dong:2020okt} we applied the QCD sum rule method to study the $s s \bar s \bar s$ tetraquark states of $J^{PC} = 0^{-+}$ by investigating two independent currents without derivatives:
\begin{eqnarray}
\nonumber \eta_{1} &=& (s_a^T C s_b) (\bar{s}_a \gamma_5 C \bar{s}_b^T) + (s_a^T C \gamma_5 s_b) (\bar{s}_a C \bar{s}_b^T)  \, ,
\\ \eta_{2} &=& (s_a^T C \sigma_{\mu\nu} s_b) (\bar{s}_a \sigma^{\mu\nu} \gamma_5 C \bar{s}_b^T) \, .
\end{eqnarray}
We took into account their mixing and further constructed two non-correlated currents, based on which we calculated the masses to be $2.51^{+0.15}_{-0.12}$~GeV and $3.14^{+0.39}_{-0.26}$~GeV. The former one can be used to explain either the $X(2370)$ or $X(2500)$ as the $P$-wave $s s \bar s \bar s$ tetraquark state of $J^{PC} = 0^{-+}$.

In this study we use the $s s \bar s \bar s$ tetraquark currents with derivatives to perform QCD sum rule analyses. As shown in Fig.~\ref{fig:category}, there are three $P$-wave $s s \bar s \bar s$ tetraquark states of $J^{PC} = 0^{-+}$. We construct their corresponding currents, as defined in Eqs.~(\ref{def:current5},\ref{def:current15},\ref{def:current16}). Clearly, the use of the covariant derivative operator when constructing interpolating currents gives us more possibilities, based on which we can better describe the internal structure of multiquark states.

The two currents, $J_{5}^{0^{-+}}$ defined in Eq.~(\ref{def:current5}) and $J_{16}^{0^{-+}}$ defined in Eq.~(\ref{def:current16}), lead to reasonable QCD sum rule results. Their masses are calculated to be $2.31^{+0.21}_{-0.26}$~GeV and $2.55^{+0.21}_{-0.23}$~GeV, respectively. These results can be used to explain both the $X(2370)$ and $X(2500)$ as the $P$-wave $s s \bar s \bar s$ tetraquark states of $J^{PC} = 0^{-+}$.

\subsection{$P$-wave states of $J^{PC} = 1^{--}$}

The $\phi(2170)$ was first observed in 2006 by the BaBar collaboration in the $\phi f_0(980)$ invariant mass spectrum~\cite{BaBar:2006gsq,BaBar:2007ptr,BaBar:2007ceh,BaBar:2011btv}, and later confirmed in the BESII/BESIII~\cite{BES:2007sqy,BESIII:2014ybv,BESIII:2017qkh,BESIII:2018ldc,BESIII:2019ebn,BESIII:2020vtu,BESIII:2020gnc,BESIII:2020xmw,BESIII:2021yam,BESIII:2021bjn,BESIII:2022wxz} and Belle~\cite{Belle:2008kuo} experiments. Besides, there might exist another structure in the $\phi f_0(980)$ invariant mass spectrum at around 2.4 GeV, whose evidences were observed in the BaBar~\cite{BaBar:2007ptr}, BESII/BESIII~\cite{BES:2007sqy,BESIII:2014ybv}, and Belle~\cite{Belle:2008kuo,Shen:2009mr} experiments.

Previously in Refs.~\cite{Chen:2018kuu,Chen:2008ej} we applied the QCD sum rule method to study the $s s \bar s \bar s$ tetraquark states of $J^{PC} = 1^{--}$ by investigating two independent currents without derivatives:
\begin{eqnarray}
\nonumber \eta_{3\mu} &=& s_a^T C \gamma_5 s_b \bar{s}_a \gamma_\mu \gamma_5 C \bar{s}_b^T - s_a^T C \gamma_\mu \gamma_5 s_b \bar{s}_a \gamma_5 C \bar{s}_b^T,
\\ \nonumber \eta_{4\mu} &=& s_a^T C \gamma^\nu s_b \bar{s}_a \sigma_{\mu\nu} C \bar{s}_b^T - s_a^T C \sigma_{\mu\nu} s_b \bar{s}_a \gamma^\nu C \bar{s}_b^T.
\\
\end{eqnarray}
We took into account their mixing and further constructed two non-correlated currents, based on which we calculated the masses to be $2.34 \pm 0.17$~GeV and $2.41 \pm 0.25$~GeV. The former one was used to explain the $\phi(2170)$ as the $S$-wave $s s \bar s \bar s$ tetraquark state of $J^{PC} = 1^{--}$, and the latter one suggests that the $\phi(2170)$ has a partner state with the mass $\Delta M = 71^{+172}_{-~48}$ MeV larger~\cite{Chen:2018kuu}.

In a recent BESIII experiment the partner state of the $\phi(2170)$, labelled as $X(2400)$, was observed in the $e^+e^- \to \phi \pi^+ \pi^-$ process with a statistical significance of $8.5\sigma$~\cite{BESIII:2021lho}. The experimental parameters of the $\phi(2170)$ and $X(2400)$ were measured to be~\cite{pdg,BESIII:2021lho}:
\begin{eqnarray}
\phi(2170)  &:& M      = 2160 \pm 80~{\rm MeV} \, ,
\\ \nonumber && \Gamma = 125 \pm 65~{\rm MeV} \, ;
\\ X(2400)     &:& M      = 2298^{+60}_{-44} \pm 6~{\rm MeV} \, ,
\\ \nonumber && \Gamma = 219^{+117}_{-112} \pm 6~{\rm MeV} \, .
\end{eqnarray}

As shown in Fig.~\ref{fig:category}, there are as many as five $P$-wave $s s \bar s \bar s$ tetraquark states of $J^{PC} = 1^{--}$. In this study we construct their corresponding currents by explicitly adding the covariant derivative operator, as defined in Eqs.~(\ref{def:current6},\ref{def:current7},\ref{def:current8},\ref{def:current17},\ref{def:current18}). Three of them lead to reasonable QCD sum rule results, and the masses are calculated to be $2.34^{+0.23}_{-0.30}$~GeV, $2.43^{+0.20}_{-0.24}$~GeV, and $2.44^{+0.20}_{-0.25}$~GeV. Similar to our previous study of Ref.~\cite{Chen:2018kuu}, these results can be used to explain both the $\phi(2170)$ and $X(2400)$ as the $P$-wave $s s \bar s \bar s$ tetraquark states of $J^{PC} = 1^{--}$.

\subsection{$P$-wave states of $J^{PC} = 1^{-+}$}

Very recently, the BESIII collaboration studied the $J/\psi \to \gamma \eta \eta^\prime$ decay process and observed the $\eta_1(1855)$ resonance with the exotic quantum number $I^GJ^{PC} = 0^+1^{-+}$ in the $\eta \eta^\prime$ invariant mass spectrum~\cite{BESIII:2022riz,BESIII:2022qzu}. Its mass and width were measured to be
\begin{eqnarray}
\eta_1(1855) &:& M = 1855 \pm 9 ^{+6}_{-1} {\rm~MeV}/c^2 \, ,
\\ \nonumber && \Gamma = 188 \pm 18 ^{+3}_{-8} {\rm~MeV} \, .
\end{eqnarray}

Previously in Refs.~\cite{Chen:2008ne,Chen:2008qw} we applied the QCD sum rule method study the $qs\bar q \bar s$ ($q=u/d$) tetraquark states of $I^GJ^{PC} = 0^+1^{-+}$ by investigating four independent currents without derivatives:
\begin{eqnarray}
\eta_{5\mu} &=& u_{a}^T C \gamma_\mu s_{b} (\bar{u}_{a} C \bar{s}_{b}^T + \bar{u}_{b} C \bar{s}_{a}^T)
\\ \nonumber  &+& u_{a}^T C s_{b} (\bar{u}_{a} \gamma_\mu C \bar{s}_{b}^T + \bar{u}_{b} \gamma_\mu C \bar{s}_{a}^T) + \{ u/\bar u \rightarrow d/\bar d  \} \, ,
\\ \eta_{6\mu} &=& u_{a}^T C \sigma_{\mu\nu} \gamma_5 s_{b} (\bar{u}_{a} \gamma^{\nu} \gamma_5 C \bar{s}_{b}^T + \bar{u}_{b} \gamma^{\nu} \gamma_5 C \bar{s}_{a}^T)
\\ \nonumber  &+& u_{a}^T C \gamma^{\nu} \gamma_5 s_{b} (\bar{u}_{a} \sigma_{\mu\nu} \gamma_5 C \bar{s}_{b}^T + \bar{u}_{b} \sigma_{\mu\nu} \gamma_5 C \bar{s}_{a}^T)
\\ \nonumber  &+& \{ u/\bar u \rightarrow d/\bar d  \} \, ,
\\ \eta_{7\mu} &=& u_{a}^T C s_{b} (\bar{u}_{a} \gamma_\mu C \bar{s}_{b}^T - \bar{u}_{b} \gamma_\mu C \bar{s}_{a}^T)
\\ \nonumber  &+& u_{a}^T C \gamma_\mu s_{b} (\bar{u}_{a} C \bar{s}_{b}^T - \bar{u}_{b} C \bar{s}_{a}^T) + \{ u/\bar u \rightarrow d/\bar d  \} \, ,
\\ \eta_{8\mu} &=& u_{a}^T C \gamma^{\nu} \gamma_5 s_{b} (\bar{u}_{a} \sigma_{\mu\nu} \gamma_5 C \bar{s}_{b}^T - \bar{u}_{b} \sigma_{\mu\nu} \gamma_5 C \bar{s}_{a}^T)
\\ \nonumber  &+& u_{a}^T C \sigma_{\mu\nu} \gamma_5 s_{b} (\bar{u}_{a} \gamma^{\nu} \gamma_5 C \bar{s}_{b}^T - \bar{u}_{b} \gamma^{\nu} \gamma_5 C \bar{s}_{a}^T)
\\ \nonumber  &+& \{ u/\bar u \rightarrow d/\bar d  \} \, .
\end{eqnarray}
We took into account the mixing of $\eta_{5\mu}$ and $\eta_{7\mu}$, and calculated the mass to be around $1.8$ $\sim$ $2.1$~GeV. This result can be used to explain the $\eta_1(1855)$ as the $P$-wave $q s \bar q \bar s$ tetraquark state of $I^GJ^{PC} = 0^+1^{-+}$. Based on this interpretation, one naturally expect the existence of the $s s \bar s \bar s$ tetraquark state of $I^GJ^{PC} = 0^+1^{-+}$.

As shown in Fig.~\ref{fig:category}, there are three $P$-wave $s s \bar s \bar s$ tetraquark states of $J^{PC} = 1^{-+}$. We construct their corresponding currents, two of which lead to reasonable QCD sum rule results. The masses extracted from the two currents, $J_{19,\alpha\beta}^{1^{-+}}$ defined in Eq.~(\ref{def:current19}) and $J_{20,\alpha\beta}^{1^{-+}}$ defined in Eq.~(\ref{def:current20}), are calculated to be $2.49^{+0.21}_{-0.25}$~GeV and $2.45^{+0.20}_{-0.25}$~GeV, respectively.

These states have the exotic quantum number $J^{PC} = 1^{-+}$, which can not be accessed by conventional $\bar q q$ mesons, so they are of particular interest. We further study the mixing effect by investigating the off-diagonal correlation function
\begin{equation}
\langle 0 | T [J_{19,\mu\nu}^{1^{-+}}(x) J_{20,\rho\sigma}^{1^{-+},\dagger} (0)] | 0 \rangle \, ,
\end{equation}
which is calculated to be zero. Therefore, the two currents $J_{19,\alpha\beta}^{1^{-+}}$ and $J_{20,\alpha\beta}^{1^{-+}}$ are non-correlated, suggesting that there might exist two almost degenerate $s s \bar s \bar s$ tetraquark states with the exotic quantum number $J^{PC} = 1^{-+}$.

To end this paper, we propose to search for the $S$- and $P$-wave fully-strange tetraquark states in the future Belle-II, BESIII, COMPASS, and GlueX experiments, etc. Besides the two-body decay channels $\eta\eta^\prime/\phi\eta/\phi\eta^\prime/\phi\phi/\phi f_0(980)$ already investigated in experiments, we propose to examine the two-body channels $\eta^\prime\eta^\prime/\eta f_0(980)/\eta^\prime f_0(980)/f_0(980)f_0(980)$ and the relevant three-body channels.

\section*{Acknowledgments}
%

We thank Wei Chen, Er-Liang Cui, Wen-Biao Yan, and Shi-Lin Zhu for useful discussions.
This project is supported by
the National Natural Science Foundation of China under Grant No.~12075019,
the Jiangsu Provincial Double-Innovation Program under Grant No.~JSSCRC2021488,
and
the Fundamental Research Funds for the Central Universities.

\appendix

\begin{widetext}
\section{Spectral densities}
\label{app:ope}

In this appendix we list the QCD sum rule equations extracted from the currents $J^{\cdots}_{2\cdots24,\cdots}$ defined in Eqs.~(\ref{def:current2}-\ref{def:current4}), Eqs.~(\ref{def:current5}-\ref{def:current12}), and Eqs.~(\ref{def:current13}-\ref{def:current24}).

\begin{eqnarray}
\nonumber
\Pi_{22} &=& \int^{s_0}_{16 m_s^2} \Bigg [
{s^4 \over 15360 \pi^6}-{m_s^2 s^3 \over 256 \pi^6}
+ {\langle g_s^2 GG \rangle \over 3072 \pi^6}s^2
+ \Big( -{3\langle g_s^2 GG \rangle m_s^2 \over 512 \pi^6 }
- {m_s \langle g_s \bar s \sigma G s \rangle \over 32 \pi^4}
+{\langle \bar s s \rangle^2 \over 6 \pi^2}\Big)s
\\ \nonumber &&
+{\langle g_s^2 GG \rangle m_s \langle \bar s s \rangle \over 64 \pi^4}
+{7m_s^2\langle \bar s s \rangle^2\over6\pi^2 }
+{\langle \bar s s \rangle \langle g_s \bar s \sigma G s \rangle \over 12 \pi^2} \Bigg ] e^{-s/M^2} ds
+\Big( { \langle g_s^2 GG \rangle m_s \langle g_s \bar s \sigma G s \rangle \over 384\pi^4}
-{\langle g_s^2 GG \rangle \langle \bar s s \rangle^2 \over 144\pi^2}
\\ \nonumber &&
+{5m_s^2 \langle \bar s s \rangle \langle g_s \bar s \sigma G s \rangle \over 8\pi^2}
-{4 m_s \langle \bar s s \rangle^3 \over 3}\Big)
+ {1 \over M_B^2} \Big(-{ \langle g_s^2 GG \rangle m_s^2 \langle \bar s s \rangle^2 \over 576\pi^2}
+{\langle g_s^2 GG \rangle \langle \bar s s \rangle \langle g_s \bar s \sigma G s \rangle \over 288\pi^2}
\\ &&
-{5m_s^2 \langle g_s \bar s \sigma G s \rangle^2\over48\pi^2}
+{7m_s\langle \bar s s \rangle^2 \langle g_s \bar s \sigma G s \rangle \over6} \Big) \, ,
\\\nonumber
\Pi_{33} &=& \int^{s_0}_{16 m_s^2} \Bigg [
{s^4 \over 12288 \pi^6}-{m_s^2 s^3 \over 2560 \pi^6}
+ \Big( {\langle g_s^2 GG \rangle \over 18432 \pi^6}
-{13 m_s \langle \bar s s \rangle \over 96 \pi^4 }\Big )s^2
+ \Big( -{\langle g_s^2 GG \rangle m_s^2 \over 2304 \pi^6 }
- {155 m_s \langle g_s \bar s \sigma G s \rangle \over 576 \pi^4}
\\ \nonumber &&
+{25\langle \bar s s \rangle^2 \over 36 \pi^2}\Big)s
-{13m_s^2 \langle \bar s s \rangle^2 \over 8 \pi^2}
+{31\langle \bar s s \rangle \langle g_s \bar s \sigma G s \rangle \over 48 \pi^2} \Bigg ] e^{-s/M^2} ds
+\Big({11m_s^2 \langle \bar s s \rangle \langle g_s \bar s \sigma G s \rangle \over 48\pi^2}
-{14 m_s \langle \bar s s \rangle^3 \over 9}
\\  &&
+{\langle g_s \bar s \sigma G s \rangle^2 \over 18 \pi^2}\Big)
+ {1 \over M_B^2} \Big(-{ \langle g_s^2 GG \rangle m_s^2 \langle \bar s s \rangle^2 \over 576\pi^2}
+{13m_s^2 \langle g_s \bar s \sigma G s \rangle^2\over96\pi^2}
+{11m_s\langle \bar s s \rangle^2 \langle g_s \bar s \sigma G s \rangle \over36} \Big) \, ,
\\\nonumber
\Pi_{44} &=& \int^{s_0}_{16 m_s^2} \Bigg [
{s^4 \over 86016 \pi^6}-{m_s^2 s^3 \over 2880 \pi^6}
+ \Big( -{11\langle g_s^2 GG \rangle \over 122880 \pi^6}
-{3 m_s \langle \bar s s \rangle \over 320 \pi^4 }\Big )s^2
+ \Big( {5\langle g_s^2 GG \rangle m_s^2 \over 9216 \pi^6 }
- {7 m_s \langle g_s \bar s \sigma G s \rangle \over 288 \pi^4}
\\ \nonumber &&
+{\langle \bar s s \rangle^2 \over 18 \pi^2} \Big)s
+{11\langle g_s^2 GG \rangle m_s \langle \bar s s \rangle \over 6912\pi^4}
+{5m_s^2 \langle \bar s s \rangle^2 \over 24 \pi^2}
+{7\langle \bar s s \rangle \langle g_s \bar s \sigma G s \rangle \over 144 \pi^2} \Bigg ] e^{-s/M^2} ds
+\Big( -{ \langle g_s^2 GG \rangle \langle \bar s s \rangle^2\over 432 \pi^2 }
\\ \nonumber &&
-{4 m_s \langle \bar s s \rangle^3 \over 9}
+{\langle g_s^2 GG \rangle m_s \langle g_s \bar s \sigma G s \rangle\over1152\pi^4}
+{m_s^2 \langle \bar s s \rangle \langle g_s \bar s \sigma G s \rangle \over 4 \pi^2}\Big)
+ {1 \over M_B^2} \Big({5 \langle g_s^2 GG \rangle m_s^2 \langle \bar s s \rangle^2 \over 3457\pi^2}
\\  &&
+{5m_s^2 \langle g_s \bar s \sigma G s \rangle^2\over288\pi^2}
+{5m_s\langle \bar s s \rangle^2 \langle g_s \bar s \sigma G s \rangle \over54} \Big) \, ,
\\\nonumber
\Pi_{55} &=& \int^{s_0}_{16 m_s^2} \Bigg [
{7s^5 \over 307200 \pi^6}
-{m_s^2 s^4 \over 2560 \pi^6}
-{m_s\langle \bar s s \rangle \over 32 \pi^4 }s^3
+ \Big( -{\langle g_s^2 GG \rangle m_s^2 \over 512 \pi^6}
+{\langle \bar s s \rangle^2 \over 4 \pi^2}
-{m_s\langle g_s \bar s \sigma G s \rangle \over 16 \pi^4}\Big )s^2
\\ \nonumber &&
+ \Big({\langle g_s^2 GG \rangle m_s \langle \bar s s \rangle \over 96 \pi^4}
-{19m_s^2\langle \bar s s \rangle^2 \over 8 \pi^2}
- {\langle \bar s s \rangle \langle g_s \bar s \sigma G s \rangle \over 4 \pi^2}\Big)s
+{\langle g_s^2 GG \rangle\langle \bar s s \rangle^2 \over 36 \pi^2}
-{\langle g_s^2 GG \rangle m_s \langle g_s \bar s \sigma G s \rangle \over 96 \pi^4}
\\ \nonumber &&
-{3m_s^2\langle \bar s s \rangle \langle g_s \bar s \sigma G s \rangle\over2\pi^2 }
-{ \langle g_s \bar s \sigma G s \rangle^2 \over 4 \pi^2} \Bigg ] e^{-s/M^2} ds
+\Big( -{ \langle g_s^2 GG \rangle m_s^2 \langle \bar s s \rangle^2 \over 48\pi^2}
+{7 \langle g_s^2 GG \rangle \langle \bar s s \rangle \langle g_s \bar s \sigma G s \rangle \over 288 \pi^2}
\\  &&
+{17m_s \langle \bar s s \rangle^2 \langle g_s \bar s \sigma G s \rangle \over 6}
-{5m_s^2 \langle g_s \bar s \sigma G s \rangle^2 \over 16\pi^2}\Big) \, ,
\\\nonumber
\Pi_{66} &=& \int^{s_0}_{16 m_s^2} \Bigg [
{s^5 \over 358400 \pi^6}
-{m_s^2 s^4 \over 7680 \pi^6}
+\Big(-{\langle g_s^2 GG \rangle \over 61440 \pi^6}
-{m_s \langle \bar s s \rangle \over 480 \pi^4 }\Big)s^3
+{\langle g_s^2 GG \rangle m_s^2 \over 6144 \pi^6}s^2
\\ \nonumber &&
+ \Big( {\langle g_s^2 GG \rangle m_s \langle \bar s s \rangle \over 1152 \pi^4}
-{m_s^2\langle \bar s s \rangle^2 \over 36 \pi^2}
- {\langle \bar s s \rangle \langle g_s \bar s \sigma G s \rangle \over 6 \pi^2}\Big)s
+{m_s^2 \langle g_s \bar s \sigma G s \rangle \langle \bar s s \rangle \over 3 \pi^2}
-{ \langle g_s \bar s \sigma G s \rangle^2 \over 8 \pi^2} \Bigg ] e^{-s/M^2} ds
\\  &&
+\Big({ \langle g_s^2 GG \rangle m_s^2 \langle \bar s s \rangle^2 \over 576\pi^2}
+{ \langle g_s^2 GG \rangle \langle \bar s s \rangle \langle g_s \bar s \sigma G s \rangle \over 288 \pi^2}
+{4m_s \langle \bar s s \rangle^2 \langle g_s \bar s \sigma G s \rangle \over 9}
+{m_s^2 \langle g_s \bar s \sigma G s \rangle^2 \over 6\pi^2}\Big) \, ,
\\\nonumber
\Pi_{77} &=& \int^{s_0}_{16 m_s^2} \Bigg [
{31 s^5 \over 3225600 \pi^6}
-{m_s^2 s^4 \over 1152 \pi^6}
+\Big({\langle g_s^2 GG \rangle \over 9216 \pi^6}
+{m_s \langle \bar s s \rangle \over 160 \pi^4 }\Big)s^3
+\Big(-{23 \langle g_s^2 GG \rangle m_s^2 \over 9216 \pi^6}
+{m_s \langle g_s \bar s \sigma G s \rangle \over 192 \pi^4}\Big)s^2
\\ \nonumber &&
+ \Big({\langle g_s^2 GG \rangle m_s \langle \bar s s \rangle \over 108 \pi^4}
+{7 m_s^2 \langle \bar s s \rangle^2 \over 12 \pi^2}
- {\langle \bar s s \rangle \langle g_s \bar s \sigma G s \rangle \over 6 \pi^2}\Big)s
-{ \langle g_s \bar s \sigma G s \rangle^2 \over 12 \pi^2} \Bigg ] e^{-s/M^2} ds
\\ &&
+\Big({ \langle g_s^2 GG \rangle m_s^2 \langle \bar s s \rangle^2 \over 432\pi^2}
+{ \langle g_s^2 GG \rangle \langle \bar s s \rangle \langle g_s \bar s \sigma G s \rangle \over 144 \pi^2}
+{4m_s \langle \bar s s \rangle^2 \langle g_s \bar s \sigma G s \rangle \over 3}
-{m_s^2 \langle g_s \bar s \sigma G s \rangle^2 \over 24\pi^2}\Big) \, ,
\\\nonumber
\Pi_{88} &=& \int^{s_0}_{16 m_s^2} \Bigg [
{19 s^5 \over 5734400 \pi^6}
-{23 m_s^2 s^4 \over 92160 \pi^6}
+\Big({7\langle g_s^2 GG \rangle \over 737280 \pi^6}
+{23 m_s \langle \bar s s \rangle \over 7680 \pi^4 }\Big)s^3
+\Big({25 \langle g_s^2 GG \rangle m_s^2 \over 147456 \pi^6}
-{\langle \bar s s \rangle^2 \over 72 \pi^2 }
\\ \nonumber &&
+{37 m_s \langle g_s \bar s \sigma G s \rangle \over 9216 \pi^4}\Big)s^2
+ \Big(-{7\langle g_s^2 GG \rangle m_s \langle \bar s s \rangle \over 6912 \pi^4}
+{11 m_s^2 \langle \bar s s \rangle^2 \over 192 \pi^2}
- {25 \langle \bar s s \rangle \langle g_s \bar s \sigma G s \rangle \over 216 \pi^2}\Big)s
\\ \nonumber &&
-{m_s^2 \langle \bar s s \rangle \langle g_s \bar s \sigma G s \rangle \over 64\pi^2}
-{17\langle g_s \bar s \sigma G s \rangle^2 \over 384 \pi^2} \Bigg ] e^{-s/M^2} ds
+\Big(-{13\langle g_s^2 GG \rangle m_s^2 \langle \bar s s \rangle^2 \over 6912\pi^2}
\\  &&
+{13\langle g_s^2 GG \rangle \langle \bar s s \rangle \langle g_s \bar s \sigma G s \rangle \over 6912 \pi^2}
+{10 m_s \langle \bar s s \rangle^2 \langle g_s \bar s \sigma G s \rangle \over 27}
-{55 m_s^2 \langle g_s \bar s \sigma G s \rangle^2 \over 1152\pi^2}\Big) \, ,
\\\nonumber
\Pi_{99} &=& \int^{s_0}_{16 m_s^2} \Bigg [
{ s^5 \over 516096 \pi^6}
-{7 m_s^2 s^4 \over 46080 \pi^6}
+\Big({23\langle g_s^2 GG \rangle \over 737280 \pi^6}
+{ m_s \langle \bar s s \rangle \over 384 \pi^4 }\Big)s^3
+\Big(-{ \langle g_s^2 GG \rangle m_s^2 \over 4096 \pi^6}
-{\langle \bar s s \rangle^2 \over 72 \pi^2 }
\\ \nonumber &&
+{53 m_s \langle g_s \bar s \sigma G s \rangle \over 9216 \pi^4}\Big)s^2
+ \Big({\langle g_s^2 GG \rangle m_s \langle \bar s s \rangle \over 13824 \pi^4}
+{5 m_s^2 \langle \bar s s \rangle^2 \over 144 \pi^2}
- { \langle \bar s s \rangle \langle g_s \bar s \sigma G s \rangle \over 12 \pi^2}\Big)s
+{m_s^2 \langle \bar s s \rangle \langle g_s \bar s \sigma G s \rangle \over 64\pi^2}
\\ \nonumber &&
-{3\langle g_s \bar s \sigma G s \rangle^2 \over 128 \pi^2} \Bigg ] e^{-s/M^2} ds
+\Big(-{5\langle g_s^2 GG \rangle m_s^2 \langle \bar s s \rangle^2 \over 2304\pi^2}
+{\langle g_s^2 GG \rangle \langle \bar s s \rangle \langle g_s \bar s \sigma G s \rangle \over 1152 \pi^2}
\\  &&
+{17 m_s \langle \bar s s \rangle^2 \langle g_s \bar s \sigma G s \rangle \over 108}
-{7 m_s^2 \langle g_s \bar s \sigma G s \rangle^2 \over 576\pi^2}\Big) \, ,
\\\nonumber
\Pi_{1010} &=& \int^{s_0}_{16 m_s^2} \Bigg [
{3 s^5 \over 358400 \pi^6}
-{m_s^2 s^4 \over 2304 \pi^6}
+\Big(-{143\langle g_s^2 GG \rangle \over 1658880 \pi^6}
-{ m_s \langle \bar s s \rangle \over 2880 \pi^4 }\Big)s^3
+\Big({7\langle g_s^2 GG \rangle m_s^2 \over 11520 \pi^6}
+{\langle \bar s s \rangle^2 \over 36 \pi^2 }
\\ \nonumber &&
-{151 m_s \langle g_s \bar s \sigma G s \rangle \over 11520 \pi^4}\Big)s^2
+ \Big(-{\langle g_s^2 GG \rangle m_s \langle \bar s s \rangle \over 3456 \pi^4}
-{17 m_s^2 \langle \bar s s \rangle^2 \over 36 \pi^2}
- {41\langle \bar s s \rangle \langle g_s \bar s \sigma G s \rangle \over 432 \pi^2}\Big)s
+{\langle g_s^2 GG \rangle \langle \bar s s \rangle^2 \over 1296 \pi^2}
\\ \nonumber &&
-{\langle g_s^2 GG \rangle m_s \langle g_s \bar s \sigma G s \rangle \over 1152\pi^2}
-{197 m_s^2 \langle \bar s s \rangle \langle g_s \bar s \sigma G s \rangle \over 432\pi^2}
-{37\langle g_s \bar s \sigma G s \rangle^2 \over 864 \pi^2} \Bigg ] e^{-s/M^2} ds
+\Big(-{5\langle g_s^2 GG \rangle m_s^2 \langle \bar s s \rangle^2 \over 1728\pi^2}
\\ &&
+{\langle g_s^2 GG \rangle \langle \bar s s \rangle \langle g_s \bar s \sigma G s \rangle \over 432 \pi^2}
+{19 m_s \langle \bar s s \rangle^2 \langle g_s \bar s \sigma G s \rangle \over 36}
-{ m_s^2 \langle g_s \bar s \sigma G s \rangle^2 \over 8\pi^2}\Big) \, ,
\\\nonumber
\Pi_{1111} &=& \int^{s_0}_{16 m_s^2} \Bigg [
{13 s^5 \over 1075200 \pi^6}
-{23 m_s^2 s^4 \over 53760 \pi^6}
+\Big(-{\langle g_s^2 GG \rangle \over 138240 \pi^6}
-{83 m_s \langle \bar s s \rangle \over 4302 \pi^4 }\Big)s^3
+\Big({\langle g_s^2 GG \rangle m_s^2 \over 3072 \pi^6}
+{8\langle \bar s s \rangle^2 \over 45 \pi^2 }
\\ \nonumber &&
-{637 m_s \langle g_s \bar s \sigma G s \rangle \over 11520 \pi^4}\Big)s^2
+ \Big(-{\langle g_s^2 GG \rangle m_s \langle \bar s s \rangle \over 864 \pi^4}
-{41 m_s^2 \langle \bar s s \rangle^2 \over 36 \pi^2}
+ {37 \langle \bar s s \rangle \langle g_s \bar s \sigma G s \rangle \over 144 \pi^2}\Big)s
+{\langle g_s^2 GG \rangle \langle \bar s s \rangle^2 \over 432 \pi^2}
\\ \nonumber &&
+{\langle g_s^2 GG \rangle m_s \langle g_s \bar s \sigma G s \rangle \over 1152\pi^4}
-{133 m_s^2 \langle \bar s s \rangle \langle g_s \bar s \sigma G s \rangle \over 216\pi^2}
+{5\langle g_s \bar s \sigma G s \rangle^2 \over 432 \pi^2} \Bigg ] e^{-s/M^2} ds
\\ &&
+\Big(-{\langle g_s^2 GG \rangle m_s^2 \langle \bar s s \rangle^2 \over 432\pi^2}
+{\langle g_s^2 GG \rangle \langle \bar s s \rangle \langle g_s \bar s \sigma G s \rangle \over 288 \pi^2}
+{55 m_s \langle \bar s s \rangle^2 \langle g_s \bar s \sigma G s \rangle \over 36}
-{49 m_s^2 \langle g_s \bar s \sigma G s \rangle^2 \over 144\pi^2}\Big) \, ,
\\\nonumber
\Pi_{1212} &=& \int^{s_0}_{16 m_s^2} \Bigg [
{ s^5 \over 691200 \pi^6}
-{5 m_s^2 s^4 \over 64512 \pi^6}
+\Big(-{19\langle g_s^2 GG \rangle \over 967680 \pi^6}
-{ m_s \langle \bar s s \rangle \over 1890 \pi^4 }\Big)s^3
+\Big({\langle g_s^2 GG \rangle m_s^2 \over 2880 \pi^6}
-{7 m_s \langle g_s \bar s \sigma G s \rangle \over 7680 \pi^4}\Big)s^2
\\ \nonumber &&
+ \Big(-{\langle g_s^2 GG \rangle m_s \langle \bar s s \rangle \over 1920 \pi^4}
+{ m_s^2 \langle \bar s s \rangle^2 \over 30 \pi^2}
- {18 \langle \bar s s \rangle \langle g_s \bar s \sigma G s \rangle \over 18 \pi^2}\Big)s
-{ m_s^2 \langle \bar s s \rangle \langle g_s \bar s \sigma G s \rangle \over 288\pi^2}
-{55\langle g_s \bar s \sigma G s \rangle^2 \over 1728 \pi^2} \Bigg ] e^{-s/M^2} ds
\\  &&
+\Big({\langle g_s^2 GG \rangle \langle \bar s s \rangle \langle g_s \bar s \sigma G s \rangle \over 432 \pi^2}
+{4 m_s \langle \bar s s \rangle^2 \langle g_s \bar s \sigma G s \rangle \over 9}
-{ m_s^2 \langle g_s \bar s \sigma G s \rangle^2 \over 12\pi^2}\Big) \, ,
\\\nonumber
\Pi_{1313} &=& \int^{s_0}_{16 m_s^2} \Bigg [
{ m_s^2 s^4 \over 3840 \pi^6}
-{ m_s^4 s^3 \over 96 \pi^6 }
+\Big(-{\langle g_s^2 GG \rangle m_s^2 \over 1536 \pi^6}
+{ m_s \langle g_s \bar s \sigma G s \rangle \over 24 \pi^4}\Big)s^2
+ \Big({m_s^2 \langle \bar s s \rangle^2 \over 3 \pi^2}
-{\langle \bar s s \rangle \langle g_s \bar s \sigma G s \rangle \over 6 \pi^2}\Big)s
\\ \nonumber &&
-{\langle g_s^2 GG \rangle m_s \langle g_s \bar s \sigma G s \rangle \over 192\pi^4}
+{m_s^2 \langle \bar s s \rangle \langle g_s \bar s \sigma G s \rangle \over 3 \pi^2}
-{\langle g_s \bar s \sigma G s \rangle^2 \over 24 \pi^2} \Bigg ] e^{-s/M^2} ds
\\ &&
+\Big({\langle g_s^2 GG \rangle m_s^2 \langle \bar s s \rangle^2 \over 144\pi^2}
+{\langle g_s^2 GG \rangle \langle \bar s s \rangle \langle g_s \bar s \sigma G s \rangle \over 288 \pi^2}
+{4 m_s \langle \bar s s \rangle^2 \langle g_s \bar s \sigma G s \rangle \over 3}
-{5 m_s^2 \langle g_s \bar s \sigma G s \rangle^2 \over 24\pi^2}\Big) \, ,
\\\nonumber
\Pi_{1414} &=& \int^{s_0}_{16 m_s^2} \Bigg [
{ s^5 \over 102400 \pi^6}
-{m_s^2 s^4 \over 1280 \pi^6}
+{\langle g_s^2 GG \rangle s^3\over 12288 \pi^6 }
+\Big(-{\langle g_s^2 GG \rangle m_s^2 \over 768 \pi^6}
+{\langle \bar s s \rangle^2\over 12 \pi^2}
-{3 m_s \langle g_s \bar s \sigma G s \rangle \over 64 \pi^4}\Big)s^2
\\ \nonumber &&
+ \Big(-{\langle g_s^2 GG \rangle m_s \langle \bar s s \rangle \over 384 \pi^4}
-{ m_s^2 \langle \bar s s \rangle^2 \over 24 \pi^2}
+ {7 \langle \bar s s \rangle \langle g_s \bar s \sigma G s \rangle \over 48 \pi^2}\Big)s
+{\langle g_s^2 GG \rangle \langle \bar s s \rangle^2 \over72\pi^2}
\\ \nonumber &&
-{\langle g_s^2 GG \rangle m_s \langle g_s \bar s \sigma G s \rangle\over96\pi^4}
-{3 m_s^2 \langle \bar s s \rangle \langle g_s \bar s \sigma G s \rangle \over 2\pi^2}
-{\langle g_s \bar s \sigma G s \rangle^2 \over 48 \pi^2} \Bigg ] e^{-s/M^2} ds
\\  &&
+\Big({\langle g_s^2 GG \rangle m_s^2 \langle \bar s s \rangle^2\over 192 \pi^2}
+{\langle g_s^2 GG \rangle \langle \bar s s \rangle \langle g_s \bar s \sigma G s \rangle \over 144 \pi^2}
+{29 m_s \langle \bar s s \rangle^2 \langle g_s \bar s \sigma G s \rangle \over 36}
-{7 m_s^2 \langle g_s \bar s \sigma G s \rangle^2 \over 48\pi^2}\Big) \, ,
\\\nonumber
\Pi_{1515} &=& \int^{s_0}_{16 m_s^2} \Bigg [
{ m_s^2 s^4 \over 3840 \pi^6}
-{ m_s^4 s^3 \over 96 \pi^6 }
+\Big(-{\langle g_s^2 GG \rangle m_s^2 \over 1536 \pi^6}
+{ m_s \langle g_s \bar s \sigma G s \rangle \over 24 \pi^4}\Big)s^2
+ \Big({m_s^2 \langle \bar s s \rangle^2 \over 3 \pi^2}
-{\langle \bar s s \rangle \langle g_s \bar s \sigma G s \rangle \over 6 \pi^2}\Big)s
\\ \nonumber &&
-{\langle g_s^2 GG \rangle m_s \langle g_s \bar s \sigma G s \rangle \over 192\pi^4}
+{m_s^2 \langle \bar s s \rangle \langle g_s \bar s \sigma G s \rangle \over 3 \pi^2}
-{\langle g_s \bar s \sigma G s \rangle^2 \over 24 \pi^2} \Bigg ] e^{-s/M^2} ds
\\ &&
+\Big({\langle g_s^2 GG \rangle m_s^2 \langle \bar s s \rangle^2 \over 144\pi^2}
+{\langle g_s^2 GG \rangle \langle \bar s s \rangle \langle g_s \bar s \sigma G s \rangle \over 288 \pi^2}
+{4 m_s \langle \bar s s \rangle^2 \langle g_s \bar s \sigma G s \rangle \over 3}
-{5 m_s^2 \langle g_s \bar s \sigma G s \rangle^2 \over 24\pi^2}\Big) \, ,
\\\nonumber
\Pi_{1616} &=& \int^{s_0}_{16 m_s^2} \Bigg [
{ s^5 \over 102400 \pi^6}
-{m_s^2 s^4 \over 1280 \pi^6}
+{\langle g_s^2 GG \rangle s^3 \over 12288 \pi^6 }
+\Big(-{\langle g_s^2 GG \rangle m_s^2 \over 512 \pi^6}
+{\langle \bar s s \rangle^2\over 12 \pi^2}
-{3 m_s \langle g_s \bar s \sigma G s \rangle \over 64 \pi^4}\Big)s^2
\\ \nonumber &&
+ \Big({\langle g_s^2 GG \rangle m_s \langle \bar s s \rangle \over 384 \pi^4}
-{ m_s^2 \langle \bar s s \rangle^2 \over 24 \pi^2}
+ {3 \langle \bar s s \rangle \langle g_s \bar s \sigma G s \rangle \over 16 \pi^2}\Big)s
+{\langle g_s^2 GG \rangle \langle \bar s s \rangle^2 \over72\pi^2}
-{\langle g_s^2 GG \rangle m_s \langle g_s \bar s \sigma G s \rangle\over96\pi^4}
\\ \nonumber &&
-{17 m_s^2 \langle \bar s s \rangle \langle g_s \bar s \sigma G s \rangle \over 12\pi^2}
+{\langle g_s \bar s \sigma G s \rangle^2 \over 48 \pi^2} \Bigg ] e^{-s/M^2} ds
+\Big({\langle g_s^2 GG \rangle m_s^2 \langle \bar s s \rangle^2\over 192 \pi^2}
\\ &&
+{\langle g_s^2 GG \rangle \langle \bar s s \rangle \langle g_s \bar s \sigma G s \rangle \over 96 \pi^2}
+{3 m_s \langle \bar s s \rangle^2 \langle g_s \bar s \sigma G s \rangle \over 4}
-{m_s^2 \langle g_s \bar s \sigma G s \rangle^2 \over 6\pi^2}\Big) \, ,
\\\nonumber
\Pi_{1717} &=& \int^{s_0}_{16 m_s^2} \Bigg [
{ 17 s^5 \over 6451200 \pi^6}
-{m_s^2 s^4 \over 5760 \pi^6}
+\Big(-{\langle g_s^2 GG \rangle \over 61440 \pi^6 }
-{m_s\langle \bar s s \rangle\over 720 \pi^4}\Big)s^3
+\Big({\langle g_s^2 GG \rangle m_s^2 \over 12288 \pi^6}
+{\langle \bar s s \rangle^2\over 36 \pi^2}
\\ \nonumber &&
-{31 m_s \langle g_s \bar s \sigma G s \rangle \over 4608 \pi^4}\Big)s^2
+ \Big({\langle g_s^2 GG \rangle m_s \langle \bar s s \rangle \over 768 \pi^4}
-{5 m_s^2 \langle \bar s s \rangle^2 \over 36 \pi^2}
+ { \langle \bar s s \rangle \langle g_s \bar s \sigma G s \rangle \over 27 \pi^2}\Big)s
\\ \nonumber &&
-{5 \langle g_s^2 GG \rangle \langle \bar s s \rangle^2 \over3456\pi^2}
-{\langle g_s^2 GG \rangle m_s \langle g_s \bar s \sigma G s \rangle\over1536\pi^4}
-{17 m_s^2 \langle \bar s s \rangle \langle g_s \bar s \sigma G s \rangle \over 144\pi^2}
-{\langle g_s \bar s \sigma G s \rangle^2 \over 288 \pi^2} \Bigg ] e^{-s/M^2} ds
\\  &&
+\Big(-{\langle g_s^2 GG \rangle m_s^2 \langle \bar s s \rangle^2\over 3456 \pi^2}
+{5\langle g_s^2 GG \rangle \langle \bar s s \rangle \langle g_s \bar s \sigma G s \rangle \over 6912 \pi^2}
+{ m_s \langle \bar s s \rangle^2 \langle g_s \bar s \sigma G s \rangle \over 6}
-{m_s^2 \langle g_s \bar s \sigma G s \rangle^2 \over 16\pi^2}\Big) \, ,
\\\nonumber
\Pi_{1818} &=& \int^{s_0}_{16 m_s^2} \Bigg [
{ 13 s^5 \over 12902400 \pi^6}
-{m_s^2 s^4 \over 18432 \pi^6}
+\Big({\langle g_s^2 GG \rangle \over 368640 \pi^6 }
-{m_s\langle \bar s s \rangle\over 960 \pi^4}\Big)s^3
+\Big(-{\langle g_s^2 GG \rangle m_s^2 \over 73728 \pi^6}
+{\langle \bar s s \rangle^2\over 72 \pi^2}
\\ \nonumber &&
-{11 m_s \langle g_s \bar s \sigma G s \rangle \over 3072 \pi^4}\Big)s^2
+ \Big(-{\langle g_s^2 GG \rangle m_s \langle \bar s s \rangle \over 1152 \pi^4}
-{ m_s^2 \langle \bar s s \rangle^2 \over 12 \pi^2}
+ {47 \langle \bar s s \rangle \langle g_s \bar s \sigma G s \rangle \over 1728 \pi^2}\Big)s
\\ \nonumber &&
+{5 \langle g_s^2 GG \rangle \langle \bar s s \rangle^2 \over3456\pi^2}
-{\langle g_s^2 GG \rangle m_s \langle g_s \bar s \sigma G s \rangle\over1152\pi^4}
-{7 m_s^2 \langle \bar s s \rangle \langle g_s \bar s \sigma G s \rangle \over 72\pi^2}
+{5\langle g_s \bar s \sigma G s \rangle^2 \over 1152 \pi^2} \Bigg ] e^{-s/M^2} ds
\\ &&
+\Big(-{\langle g_s^2 GG \rangle m_s^2 \langle \bar s s \rangle^2\over 1728 \pi^2}
+{\langle g_s^2 GG \rangle \langle \bar s s \rangle \langle g_s \bar s \sigma G s \rangle \over 2304 \pi^2}
+{61 m_s \langle \bar s s \rangle^2 \langle g_s \bar s \sigma G s \rangle \over 432}
-{29 m_s^2 \langle g_s \bar s \sigma G s \rangle^2 \over 576\pi^2}\Big) \, ,
\\\nonumber
\Pi_{1919} &=& \int^{s_0}_{16 m_s^2} \Bigg [
{ 17 s^5 \over 6451200 \pi^6}
-{m_s^2 s^4 \over 5760 \pi^6}
+\Big(-{\langle g_s^2 GG \rangle \over 61440 \pi^6 }
-{m_s\langle \bar s s \rangle\over 720 \pi^4}\Big)s^3
+\Big({13\langle g_s^2 GG \rangle m_s^2 \over 36864 \pi^6}
+{\langle \bar s s \rangle^2\over 36 \pi^2}
\\ \nonumber &&
-{41 m_s \langle g_s \bar s \sigma G s \rangle \over 4608 \pi^4}\Big)s^2
+ \Big(-{11\langle g_s^2 GG \rangle m_s \langle \bar s s \rangle \over 6912 \pi^4}
-{5 m_s^2 \langle \bar s s \rangle^2 \over 36 \pi^2}
+ {13 \langle \bar s s \rangle \langle g_s \bar s \sigma G s \rangle \over 216 \pi^2}\Big)s
\\ \nonumber &&
+{5 \langle g_s^2 GG \rangle \langle \bar s s \rangle^2 \over3456\pi^2}
-{\langle g_s^2 GG \rangle m_s \langle g_s \bar s \sigma G s \rangle\over1536\pi^4}
-{23 m_s^2 \langle \bar s s \rangle \langle g_s \bar s \sigma G s \rangle \over 144\pi^2}
+{\langle g_s \bar s \sigma G s \rangle^2 \over 72 \pi^2} \Bigg ] e^{-s/M^2} ds
\\ &&
+\Big(-{\langle g_s^2 GG \rangle m_s^2 \langle \bar s s \rangle^2\over 3456 \pi^2}
+{5\langle g_s^2 GG \rangle \langle \bar s s \rangle \langle g_s \bar s \sigma G s \rangle \over 6912 \pi^2}
+{ m_s \langle \bar s s \rangle^2 \langle g_s \bar s \sigma G s \rangle \over 6}
-{ m_s^2 \langle g_s \bar s \sigma G s \rangle^2 \over 16\pi^2}\Big) \, ,
\\\nonumber
\Pi_{2020} &=& \int^{s_0}_{16 m_s^2} \Bigg [
{ 13 s^5 \over 12902400 \pi^6}
-{m_s^2 s^4 \over 18432 \pi^6}
+\Big({\langle g_s^2 GG \rangle \over 368640 \pi^6 }
-{m_s\langle \bar s s \rangle\over 960 \pi^4}\Big)s^3
+\Big({\langle g_s^2 GG \rangle m_s^2 \over 73728 \pi^6}
+{\langle \bar s s \rangle^2\over 72 \pi^2}
\\ \nonumber &&
-{35 m_s \langle g_s \bar s \sigma G s \rangle \over 9216 \pi^4}\Big)s^2
+ \Big(-{\langle g_s^2 GG \rangle m_s \langle \bar s s \rangle \over 864 \pi^4}
-{ m_s^2 \langle \bar s s \rangle^2 \over 12 \pi^2}
+ {55 \langle \bar s s \rangle \langle g_s \bar s \sigma G s \rangle \over 1728 \pi^2}\Big)s
\\ \nonumber &&
+{7 \langle g_s^2 GG \rangle \langle \bar s s \rangle^2 \over3456\pi^2}
-{\langle g_s^2 GG \rangle m_s \langle g_s \bar s \sigma G s \rangle\over1152\pi^4}
-{31 m_s^2 \langle \bar s s \rangle \langle g_s \bar s \sigma G s \rangle \over 288\pi^2}
+{11\langle g_s \bar s \sigma G s \rangle^2 \over 1152 \pi^2} \Bigg ] e^{-s/M^2} ds
\\ &&
+\Big(-{\langle g_s^2 GG \rangle m_s^2 \langle \bar s s \rangle^2\over 1728 \pi^2}
+{7\langle g_s^2 GG \rangle \langle \bar s s \rangle \langle g_s \bar s \sigma G s \rangle \over 6912 \pi^2}
+{ 61m_s \langle \bar s s \rangle^2 \langle g_s \bar s \sigma G s \rangle \over 432}
-{31 m_s^2 \langle g_s \bar s \sigma G s \rangle^2 \over 576\pi^2}\Big) \, ,
\\\nonumber
\Pi_{2121} &=& \int^{s_0}_{16 m_s^2} \Bigg [
{ s^5 \over 307200 \pi^6}
-{11 m_s^2 s^4 \over 64512 \pi^6}
+\Big(-{7\langle g_s^2 GG \rangle \over 3317760 \pi^6 }
-{23 m_s\langle \bar s s \rangle\over 8640 \pi^4}\Big)s^3
+\Big(-{7\langle g_s^2 GG \rangle m_s^2 \over 184320 \pi^6}
+{\langle \bar s s \rangle^2\over 30 \pi^2}
\\ \nonumber &&
-{7 m_s \langle g_s \bar s \sigma G s \rangle \over 960 \pi^4}\Big)s^2
+ \Big(-{\langle g_s^2 GG \rangle m_s \langle \bar s s \rangle \over 864 \pi^4}
-{ m_s^2 \langle \bar s s \rangle^2 \over 24 \pi^2}
+ {7 \langle \bar s s \rangle \langle g_s \bar s \sigma G s \rangle \over 288 \pi^2}\Big)s
\\ \nonumber &&
+{\langle g_s^2 GG \rangle \langle \bar s s \rangle^2 \over864\pi^2}
-{17 \langle g_s^2 GG \rangle m_s \langle g_s \bar s \sigma G s \rangle\over6912\pi^4}
-{25 m_s^2 \langle \bar s s \rangle \langle g_s \bar s \sigma G s \rangle \over 216\pi^2}
-{17\langle g_s \bar s \sigma G s \rangle^2 \over 864 \pi^2} \Bigg ] e^{-s/M^2} ds
\\ &&
+\Big({\langle g_s^2 GG \rangle m_s^2 \langle \bar s s \rangle^2\over 3456 \pi^2}
-{5\langle g_s^2 GG \rangle \langle \bar s s \rangle \langle g_s \bar s \sigma G s \rangle \over 3456 \pi^2}
+{ 7m_s \langle \bar s s \rangle^2 \langle g_s \bar s \sigma G s \rangle \over 9}
-{43 m_s^2 \langle g_s \bar s \sigma G s \rangle^2 \over 288\pi^2}\Big) \, ,
\\\nonumber
\Pi_{2222} &=& \int^{s_0}_{16 m_s^2} \Bigg [
{ 3 s^5 \over 1433600 \pi^6}
-{11m_s^2 s^4 \over 92160 \pi^6}
+\Big(-{\langle g_s^2 GG \rangle \over 3317760 \pi^6 }
-{11m_s\langle \bar s s \rangle\over 5760 \pi^4}\Big)s^3
+\Big({\langle g_s^2 GG \rangle m_s^2 \over 11520 \pi^6}
+{\langle \bar s s \rangle^2\over 36 \pi^2}
\\ \nonumber &&
-{209 m_s \langle g_s \bar s \sigma G s \rangle \over 23040 \pi^4}\Big)s^2
+ \Big(-{7\langle g_s^2 GG \rangle m_s \langle \bar s s \rangle \over 3456 \pi^4}
-{7 m_s^2 \langle \bar s s \rangle^2 \over 48 \pi^2}
+ { \langle \bar s s \rangle \langle g_s \bar s \sigma G s \rangle \over 16 \pi^2}\Big)s
\\ \nonumber &&
+{\langle g_s^2 GG \rangle \langle \bar s s \rangle^2 \over324\pi^2}
-{7\langle g_s^2 GG \rangle m_s \langle g_s \bar s \sigma G s \rangle\over3456\pi^4}
-{7 m_s^2 \langle \bar s s \rangle \langle g_s \bar s \sigma G s \rangle \over 27\pi^2}
+{19\langle g_s \bar s \sigma G s \rangle^2 \over 1728 \pi^2} \Bigg ] e^{-s/M^2} ds
\\ &&
+\Big(-{\langle g_s^2 GG \rangle m_s^2 \langle \bar s s \rangle^2\over 3456 \pi^2}
+{5\langle g_s^2 GG \rangle \langle \bar s s \rangle \langle g_s \bar s \sigma G s \rangle \over 3456 \pi^2}
+{ 59m_s \langle \bar s s \rangle^2 \langle g_s \bar s \sigma G s \rangle \over 216}
-{ m_s^2 \langle g_s \bar s \sigma G s \rangle^2 \over 12\pi^2}\Big) \, ,
\\\nonumber
\Pi_{2323} &=& \int^{s_0}_{16 m_s^2} \Bigg [
{ s^5 \over 307200 \pi^6}
-{11 m_s^2 s^4 \over 64512 \pi^6}
+\Big(-{7\langle g_s^2 GG \rangle \over 3317760 \pi^6 }
-{23 m_s\langle \bar s s \rangle\over 8640 \pi^4}\Big)s^3
+\Big(-{7\langle g_s^2 GG \rangle m_s^2 \over 184320 \pi^6}
+{\langle \bar s s \rangle^2\over 30 \pi^2}
\\ \nonumber &&
-{7 m_s \langle g_s \bar s \sigma G s \rangle \over 960 \pi^4}\Big)s^2
+ \Big(-{\langle g_s^2 GG \rangle m_s \langle \bar s s \rangle \over 864 \pi^4}
-{ m_s^2 \langle \bar s s \rangle^2 \over 24 \pi^2}
+ {17 \langle \bar s s \rangle \langle g_s \bar s \sigma G s \rangle \over 288 \pi^2}\Big)s
\\ \nonumber &&
+{13\langle g_s^2 GG \rangle \langle \bar s s \rangle^2 \over2592\pi^2}
-{17 \langle g_s^2 GG \rangle m_s \langle g_s \bar s \sigma G s \rangle\over6912\pi^4}
-{35 m_s^2 \langle \bar s s \rangle \langle g_s \bar s \sigma G s \rangle \over 216\pi^2}
+{23\langle g_s \bar s \sigma G s \rangle^2 \over 864 \pi^2} \Bigg ] e^{-s/M^2} ds
\\&&
+\Big({\langle g_s^2 GG \rangle m_s^2 \langle \bar s s \rangle^2\over 3456 \pi^2}
+{5\langle g_s^2 GG \rangle \langle \bar s s \rangle \langle g_s \bar s \sigma G s \rangle \over 1152 \pi^2}
+{7 m_s \langle \bar s s \rangle^2 \langle g_s \bar s \sigma G s \rangle \over 9}
-{53 m_s^2 \langle g_s \bar s \sigma G s \rangle^2 \over 288\pi^2}\Big) \, ,
\\\nonumber
\Pi_{2424} &=& \int^{s_0}_{16 m_s^2} \Bigg [
{3 s^5 \over 1433600 \pi^6}
-{11 m_s^2 s^4 \over 92160 \pi^6}
+\Big(-{\langle g_s^2 GG \rangle \over 3317760 \pi^6 }
-{11 m_s\langle \bar s s \rangle\over 5760 \pi^4}\Big)s^3
+\Big({\langle g_s^2 GG \rangle m_s^2 \over 11520 \pi^6}
+{\langle \bar s s \rangle^2\over 36 \pi^2}
\\ \nonumber &&
-{209 m_s \langle g_s \bar s \sigma G s \rangle \over 23040 \pi^4}\Big)s^2
+ \Big(-{7\langle g_s^2 GG \rangle m_s \langle \bar s s \rangle \over 3456 \pi^4}
-{ 7m_s^2 \langle \bar s s \rangle^2 \over 48 \pi^2}
+ { \langle \bar s s \rangle \langle g_s \bar s \sigma G s \rangle \over 16 \pi^2}\Big)s
\\ \nonumber &&
+{\langle g_s^2 GG \rangle \langle \bar s s \rangle^2 \over324\pi^2}
-{7 \langle g_s^2 GG \rangle m_s \langle g_s \bar s \sigma G s \rangle\over3456\pi^4}
-{7 m_s^2 \langle \bar s s \rangle \langle g_s \bar s \sigma G s \rangle \over 27\pi^2}
+{19\langle g_s \bar s \sigma G s \rangle^2 \over 1728 \pi^2} \Bigg ] e^{-s/M^2} ds
\\&&
+\Big(-{\langle g_s^2 GG \rangle m_s^2 \langle \bar s s \rangle^2\over 3456 \pi^2}
+{5\langle g_s^2 GG \rangle \langle \bar s s \rangle \langle g_s \bar s \sigma G s \rangle \over 3456 \pi^2}
+{59 m_s \langle \bar s s \rangle^2 \langle g_s \bar s \sigma G s \rangle \over 216}
-{ m_s^2 \langle g_s \bar s \sigma G s \rangle^2 \over 12\pi^2}\Big) \, .
\end{eqnarray}

\end{widetext}

%

%

\end{document}